\def\standardrisposta{s }\def\reducedrisposta{r }
\def\mplarisposta{mpla }\def\zerorisposta{z }\def\bigrisposta{big }
\def\doublerisposta{d }\def\cartarisposta{e }\def\amsrisposta{y }
\newcount\ingrandimento \newcount\sinnota \newcount\dimnota
\newcount\unoduecol \newdimen\collhsize \newdimen\tothsize
\newdimen\fullhsize \newcount\controllorisposta \sinnota=1
\newskip\infralinea  \global\controllorisposta=0
\immediate\write16 { ********  Welcome to PANDA macros (Plain TeX,
AP, 1991) ******** }
\immediate\write16 { You'll have to answer a few questions in
lowercase.}
\message{>  Do you want it in double-page (d), reduced (r)
or standard format (s) ? }\read-1 to\risposta
\message{>  Do you want it in USA A4 (u) or EUROPEAN A4
(e) paper size ? }\read-1 to\srisposta
%
%
%
%
\def\arisposta{y }
\ifx\risposta\standardrisposta \ingrandimento=1200
\message {>> This will come out UNREDUCED << }
\dimnota=2 \unoduecol=1 \global\controllorisposta=1 \fi
\ifx\risposta\bigrisposta \ingrandimento=1440
\message {>> This will come out ENLARGED << }
\dimnota=2 \unoduecol=1 \global\controllorisposta=1 \fi
\ifx\risposta\reducedrisposta \ingrandimento=1095 \dimnota=1
\unoduecol=1  \global\controllorisposta=1
\message {>> This will come out REDUCED << } \fi
\ifx\risposta\doublerisposta \ingrandimento=1000 \dimnota=2
\unoduecol=2  \message {>> You must print this in
LANDSCAPE orientation << } \global\controllorisposta=1 \fi
\ifx\risposta\mplarisposta \ingrandimento=1000 \dimnota=1
\message {>> Mod. Phys. Lett. A format << }
\unoduecol=1 \global\controllorisposta=1 \fi
\ifx\risposta\zerorisposta \ingrandimento=1000 \dimnota=2
\message {>> Zero Magnification format << }
\unoduecol=1 \global\controllorisposta=1 \fi
\ifnum\controllorisposta=0  \ingrandimento=1200
\message {>>> ERROR IN INPUT, I ASSUME STANDARD
UNREDUCED FORMAT <<< }  \dimnota=2 \unoduecol=1 \fi
\magnification=\ingrandimento
%
%
%
%
\newdimen\eucolumnsize \newdimen\eudoublehsize \newdimen\eudoublevsize
\newdimen\uscolumnsize \newdimen\usdoublehsize \newdimen\usdoublevsize
\newdimen\eusinglehsize \newdimen\eusinglevsize \newdimen\ussinglehsize
\newskip\standardbaselineskip \newdimen\ussinglevsize
\newskip\reducedbaselineskip \newskip\doublebaselineskip
\newskip\bigbaselineskip
\eucolumnsize=12.0truecm    
\eudoublehsize=25.5truecm   
\eudoublevsize=6.5truein    
\uscolumnsize=4.4truein     
\usdoublehsize=9.4truein    
\usdoublevsize=6.8truein    
\eusinglehsize=6.3truein    
\eusinglevsize=24truecm     
\ussinglehsize=6.5truein    
\ussinglevsize=8.9truein    
\bigbaselineskip=18pt plus.2pt       
\standardbaselineskip=16pt plus.2pt  
\reducedbaselineskip=14pt plus.2pt   
\doublebaselineskip=12pt plus.2pt    
%
%
\def\Portoffset{}
\def\Landoffset{\hoffset=-.140truein}
\ifx\risposta\mplarisposta \def\Portoffset{\hoffset=1.9truecm
\voffset=1.4truecm} \fi
%
%
\def\Landspec{}
\tolerance=10000
\parskip=0pt plus2pt  \leftskip=0pt \rightskip=0pt
%
%
\ifx\risposta\bigrisposta      \infralinea=\bigbaselineskip \fi
\ifx\risposta\standardrisposta \infralinea=\standardbaselineskip \fi
\ifx\risposta\reducedrisposta  \infralinea=\reducedbaselineskip \fi
\ifx\risposta\doublerisposta   \infralinea=\doublebaselineskip \fi
\ifx\risposta\mplarisposta     \infralinea=13pt \fi
\ifx\risposta\zerorisposta     \infralinea=12pt plus.2pt\fi
\ifnum\controllorisposta=0    \infralinea=\standardbaselineskip \fi
\ifx\risposta\doublerisposta   \Landoffset \else \Portoffset \fi
\ifx\risposta\doublerisposta \ifx\srisposta\cartarisposta
\tothsize=\eudoublehsize \collhsize=\eucolumnsize
\vsize=\eudoublevsize  \else  \tothsize=\usdoublehsize
\collhsize=\uscolumnsize \vsize=\usdoublevsize \fi \else
\ifx\srisposta\cartarisposta \tothsize=\eusinglehsize
\vsize=\eusinglevsize \else  \tothsize=\ussinglehsize
\vsize=\ussinglevsize \fi \collhsize=4.4truein \fi
\ifx\risposta\mplarisposta \tothsize=5.0truein
\vsize=7.8truein \collhsize=4.4truein \fi
%
%
%
%
\newcount\contaeuler \newcount\contacyrill \newcount\contaams
\newcount\contasym
\font\ninerm=cmr9  \font\eightrm=cmr8  \font\sixrm=cmr6
\font\ninei=cmmi9  \font\eighti=cmmi8  \font\sixi=cmmi6
\font\ninesy=cmsy9  \font\eightsy=cmsy8  \font\sixsy=cmsy6
\font\ninebf=cmbx9  \font\eightbf=cmbx8  \font\sixbf=cmbx6
\font\ninett=cmtt9  \font\eighttt=cmtt8  \font\nineit=cmti9
\font\eightit=cmti8 \font\ninesl=cmsl9  \font\eightsl=cmsl8
\skewchar\ninei='177 \skewchar\eighti='177 \skewchar\sixi='177
\skewchar\ninesy='60 \skewchar\eightsy='60 \skewchar\sixsy='60
\hyphenchar\ninett=-1 \hyphenchar\eighttt=-1 \hyphenchar\tentt=-1
\def\bfmath{\cmmib}                 
\font\tencmmib=cmmib10  \newfam\cmmibfam  \skewchar\tencmmib='177
\font\tencmbsy=cmbsy10  \newfam\cmbsyfam  \skewchar\tencmbsy='60
\def\scaps{\cmcsc}                 
\font\tencmcsc=cmcsc10  \newfam\cmcscfam
\ifnum\ingrandimento=1095 
 
\font\bfone=cmbx10 at 10.95pt

\font\capsone=cmcsc10 at 10.95pt 

\else  
 
\font\bfone=cmbx10 at 12pt

\font\capsone=cmcsc10 at 12pt 
\fi
\def\chapterfont#1{\xdef\ttaarr{#1}}
\def\sectionfont#1{\xdef\ppaarr{#1}}
\def\ttaarr{\bf}                
\def\ppaarr{\sl}                

%
%
%
\newfam\eufmfam \newfam\msamfam \newfam\msbmfam \newfam\eufbfam
\def\Loadeulerfonts{\global\contaeuler=1 \ifx\arisposta\amsrisposta
\font\teneufm=eufm10              
\font\eighteufm=eufm8 \font\nineeufm=eufm9 \font\sixeufm=eufm6
\font\seveneufm=eufm7  \font\fiveeufm=eufm5
\font\teneufb=eufb10              
\font\eighteufb=eufb8 \font\nineeufb=eufb9 \font\sixeufb=eufb6
\font\seveneufb=eufb7  \font\fiveeufb=eufb5
\font\teneurm=eurm10              
\font\eighteurm=eurm8 \font\nineeurm=eurm9
\font\teneurb=eurb10              
\font\eighteurb=eurb8 \font\nineeurb=eurb9
\font\teneusm=eusm10              
\font\eighteusm=eusm8 \font\nineeusm=eusm9
\font\teneusb=eusb10              
\font\eighteusb=eusb8 \font\nineeusb=eusb9
\else \def\eufm{\tt} \def\eufb{\tt} \def\eurm{\tt} \def\eurb{\tt}
\def\eusm{\tt} \def\eusb{\tt}    \fi}
\def\loadamsmath{\global\contaams=1 \ifx\arisposta\amsrisposta
\font\tenmsam=msam10 \font\ninemsam=msam9 \font\eightmsam=msam8
\font\sevenmsam=msam7 \font\sixmsam=msam6 \font\fivemsam=msam5
\font\tenmsbm=msbm10 \font\ninemsbm=msbm9 \font\eightmsbm=msbm8
\font\sevenmsbm=msbm7 \font\sixmsbm=msbm6 \font\fivemsbm=msbm5
\else \def\msbm{\bf} \fi \def\Bbb{\msbm} \def\symbl{\msam} \tenpoint}
\def\loadcyrill{\global\contacyrill=1 \ifx\arisposta\amsrisposta
\font\tenwncyr=wncyr10 \font\ninewncyr=wncyr9 \font\eightwncyr=wncyr8
\font\tenwncyb=wncyr10 \font\ninewncyb=wncyr9 \font\eightwncyb=wncyr8
\font\tenwncyi=wncyr10 \font\ninewncyi=wncyr9 \font\eightwncyi=wncyr8
\else \def\cyrill{\sl} \def\cyrilb{\sl} \def\cyrili{\sl} \fi\tenpoint}
\catcode`\@=11
\def\undefine#1{\let#1\undefined}
\def\newsymbol#1#2#3#4#5{\let\next@\relax
 \ifnum#2=\@ne\let\next@\msafam@\else
 \ifnum#2=\tw@\let\next@\msbfam@\fi\fi
 \mathchardef#1="#3\next@#4#5}
\def\mathhexbox@#1#2#3{\relax
 \ifmmode\mathpalette{}{\m@th\mathchar"#1#2#3}%
 \else\leavevmode\hbox{$\m@th\mathchar"#1#2#3$}\fi}
\def\hexnumber@#1{\ifcase#1 0\or 1\or 2\or 3\or 4\or 5\or 6\or 7\or 8\or
9\or A\or B\or C\or D\or E\or F\fi}
\edef\msafam@{\hexnumber@\msamfam}
\edef\msbfam@{\hexnumber@\msbmfam}
\mathchardef\dabar@"0\msafam@39
\catcode`\@=12
\def\loadamssym{\ifx\arisposta\amsrisposta  \ifnum\contaams=1
\global\contasym=1
\catcode`\@=11
\def\dashrightarrow{\mathrel{\dabar@\dabar@\mathchar"0\msafam@4B}}
\def\dashleftarrow{\mathrel{\mathchar"0\msafam@4C\dabar@\dabar@}}
\let\dasharrow\dashrightarrow
\def\ulcorner{\delimiter"4\msafam@70\msafam@70 }
\def\urcorner{\delimiter"5\msafam@71\msafam@71 }
\def\llcorner{\delimiter"4\msafam@78\msafam@78 }
\def\lrcorner{\delimiter"5\msafam@79\msafam@79 }
\def\yen{{\mathhexbox@\msafam@55}}
\def\checkmark{{\mathhexbox@\msafam@58 }}
\def\circledR{{\mathhexbox@\msafam@72 }}
\def\maltese{{\mathhexbox@\msafam@7A }}
\catcode`\@=12
\input amssym.tex     \else
\message{Panda error - First you have to use loadamsmath !!!!} \fi
\else \message{Panda error - You need the AMSFonts for these symbols
!!!!}\fi}
\ifx\arisposta\amsrisposta
\font\sevenex=cmex7               
\font\eightex=cmex8  \font\nineex=cmex9
\font\ninecmmib=cmmib9   \font\eightcmmib=cmmib8
\font\sevencmmib=cmmib7 \font\sixcmmib=cmmib6
\font\fivecmmib=cmmib5   \skewchar\ninecmmib='177
\skewchar\eightcmmib='177  \skewchar\sevencmmib='177
\skewchar\sixcmmib='177   \skewchar\fivecmmib='177
\font\ninecmbsy=cmbsy9    \font\eightcmbsy=cmbsy8
\font\sevencmbsy=cmbsy7  \font\sixcmbsy=cmbsy6
\font\fivecmbsy=cmbsy5   \skewchar\ninecmbsy='60
\skewchar\eightcmbsy='60  \skewchar\sevencmbsy='60
\skewchar\sixcmbsy='60    \skewchar\fivecmbsy='60
\font\ninecmcsc=cmcsc9    \font\eightcmcsc=cmcsc8     \else
\def\cmmib{\fam\cmmibfam\tencmmib}\textfont\cmmibfam=\tencmmib
\scriptfont\cmmibfam=\tencmmib \scriptscriptfont\cmmibfam=\tencmmib
\def\cmbsy{\fam\cmbsyfam\tencmbsy} \textfont\cmbsyfam=\tencmbsy
\scriptfont\cmbsyfam=\tencmbsy \scriptscriptfont\cmbsyfam=\tencmbsy
\scriptfont\cmcscfam=\tencmcsc \scriptscriptfont\cmcscfam=\tencmcsc
\def\cmcsc{\fam\cmcscfam\tencmcsc} \textfont\cmcscfam=\tencmcsc \fi
\catcode`@=11
\newskip\ttglue
\gdef\tenpoint{\def\rm{\fam0\tenrm}
  \textfont0=\tenrm \scriptfont0=\sevenrm \scriptscriptfont0=\fiverm
  \textfont1=\teni \scriptfont1=\seveni \scriptscriptfont1=\fivei
  \textfont2=\tensy \scriptfont2=\sevensy \scriptscriptfont2=\fivesy
  \textfont3=\tenex \scriptfont3=\tenex \scriptscriptfont3=\tenex
  \def\mcal{\fam2 \tensy}  \def\mmit{\fam1 \teni}
  \textfont\itfam=\tenit \def\it{\fam\itfam\tenit}
  \textfont\slfam=\tensl \def\sl{\fam\slfam\tensl}
  \textfont\ttfam=\tentt \scriptfont\ttfam=\eighttt
  \scriptscriptfont\ttfam=\eighttt  \def\tt{\fam\ttfam\tentt}
  \textfont\bffam=\tenbf \scriptfont\bffam=\sevenbf
  \scriptscriptfont\bffam=\fivebf \def\bf{\fam\bffam\tenbf}
     \ifx\arisposta\amsrisposta    \ifnum\contaeuler=1
  \textfont\eufmfam=\teneufm \scriptfont\eufmfam=\seveneufm
  \scriptscriptfont\eufmfam=\fiveeufm \def\eufm{\fam\eufmfam\teneufm}
  \textfont\eufbfam=\teneufb \scriptfont\eufbfam=\seveneufb
  \scriptscriptfont\eufbfam=\fiveeufb \def\eufb{\fam\eufbfam\teneufb}
  \def\eurm{\teneurm} \def\eurb{\teneurb} \def\eusm{\teneusm}
  \def\eusb{\teneusb}    \fi    \ifnum\contaams=1
  \textfont\msamfam=\tenmsam \scriptfont\msamfam=\sevenmsam
  \scriptscriptfont\msamfam=\fivemsam \def\msam{\fam\msamfam\tenmsam}
  \textfont\msbmfam=\tenmsbm \scriptfont\msbmfam=\sevenmsbm
  \scriptscriptfont\msbmfam=\fivemsbm \def\msbm{\fam\msbmfam\tenmsbm}
     \fi      \ifnum\contacyrill=1     \def\cyrill{\tenwncyr}
  \def\cyrilb{\tenwncyb}  \def\cyrili{\tenwncyi}         \fi
  \textfont3=\tenex \scriptfont3=\sevenex \scriptscriptfont3=\sevenex
  \def\cmmib{\fam\cmmibfam\tencmmib} \scriptfont\cmmibfam=\sevencmmib
  \textfont\cmmibfam=\tencmmib  \scriptscriptfont\cmmibfam=\fivecmmib
  \def\cmbsy{\fam\cmbsyfam\tencmbsy} \scriptfont\cmbsyfam=\sevencmbsy
  \textfont\cmbsyfam=\tencmbsy  \scriptscriptfont\cmbsyfam=\fivecmbsy
  \def\cmcsc{\fam\cmcscfam\tencmcsc} \scriptfont\cmcscfam=\eightcmcsc
  \textfont\cmcscfam=\tencmcsc \scriptscriptfont\cmcscfam=\eightcmcsc
     \fi            \tt \ttglue=.5em plus.25em minus.15em
  \normalbaselineskip=12pt
  \setbox\strutbox=\hbox{\vrule height8.5pt depth3.5pt width0pt}
  \let\sc=\eightrm \let\big=\tenbig   \normalbaselines
  \baselineskip=\infralinea  \rm}
\gdef\ninepoint{\def\rm{\fam0\ninerm}
  \textfont0=\ninerm \scriptfont0=\sixrm \scriptscriptfont0=\fiverm
  \textfont1=\ninei \scriptfont1=\sixi \scriptscriptfont1=\fivei
  \textfont2=\ninesy \scriptfont2=\sixsy \scriptscriptfont2=\fivesy
  \textfont3=\tenex \scriptfont3=\tenex \scriptscriptfont3=\tenex
  \def\mcal{\fam2 \ninesy}  \def\mmit{\fam1 \ninei}
  \textfont\itfam=\nineit \def\it{\fam\itfam\nineit}
  \textfont\slfam=\ninesl \def\sl{\fam\slfam\ninesl}
  \textfont\ttfam=\ninett \scriptfont\ttfam=\eighttt
  \scriptscriptfont\ttfam=\eighttt \def\tt{\fam\ttfam\ninett}
  \textfont\bffam=\ninebf \scriptfont\bffam=\sixbf
  \scriptscriptfont\bffam=\fivebf \def\bf{\fam\bffam\ninebf}
     \ifx\arisposta\amsrisposta  \ifnum\contaeuler=1
  \textfont\eufmfam=\nineeufm \scriptfont\eufmfam=\sixeufm
  \scriptscriptfont\eufmfam=\fiveeufm \def\eufm{\fam\eufmfam\nineeufm}
  \textfont\eufbfam=\nineeufb \scriptfont\eufbfam=\sixeufb
  \scriptscriptfont\eufbfam=\fiveeufb \def\eufb{\fam\eufbfam\nineeufb}
  \def\eurm{\nineeurm} \def\eurb{\nineeurb} \def\eusm{\nineeusm}
  \def\eusb{\nineeusb}     \fi   \ifnum\contaams=1
  \textfont\msamfam=\ninemsam \scriptfont\msamfam=\sixmsam
  \scriptscriptfont\msamfam=\fivemsam \def\msam{\fam\msamfam\ninemsam}
  \textfont\msbmfam=\ninemsbm \scriptfont\msbmfam=\sixmsbm
  \scriptscriptfont\msbmfam=\fivemsbm \def\msbm{\fam\msbmfam\ninemsbm}
     \fi       \ifnum\contacyrill=1     \def\cyrill{\ninewncyr}
  \def\cyrilb{\ninewncyb}  \def\cyrili{\ninewncyi}         \fi
  \textfont3=\nineex \scriptfont3=\sevenex \scriptscriptfont3=\sevenex
  \def\cmmib{\fam\cmmibfam\ninecmmib}  \textfont\cmmibfam=\ninecmmib
  \scriptfont\cmmibfam=\sixcmmib \scriptscriptfont\cmmibfam=\fivecmmib
  \def\cmbsy{\fam\cmbsyfam\ninecmbsy}  \textfont\cmbsyfam=\ninecmbsy
  \scriptfont\cmbsyfam=\sixcmbsy \scriptscriptfont\cmbsyfam=\fivecmbsy
  \def\cmcsc{\fam\cmcscfam\ninecmcsc} \scriptfont\cmcscfam=\eightcmcsc
  \textfont\cmcscfam=\ninecmcsc \scriptscriptfont\cmcscfam=\eightcmcsc
     \fi            \tt \ttglue=.5em plus.25em minus.15em
  \normalbaselineskip=11pt
  \setbox\strutbox=\hbox{\vrule height8pt depth3pt width0pt}
  \let\sc=\sevenrm \let\big=\ninebig \normalbaselines\rm}
\gdef\eightpoint{\def\rm{\fam0\eightrm}
  \textfont0=\eightrm \scriptfont0=\sixrm \scriptscriptfont0=\fiverm
  \textfont1=\eighti \scriptfont1=\sixi \scriptscriptfont1=\fivei
  \textfont2=\eightsy \scriptfont2=\sixsy \scriptscriptfont2=\fivesy
  \textfont3=\tenex \scriptfont3=\tenex \scriptscriptfont3=\tenex
  \def\mcal{\fam2 \eightsy}  \def\mmit{\fam1 \eighti}
  \textfont\itfam=\eightit \def\it{\fam\itfam\eightit}
  \textfont\slfam=\eightsl \def\sl{\fam\slfam\eightsl}
  \textfont\ttfam=\eighttt \scriptfont\ttfam=\eighttt
  \scriptscriptfont\ttfam=\eighttt \def\tt{\fam\ttfam\eighttt}
  \textfont\bffam=\eightbf \scriptfont\bffam=\sixbf
  \scriptscriptfont\bffam=\fivebf \def\bf{\fam\bffam\eightbf}
     \ifx\arisposta\amsrisposta   \ifnum\contaeuler=1
  \textfont\eufmfam=\eighteufm \scriptfont\eufmfam=\sixeufm
  \scriptscriptfont\eufmfam=\fiveeufm \def\eufm{\fam\eufmfam\eighteufm}
  \textfont\eufbfam=\eighteufb \scriptfont\eufbfam=\sixeufb
  \scriptscriptfont\eufbfam=\fiveeufb \def\eufb{\fam\eufbfam\eighteufb}
  \def\eurm{\eighteurm} \def\eurb{\eighteurb} \def\eusm{\eighteusm}
  \def\eusb{\eighteusb}       \fi    \ifnum\contaams=1
  \textfont\msamfam=\eightmsam \scriptfont\msamfam=\sixmsam
  \scriptscriptfont\msamfam=\fivemsam \def\msam{\fam\msamfam\eightmsam}
  \textfont\msbmfam=\eightmsbm \scriptfont\msbmfam=\sixmsbm
  \scriptscriptfont\msbmfam=\fivemsbm \def\msbm{\fam\msbmfam\eightmsbm}
     \fi       \ifnum\contacyrill=1     \def\cyrill{\eightwncyr}
  \def\cyrilb{\eightwncyb}  \def\cyrili{\eightwncyi}         \fi
  \textfont3=\eightex \scriptfont3=\sevenex \scriptscriptfont3=\sevenex
  \def\cmmib{\fam\cmmibfam\eightcmmib}  \textfont\cmmibfam=\eightcmmib
  \scriptfont\cmmibfam=\sixcmmib \scriptscriptfont\cmmibfam=\fivecmmib
  \def\cmbsy{\fam\cmbsyfam\eightcmbsy}  \textfont\cmbsyfam=\eightcmbsy
  \scriptfont\cmbsyfam=\sixcmbsy \scriptscriptfont\cmbsyfam=\fivecmbsy
  \def\cmcsc{\fam\cmcscfam\eightcmcsc} \scriptfont\cmcscfam=\eightcmcsc
  \textfont\cmcscfam=\eightcmcsc \scriptscriptfont\cmcscfam=\eightcmcsc
     \fi             \tt \ttglue=.5em plus.25em minus.15em
  \normalbaselineskip=9pt
  \setbox\strutbox=\hbox{\vrule height7pt depth2pt width0pt}
  \let\sc=\sixrm \let\big=\eightbig \normalbaselines\rm }
\gdef\tenbig#1{{\hbox{$\left#1\vbox to8.5pt{}\right.\n@space$}}}
\gdef\ninebig#1{{\hbox{$\textfont0=\tenrm\textfont2=\tensy
   \left#1\vbox to7.25pt{}\right.\n@space$}}}
\gdef\eightbig#1{{\hbox{$\textfont0=\ninerm\textfont2=\ninesy
   \left#1\vbox to6.5pt{}\right.\n@space$}}}
\def\alternativefont#1#2{\ifx\arisposta\amsrisposta \relax \else
\xdef#1{#2} \fi}
\global\contaeuler=0 \global\contacyrill=0 \global\contaams=0
%
%
%
%
\newbox\fotlinebb \newbox\hedlinebb \newbox\leftcolumn
\gdef\makeheadline{\vbox to 0pt{\vskip-22.5pt
     \fullline{\vbox to8.5pt{}\the\headline}\vss}\nointerlineskip}
\gdef\makehedlinebb{\vbox to 0pt{\vskip-22.5pt
     \fullline{\vbox to8.5pt{}\copy\hedlinebb\hfil
     \line{\hfill\the\headline\hfill}}\vss} \nointerlineskip}
\gdef\makefootline{\baselineskip=24pt \fullline{\the\footline}}
\gdef\makefotlinebb{\baselineskip=24pt
    \fullline{\copy\fotlinebb\hfil\line{\hfill\the\footline\hfill}}}
\gdef\doubleformat{\shipout\vbox{\Landspec\makehedlinebb
     \fullline{\box\leftcolumn\hfil\columnbox}\makefotlinebb}
     \advancepageno}
\gdef\columnbox{\leftline{\pagebody}}
\gdef\line#1{\hbox to\hsize{\hskip\leftskip#1\hskip\rightskip}}
\gdef\fullline#1{\hbox to\fullhsize{\hskip\leftskip{#1}%
\hskip\rightskip}}
\gdef\footnote#1{\let\@sf=\empty
	 \ifhmode\edef\#sf{\spacefactor=\the\spacefactor}\/\fi
	 #1\@sf\vfootnote{#1}}
\gdef\vfootnote#1{\insert\footins\bgroup
	 \ifnum\dimnota=1  \eightpoint\fi
	 \ifnum\dimnota=2  \ninepoint\fi
	 \ifnum\dimnota=0  \tenpoint\fi
	 \interlinepenalty=\interfootnotelinepenalty
	 \splittopskip=\ht\strutbox
	 \splitmaxdepth=\dp\strutbox \floatingpenalty=20000
	 \leftskip=\oldssposta \rightskip=\olddsposta
	 \spaceskip=0pt \xspaceskip=0pt
	 \ifnum\sinnota=0   \textindent{#1}\fi
	 \ifnum\sinnota=1   \item{#1}\fi
	 \footstrut\futurelet\next\fo@t}
\gdef\fo@t{\ifcat\bgroup\noexpand\next \let\next\f@@t
	     \else\let\next\f@t\fi \next}
\gdef\f@@t{\bgroup\aftergroup\@foot\let\next}
\gdef\f@t#1{#1\@foot} \gdef\@foot{\strut\egroup}
\gdef\footstrut{\vbox to\splittopskip{}}
\skip\footins=\bigskipamount
\count\footins=1000  \dimen\footins=8in
\catcode`@=12
\tenpoint
\ifnum\unoduecol=1 \hsize=\tothsize   \fullhsize=\tothsize \fi
\ifnum\unoduecol=2 \hsize=\collhsize  \fullhsize=\tothsize \fi
\global\let\lrcol=L      \ifnum\unoduecol=1
\output{\plainoutput{\ifnum\tipbnota=2 \clearnmbnota\fi}} \fi
\ifnum\unoduecol=2 \output{\if L\lrcol
     \global\setbox\leftcolumn=\columnbox
     \global\setbox\fotlinebb=\line{\hfill\the\footline\hfill}
     \global\setbox\hedlinebb=\line{\hfill\the\headline\hfill}
     \advancepageno  \global\let\lrcol=R
     \else  \doubleformat \global\let\lrcol=L \fi
     \ifnum\outputpenalty>-20000 \else\dosupereject\fi
     \ifnum\tipbnota=2\clearnmbnota\fi }\fi
\def\ifdoublepage{\ifnum\unoduecol=2 }
\gdef\yespagenumbers{\footline={\hss\tenrm\folio\hss}}
\gdef\ciao{ \ifnum\fdefcontre=1 \endfdef\fi
     \par\vfill\supereject \ifnum\unoduecol=2
     \if R\lrcol  \headline={}\nopagenumbers\null\vfill\eject
     \fi\fi \end}

\newskip\olddsposta \newskip\oldssposta
\global\oldssposta=\leftskip \global\olddsposta=\rightskip

\def\filldots{\leaders\hbox to 1em{\hss.\hss}\hfill}
\def\inquadrb#1 {\vbox {\hrule  \hbox{\vrule \vbox {\vskip .2cm
    \hbox {\ #1\ } \vskip .2cm } \vrule  }  \hrule} }
 \def\newline{\hfil\break}
\def\jump{\vskip\baselineskip} \newskip\iinnffrr
\def\sjump{\iinnffrr=\baselineskip
	  \divide\iinnffrr by 2 \vskip\iinnffrr}
\def\bjump{\vskip\baselineskip \vskip\baselineskip}
\newcount\nmbnota  \def\clearnmbnota{\global\nmbnota=0}
\newcount\tipbnota \def\letterfootnote{\global\tipbnota=1}

\def\note#1{\global\advance\nmbnota by 1 \ifnum\tipbnota=1
    \footnote{$^{\rm\nttlett}$}{#1} \else {\ifnum\tipbnota=2
    \footnote{$^{\nttsymb}$}{#1}
    \else\footnote{$^{\the\nmbnota}$}{#1}\fi}\fi}
\def\nttlett{\ifcase\nmbnota \or a\or b\or c\or d\or e\or f\or
g\or h\or i\or j\or k\or l\or m\or n\or o\or p\or q\or r\or
s\or t\or u\or v\or w\or y\or x\or z\fi}
\def\nttsymb{\ifcase\nmbnota \or\dag\or\sharp\or\ddag\or\star\or
\natural\or\flat\or\clubsuit\or\diamondsuit\or\heartsuit
\or\spadesuit\fi}   \clearnmbnota
\def\numberfootnote{\global\tipbnota=0} \numberfootnote
\def\setnote#1{\expandafter\xdef\csname#1\endcsname{
\ifnum\tipbnota=1 {\rm\nttlett} \else {\ifnum\tipbnota=2
{\nttsymb} \else \the\nmbnota\fi}\fi} }
\newcount\nbmfig  \def\clearnbmfig{\global\nbmfig=0}
\gdef\figure{\global\advance\nbmfig by 1
      {\rm fig. \the\nbmfig}}   \clearnbmfig
\def\setfig#1{\expandafter\xdef\csname#1\endcsname{fig. \the\nbmfig}}
 \def\endformula{\eqno\numero $$}
 \def\efr{\endformula}
\newcount\frmcount \def\clearfrmcount{\global\frmcount=0}
\def\numero{\global\advance\frmcount by 1   \ifnum\indappcount=0
  {\ifnum\cpcount <1 {\hbox{\rm (\the\frmcount )}}  \else
  {\hbox{\rm (\the\cpcount .\the\frmcount )}} \fi}  \else
  {\hbox{\rm (\applett .\the\frmcount )}} \fi}
\def\nfr{\nameformula}    \def\numali{\numero}
\def\nameformula#1{\global\advance\frmcount by 1%
{\ifnum\indappcount=0%
{\ifnum\cpcount<1\xdef\spzzttrra{(\the\frmcount )}%
\else\xdef\spzzttrra{(\the\cpcount .\the\frmcount )}\fi}%
\else\xdef\spzzttrra{(\applett .\the\frmcount )}\fi}%
\expandafter\xdef\csname#1\endcsname{\spzzttrra}%
\eqno{\ifnum\draftnum=0\hbox{\rm\spzzttrra}\else%
\hbox{$\buildchar{\rm\spzzttrra}{\tt\scriptscriptstyle#1}{}$}\fi}$$}
\def\nameali#1{\global\advance\frmcount by 1%
{\ifnum\indappcount=0%
{\ifnum\cpcount<1\xdef\spzzttrra{(\the\frmcount )}%
\else\xdef\spzzttrra{(\the\cpcount .\the\frmcount )}\fi}%
\else\xdef\spzzttrra{(\applett .\the\frmcount )}\fi}%
\expandafter\xdef\csname#1\endcsname{\spzzttrra}%
\ifnum\draftnum=0\hbox{\rm\spzzttrra}\else%
\hbox{$\buildchar{\rm\spzzttrra}{\tt\scriptscriptstyle#1}{}$}\fi}
\clearfrmcount
\newcount\cpcount \def\clearcpcount{\global\cpcount=0}
\newcount\subcpcount \def\clearsubcpcount{\global\subcpcount=0}
\newcount\appcount \def\clearappcount{\global\appcount=0}
\newcount\indappcount \def\clearindappcount{\indappcount=0}
\newcount\sottoparcount 

\def\applett{\ifcase\appcount  \or {A}\or {B}\or {C}\or
{D}\or {E}\or {F}\or {G}\or {H}\or {I}\or {J}\or {K}\or {L}\or
{M}\or {N}\or {O}\or {P}\or {Q}\or {R}\or {S}\or {T}\or {U}\or
{V}\or {W}\or {X}\or {Y}\or {Z}\fi    \ifnum\appcount<0
\immediate\write16 {Panda ERROR - Appendix: counter "appcount"
out of range}\fi  \ifnum\appcount>26  \immediate\write16 {Panda
ERROR - Appendix: counter "appcount" out of range}\fi}
\clearappcount  \clearindappcount \newcount\connttrre
\def\clearconnttrre{\global\connttrre=0} \newcount\countref
\def\clearcountref{\global\countref=0} \clearcountref
\def\chapter#1{\global\advance\cpcount by 1 \clearfrmcount
		 \goodbreak\null\vbox{\jump\nobreak
		 \clearsubcpcount\clearindappcount
		 \itemitem{\ttaarr\the\cpcount .\qquad}{\ttaarr #1}
		 \par\nobreak\jump\sjump}\nobreak}
\def\section#1{\global\advance\subcpcount by 1 \goodbreak\null
	       \vbox{\sjump\nobreak\ifnum\indappcount=0
		 {\ifnum\cpcount=0 {\itemitem{\ppaarr
	       .\the\subcpcount\quad\enskip\ }{\ppaarr #1}\par} \else
		 {\itemitem{\ppaarr\the\cpcount .\the\subcpcount\quad
		  \enskip\ }{\ppaarr #1} \par}  \fi}
		\else{\itemitem{\ppaarr\applett .\the\subcpcount\quad
		 \enskip\ }{\ppaarr #1}\par}\fi\nobreak\jump}\nobreak}
\clearsubcpcount
\def\appendix#1{\global\advance\appcount by 1 \clearfrmcount
		  \goodbreak\null\vbox{\jump\nobreak
		  \global\advance\indappcount by 1 \clearsubcpcount
	  \itemitem{ }{\hskip-40pt\ttaarr Appendix\ \applett :\ #1}
	     \nobreak\jump\sjump}\nobreak}
\clearappcount \clearindappcount
\def\references{\goodbreak\null\vbox{\jump\nobreak
   \itemitem{}{\ttaarr References} \nobreak\jump\sjump}\nobreak}

\def\introsumm{\clearindappcount\clearappcount\clearcpcount
		  \clearsubcpcount\goodbreak\null\vbox{\jump\nobreak
  \itemitem{}{\ttaarr Introduction and Summary} \nobreak\jump\sjump}\nobreak}
\clearcpcount\clearcountref
\def\acknowledgements{\goodbreak\null\vbox{\jump\nobreak
\itemitem{ }{\ttaarr Acknowledgements} \nobreak\jump\sjump}\nobreak}
\def\setchap#1{\ifnum\indappcount=0{\ifnum\subcpcount=0%
\xdef\spzzttrra{\the\cpcount}%
\else\xdef\spzzttrra{\the\cpcount .\the\subcpcount}\fi}
\else{\ifnum\subcpcount=0 \xdef\spzzttrra{\applett}%
\else\xdef\spzzttrra{\applett .\the\subcpcount}\fi}\fi
\expandafter\xdef\csname#1\endcsname{\spzzttrra}}
\newcount\draftnum \newcount\ppora   \newcount\ppminuti
\global\ppora=\time   \global\ppminuti=\time
\global\divide\ppora by 60  \draftnum=\ppora
\multiply\draftnum by 60    \global\advance\ppminuti by -\draftnum
\def\droggi{\number\day /\number\month /\number\year\ \the\ppora
:\the\ppminuti}     \global\draftnum=0
\def\draftcomment#1{\ifnum\draftnum=0 \relax \else
{\ {\bf ***}\ #1\ {\bf ***}\ }\fi} 
%
%
\catcode`@=11
\gdef\Ref#1{\expandafter\ifx\csname @rrxx@#1\endcsname\relax%
{\global\advance\countref by 1    \ifnum\countref>200
\immediate\write16 {Panda ERROR - Ref: maximum number of references
exceeded}  \expandafter\xdef\csname @rrxx@#1\endcsname{0}\else
\expandafter\xdef\csname @rrxx@#1\endcsname{\the\countref}\fi}\fi
\ifnum\draftnum=0 \csname @rrxx@#1\endcsname \else#1\fi}
\gdef\beginref{\ifnum\draftnum=0  \gdef\Rref{\fairef}
\gdef\endref{\scriviref} \else\relax\fi
\ifx\risposta\mplarisposta \ninepoint \fi
\baselineskip=12pt \parskip 2pt plus.2pt }
\def\Reflab#1{[#1]} \gdef\Rref#1#2{\item{\Reflab{#1}}{#2}}
\gdef\endref{\relax}  \newcount\conttemp
\gdef\fairef#1#2{\expandafter\ifx\csname @rrxx@#1\endcsname\relax
{\global\conttemp=0 \immediate\write16 {Panda ERROR - Ref: reference
[#1] undefined}} \else
{\global\conttemp=\csname @rrxx@#1\endcsname } \fi
\global\advance\conttemp by 50  \global\setbox\conttemp=\hbox{#2} }
\gdef\scriviref{\clearconnttrre\conttemp=50
\loop\ifnum\connttrre<\countref \advance\conttemp by 1
\advance\connttrre by 1
\item{\Reflab{\the\connttrre}}{\unhcopy\conttemp} \repeat}
\clearcountref \clearconnttrre
\catcode`@=12
\ifx\risposta\mplarisposta \def\Reflab#1{#1.} \letterfootnote \fi
%
%

\def\slashchar#1{\setbox0=\hbox{$#1$} \dimen0=\wd0
     \setbox1=\hbox{/} \dimen1=\wd1 \ifdim\dimen0>\dimen1
      \rlap{\hbox to \dimen0{\hfil/\hfil}} #1 \else
      \rlap{\hbox to \dimen1{\hfil$#1$\hfil}} / \fi}
\ifx\oldchi\undefined \let\oldchi=\chi
  \def\cchi{{\raise 1pt\hbox{$\oldchi$}}} \let\chi=\cchi \fi
\ifnum\contasym=1 \else \fi
 \def\del{\partial}   

\def\frac#1#2{{\textstyle{#1 \over #2}}}

\def\half{\ifinner {\scriptstyle {1 \over 2}}\else {1 \over 2} \fi}
  \def\ket#1{\vert#1\rangle}

\def\vev#1{\langle#1\rangle}

\def\simge{\rlap{\raise 2pt \hbox{$>$}}{\lower 2pt \hbox{$\sim$}}}
\def\simle{\rlap{\raise 2pt \hbox{$<$}}{\lower 2pt \hbox{$\sim$}}}

\def\buildchar#1#2#3{{\null\!\mathop{#1}\limits^{#2}_{#3}\!\null}}

\def\vbig#1#2{{\vbigd@men=#2\divide\vbigd@men by 2%
\hbox{$\left#1\vbox to \vbigd@men{}\right.\n@space$}}}

\def\noblackbox{\overfullrule=0pt} 
%
%
\newcount\fdefcontre \newcount\fdefcount \newcount\indcount
\newread\filefdef  \newread\fileftmp  \newwrite\filefdef
\newwrite\fileftmp     \def\strip #1*.A {#1}%
\def\futuredef#1{\beginfdef
\expandafter\ifx\csname#1\endcsname\relax%
{\immediate\write\fileftmp{#1*.A}%
\immediate\write16 {Panda Warning - fdef: macro "#1" on page
\the\pageno \space undefined}
\ifnum\draftnum=0 \expandafter\xdef\csname#1\endcsname{(?)}
\else \expandafter\xdef\csname#1\endcsname{(#1)}\fi
\global\advance\fdefcount by 1}\fi\csname#1\endcsname}
\def\fdef{\futuredef}
\def\beginfdef{\ifnum\fdefcontre=0
\immediate\openin\filefdef\jobname.fdef
\immediate\openout\fileftmp\jobname.ftmp
\global\fdefcontre=1  \ifeof\filefdef \immediate\write16 {Panda
WARNING - fdef: file \jobname.fdef not found, run TeX again}
\else \immediate\read\filefdef to\spzzttrra
\global\advance\fdefcount by \spzzttrra
\indcount=0 \loop\ifnum\indcount<\fdefcount
\advance\indcount by 1%
\immediate\read\filefdef to\spezttrra%
\immediate\read\filefdef to\sppzttrra%
\edef\spzzttrra{\expandafter\strip\spezttrra}%
\immediate\write\fileftmp {\spzzttrra *.A}
\expandafter\xdef\csname\spzzttrra\endcsname{\sppzttrra}%
\repeat \fi \immediate\closein\filefdef \fi}
\def\endfdef{\immediate\closeout\fileftmp   \ifnum\fdefcount>0
\immediate\openin\fileftmp \jobname.ftmp
\immediate\openout\filefdef \jobname.fdef
\immediate\write\filefdef {\the\fdefcount}   \indcount=0
\loop\ifnum\indcount<\fdefcount    \advance\indcount by 1
\immediate\read\fileftmp to\spezttrra
\edef\spzzttrra{\expandafter\strip\spezttrra}
\immediate\write\filefdef{\spzzttrra *.A}
\edef\spezttrra{\string{\csname\spzzttrra\endcsname\string}}
\iwritel\filefdef{\spezttrra}
\repeat  \immediate\closein\fileftmp \immediate\closeout\filefdef
\immediate\write16 {Panda Warning - fdef: Label(s) may have changed,
re-run TeX to get them right}\fi}
\def\iwritel#1#2{\newlinechar=-1
{\newlinechar=`\ \immediate\write#1{#2}}\newlinechar=-1}
\global\fdefcontre=0 \global\fdefcount=0 \global\indcount=0
%
%
%
\mathchardef\alpha="710B   \mathchardef\beta="710C
\mathchardef\gamma="710D   \mathchardef\delta="710E
\mathchardef\epsilon="710F   \mathchardef\zeta="7110
\mathchardef\eta="7111   \mathchardef\theta="7112
\mathchardef\iota="7113   \mathchardef\kappa="7114
\mathchardef\lambda="7115   \mathchardef\mu="7116
\mathchardef\nu="7117   \mathchardef\xi="7118
\mathchardef\pi="7119   \mathchardef\rho="711A
\mathchardef\sigma="711B   \mathchardef\tau="711C
\mathchardef\upsilon="711D   \mathchardef\phi="711E
\mathchardef\chi="711F   \mathchardef\psi="7120
\mathchardef\omega="7121   \mathchardef\varepsilon="7122
\mathchardef\vartheta="7123   \mathchardef\varpi="7124
\mathchardef\varrho="7125   \mathchardef\varsigma="7126
\mathchardef\varphi="7127
%
%
\null
%
%
%
%
%
\loadamsmath
\chapterfont{\bfone} \sectionfont{\scaps}
\noblackbox

\def\barj{\bar{\hbox{\it\j}}}
\def\Teta#1#2{\Theta\left[{}^{#1}_{#2}\right]}
\def\Tetab#1#2{\overline\Theta\left[{}^{#1}_{#2}\right]}
\def\Ical#1#2{{\cal I}\left[{}^{#1}_{#2}\right]}
\def\II#1#2{{I}\left[{}^{#1}_{#2}\right]}
\def\IIb#1#2{\overline{I}\left[{}^{#1}_{#2}\right]}
\def\GGp#1#2{{G^+}\left[{}^{#1}_{#2}\right]}
\def\GGm#1#2{{G^-}\left[{}^{#1}_{#2}\right]}
\def\eqmodone{\ \buildchar{=}{{\rm{\scriptscriptstyle MOD\ 1}}}{ }\ }
\def\eqmodtwo{\ \buildchar{=}{{\rm{\scriptscriptstyle MOD\ 2}}}{ }\ }
\def\eqope{\ \buildchar{=}{\rm\scriptscriptstyle OPE}{ }\ }
\def\modone#1{[\hskip-1.2pt[#1]\hskip-1.2pt]}
\def\wew#1{\langle\langle\, #1\, \rangle\rangle}
\def\di{{\rm d}}
\nopagenumbers
{\baselineskip=12pt
\line{\hfill NBI-HE-94-47}
\line{\hfill hep-th/9411015}
\line{\hfill October, 1994}}
{\baselineskip=14pt
\vfill
\centerline{\capsone On the Computation of One-Loop Amplitudes with}
\sjump
\centerline{\capsone External Fermions in 4D Heterotic Superstrings}
\bjump\bjump
\centerline{\scaps Andrea Pasquinucci~\footnote{$^\dagger$}{Supported
by EU grant no. ERBCHBGCT920179.} and Kaj
Roland~\footnote{$^\ddagger$}{Supported by the Carlsberg Foundation.}}
\sjump
\centerline{\sl The Niels Bohr Institute, University of Copenhagen,}
\centerline{\sl Blegdamsvej 17, DK-2100, Copenhagen, Denmark}
\bjump \vfill
\centerline{\capsone ABSTRACT}
\sjump
\noindent
We present the technical tools needed to compute
any one-loop amplitude involving external spacetime fermions in a
four-dimensional heterotic string model \`a la Kawai-Lewellen-Tye.
As an example, we compute the one-loop three-point amplitude
with one ``photon" and two external massive fermions (``electrons").
As a check of our computation, we verify that the one-loop
contribution to the Anomalous Magnetic Moment vanishes if the
model has spacetime supersymmetry, as required by the supersymmetric
sum rules.
\sjump \vfill
\pageno=0 \eject }
\yespagenumbers\pageno=1
\null\bjump
\introsumm

String theory~[\Ref{GSW}], since its beginning more than twenty years
ago, has been a very interesting arena for the development of new ideas
in theoretical high-energy physics and appears to be
the only candidate for the unification of all
elementary interactions. This being the case,
it could then seem strange that rather few computations of
one-loop amplitudes have ever been performed in string models.
There are various
reasons for this. One is that string theories are naturally
formulated at the string scale which is of the order of magnitude of the
Planck scale. Thus, the interesting phenomenology at experimentally
accessible energy scales is described by a low-energy effective field theory
for the string's massless modes.
Another reason is that computations of string amplitudes turn out
to be very long and tricky. Indeed, as of today, full one-loop
amplitudes have been explicitly computed only for external
spacetime bosons.

The purpose of this paper is to address the (mostly
technical) problem of computing loop amplitudes with external
space-time fermions in a
four-dimensional heterotic string theory.
Such computations, although not of a direct
phenomenological importance, can have several interesting applications.

Obviously, the computation of string loop amplitudes will give a
better understanding of the properties and characteristics of
string theories per se. Among the various interesting issues is the full
understanding of the analytical properties of string amplitudes
[\Ref{Hoker},\Ref{Weisberger},\Ref{Berera}],
their divergencies and the associated renormalization
[\Ref{Fischler},\Ref{Weinberg},\Ref{Tseytlin}].
Clearly a full discussion of these points would require an off-shell
formulation, or at least computation, of string amplitudes.

Another point that can have direct consequences for
phenomenology and in general for our understanding of field theory,
is that in the low energy ($\alpha^\prime\rightarrow 0$) limit, a string
theory becomes a bona fide field theory. This limit is not in any way
straightforward, since, for example, in closed string theories at each
loop order any amplitude is given by only one ``diagram''.

Interesting are, for example, the results obtained by
Kaplunovsky~[\Ref{Kaplu}]
who computed the one-loop
Yang-Mills beta function and the threshold corrections
from string theory. Moreover, Bern and Kosower [\Ref{BK}]
were able to obtain new
(simplified) Feynman-like rules for one-loop computations in pure
Yang-Mills theory.
It would obviously be interesting to extend their results to the
full QCD theory.

The necessary tools for computing loop amplitudes involving
external spacetime fermions are known in principle~[\Ref{Phong}], but
the technical difficulties involved, particularly in four dimensional
heterotic models, are quite considerable.
We choose to work with string models
constructed using free world-sheet fermions \`a la Kawai-Lewellen-Tye
(KLT)~[\Ref{KLT}] (see also [\Ref{Anto},\Ref{Bluhm}]).
The basic problem is then to compute free-fermion
correlation functions on an arbitrary
Riemann surface in the presence of spin field operators. The well known
way of doing this is by bosonization~[\Ref{FMS},\Ref{Koste}] (but for a
different approach see ref.~[\Ref{Atick}]).
But even in this case the computation of an amplitude
remains a non trivial task and in this paper we
present the technical tools needed to compute
any amplitude involving external spacetime fermions.
We restrict ourselves to one-loop amplitudes but would like to stress
that the generalization to multiloop amplitudes is straightforward.

The main technical point concerns the bosonization itself: In bosonizing
the world-sheet fermions one needs to introduce cocycles to guarantee the
correct anti-commutation relations. The cocycles play a fundamental role
in reconstructing the Lorentz algebra and the explicit Lorentz
covariance of the final result, which is lost when the amplitude
is written in bosonized form. However, as was discussed already in
ref.~[\Ref{Koste}], cocycles are in general
not uniquely defined but can be introduced in many different ways, not all of
which are physically acceptable. We determine which are
the conditions that a proper set of cocycles must satisfy and present an
explicit solution in the context of a particular (KLT) {\it toy} model.
Only given such a solution can the bosonization procedure be said to be
completely well-defined.

Anyway, this is not yet enough to reconstruct the Lorentz covariance
of the final result. Indeed it turns out that correlators involving for
example the Lorentz Ka\v{c}-Moody current $\psi^{\mu} \psi^{\nu} (z)$
involve different expressions in terms of theta functions,
depending on the values of the Lorentz indices $\mu$, $\nu$.
Then, to achieve explicit Lorentz covariance, one needs to prove
some non-trivial identities in theta functions.

We also consider the correct normalization of the string
amplitudes. We offer a general formula for the $N$-string one-loop amplitude
with the correct overall normalization. One still has to properly normalize
the vertex operators. This may be done, case by case, using the method
advocated in ref.~[\Ref{Kaj}]. An example is provided in appendix D.

For a generic amplitude, the final form at which we would arrive
still leaves to do the sum over the spin structures and
the integral over the moduli. This resembles the stage in a field theory
calculation where loop momentum integrals and
internal Lorentz algebra has been performed, leaving only an integral
over Schwinger proper times (or Feynman parameters) --- except, of course,
that in field theory we have the contribution of many diagrams.
In general neither the summation over
the spin structures nor the integral over the moduli can be done
analytically, but in simple cases it is possible to evaluate them
numerically.

As a non-trivial check of the correctness and consistency of our approach,
and in order to present the reader with a relatively {\it simple\/}
example, we explicitly compute a one-loop three-point amplitude in our
KLT toy model, involving one $U(1)$ gauge boson (a
``photon'') and two ``electrons'', that is,
spin $\frac12$ particles with mass of the order of the Planck mass and
nonzero $U(1)$ charge. One of the terms in this amplitude
gives the one-loop contribution to the Anomalous Magnetic Moment (AMM)
of the ``electron''. Recently, Ferrara and Porrati [\Ref{FP}] have
proven some Supersymmetric Sum Rules which state that in a model with
$N=1$ space-time supersymmetry the anomalous magnetic moment for particles
of spin $\frac12$ in $(0,\frac12)$ multiplets, vanishes. In other words, the
tree-level value for the gyromagnetic ratio, $g=2$, does not receive any
corrections. The toy model we have chosen to work with has the
particular property that the spectrum is either $N=1$ supersymmetric
or non-supersymmetric, depending on the values of certain parameters
defining the GSO projections. Checking the Ferrara-Porrati Sum Rules
will then provide us with a quite non-trivial check on our
computations. (To this end a crucial role is played by a
spin-structure dependent phase appearing in the superghost part of the
amplitude, as explicitly computed in appendix B. As far as we know this
phase has been accounted for only in ref.~[\Ref{PDV2}].)

It should be mentioned that, when decomposed in Lorentz structures,
the three-point amplitude has two more terms. The first term has the
same structure as the tree-level amplitude, and gives rise to various
renormalizations~[\Ref{Fischler},\Ref{Weinberg},\Ref{Tseytlin}].
Since the integral over the moduli diverges,
a proper treatment should be done in the context of an off-shell
computation and we do not consider it in the present work.
The last term has the Lorentz structure of an Electric Dipole
Moment, but it turns out not to depend on the sign of the electric
charge of the ``electron/positron''. For this reason we call it a
{\it Pseudo Electric Dipole Moment} (PEDM): On top of violating P and T,
like an ordinary electric dipole moment, this PEDM also violates C.
Then it violates CPT. But this should not be possible
since it was claimed in ref.~[\Ref{Sonoda}] (see also
[\Ref{Pott},\Ref{Us}]) that KLT string models
do not violate CPT perturbatively. Indeed we will show
that for this term in the amplitude the integral over the moduli vanishes,
leaving no contradictions. This gives us another, unexpected, non-trivial
check on our computations.

The paper is organized as follows. In the first section we
review the KLT-formalism for constructing four-dimensional heterotic string
models with free world-sheet fermions. Our notations differ somewhat
from those of ref.~[\Ref{KLT}] and furthermore, we use the Lorentz covariant
(rather than the light-cone) formulation.
We then introduce our toy model and discuss its spectrum,
the GSO projection conditions and the spacetime
supersymmetry.

In the second section we introduce the tools necessary for the
computation of arbitrary loop amplitudes involving external spacetime
fermions. Thus, we introduce the spin field vertex operators through
bosonization of the world-sheet fermions and we discuss in details how
to make a consistent and convenient choice of cocycles. In the context
of the toy model we introduce the ``electron/positron'' (and ``photon'')
vertex operators, discuss the related Dirac equation and
introduce a generalized charge conjugation matrix.

In the third section we compute the specific one-loop three-point
amplitude of two ``electrons'' and a ``photon''.
We consider the role of the Picture Changing
Operators (PCOs) and outline the various steps involved in the
computation: The evaluation of the various (world-sheet)
correlators, the appearance of the identities in theta functions needed for
obtaining a Lorentz covariant result, and the use of the GSO projection
conditions and of the Dirac equation. Finally, as the first check of
consistency, we show that the amplitude thus obtained does not depend on the
point of insertion of the PCO.

In the last section we discuss the vanishing of the Pseudo Electric
Dipole Moment and we show how in models with spacetime supersymmetry,
the Anomalous Magnetic Moment vanishes, in agreement with the
Ferrara-Porrati sum rules.

The appendices contain: A summary of notations, conventions and
useful formul\ae; some details on the computation of ghost and superghost
correlators; a discussion of the Lorentz covariant formulation of the
KLT formalism; the computation of the normalization of the
``electron/positron'' vertex
operators; and the proof of one of the identities in theta functions
required for the explicit Lorentz covariance.

\chapter{The KLT 4d Heterotic String Models}
We start out by briefly reviewing the KLT construction [\Ref{KLT}] of 4d
heterotic string models; our notations differ somewhat from those of
ref.~[\Ref{KLT}]. Also, we choose to work in the Lorentz-covariant
formulation, rather than the light-cone gauge. We choose Euclidean
signature on the space-time metric throughout, only rotating to
Minkowski space at the very end of calculations.
\section{The KLT formalism}
A 4-dimensional heterotic KLT model is described in the
Lorentz-covariant formulation by the four space-time
coordinate fields $X^\mu(z, \bar{z})$; twenty-two left-moving complex
fermions
$\bar{\psi}_{(\bar{l})}(\bar{z})$, $\bar{l} = \bar{1}, \ldots ,
\overline{22}$; eleven right-moving complex fermions
$\psi_{(l)}(z)$, $l= 0,1;2,\ldots,10$; right-moving superghosts $\beta,
\gamma$; and left- and right-moving reparametrization ghosts $\bar{b},
\bar{c}$ and $b,c$.

Corresponding to each of the right-moving
complex fermions we define two real fermions by
$$
\psi^m_{(l)} = \left\{ \frac1{\sqrt2}(\psi_{(l)} + \psi^*_{(l)} )\ ,
\ \frac1{i\sqrt2}(\psi_{(l)} - \psi^*_{(l)}) \right\}\ ,\quad m=1,2\ .
\efr
The four real fermions $\psi^{\mu}$ that transform as a space-time vector
are related to the complex fermions $\psi_{(0)}$ and $\psi_{(1)}$ by
$\psi^0 \equiv \psi_{(0)}^1$, $\psi^1 \equiv \psi_{(0)}^2$,
$\psi^2 \equiv \psi_{(1)}^1$ and $\psi^3 \equiv \psi_{(1)}^2$,
while the nine complex fermions $\psi_{(l)} (z)$, $l=2,\ldots,10$ are
called {\it internal}.

The right-movers possess $N=1$ world-sheet supersymmetry, generated by
the supercurrent
$$ T_F = T_F^{[X,\psi]} - c \partial \beta - {3 \over 2} (\partial c)
\beta + {1 \over 2} \gamma b \ , \nfr{supcurr}
where the orbital part is given by
$$T^{[X,\psi]}_F = -\frac{i}2\del X\cdot \psi -\frac{i}2
\sum_{m=1}^2 (\psi^m_{(2)}\psi^m_{(3)}\psi^m_{(4)} + \psi^m_{(5)}
\psi^m_{(6)}\psi^m_{(7)} + \psi^m_{(8)}\psi^m_{(9)}\psi^m_{(10)}) \ .
\nfr{supcur}
Notice that the supercurrent arranges the nine internal fermions into
three {\it triplets}.

Any KLT model is specified by
a certain number of basis
vectors ${\bf W}_i$ defining the set of possible boundary
conditions (spin structures)
for the fermions, and a set of parameters $k_{ij}$ defining
the GSO projection.

Each entry in the vector ${\bf W}_i$ is a rational number and
corresponds to one of the complex fermions.
Since the fermions $\psi_{(0)}$ and
$\psi_{(1)}$ (and the superghosts)
are forced to carry the same spin structure
(otherwise the supercurrent \supcurr\ would not have
well-defined boundary conditions) we include in the vectors ${\bf W}_i$
only the fermions
$\bar{\psi}_{(\bar{l})}$, $\bar{l}=\bar{1},\ldots,\overline{22}$
and $\psi_{(l)}$, $l=1,
\ldots,10$.

On the cylinder, described by a complex coordinate $z$, the boundary
conditions of the fermions are then specified as follows:
$$\eqalignno{
\psi_{(l)} (e^{2\pi i} z) = e^{2\pi i ({1 \over 2} - \alpha_l)}
\psi_{(l)} (z) \ \ \ , \ \ & l= 0,1,\ldots,10 &  \nameali{bound} \cr
\bar{\psi}_{(\bar{l})} (e^{-2\pi i} \bar{z}) = e^{-2\pi i ({1 \over 2} -
\bar{\alpha}_l)}
\bar{\psi}_{(\bar{l})} (\bar{z}) \ \ \ , \ \ & l = 1,\ldots,22 \ , \cr} $$
where $\alpha_0 \equiv \alpha_1$ and $\bar{\alpha}_{l}$ ($l=
1,\ldots,22$) and $\alpha_l$ ($l=1,\ldots,10$) are the components of the
vector
$$
{\bfmath\alpha}  =
\sum_{i=0,1,\ldots} m_i {\bf W}_i \equiv m {\bf
W} \ , \nfr{spins}
which is parametrized by integers $m_i$ taking
values in $\{0,\dots,M_i-1\}$, $M_i$ being the smallest integer such
that
$M_i {\bf W}_i$ ($i$ not summed) is a vector of integer
numbers. The Ramond (R) and Neveu-Schwarz (NS) boundary conditions
correspond
to $\alpha_{l} = 0$ and $\alpha_l = \frac12$ respectively.

Notice that in the class of models we consider, formulated in terms of
complex fermions, the requirement that the supercurrent \supcur\
has well-defined boundary conditions dictates that all the right-moving
fermions satisfy either R or NS boundary conditions and furthermore, that
$$ \alpha_1 \eqmodone \sum_{l=2}^4 \alpha_l \eqmodone \sum_{l=5}^7
\alpha_l \eqmodone \sum_{l=8}^{10} \alpha_l \nfr{supcurconst}
for any set of boundary conditions $\bfmath{\alpha}$.
For the left-moving fermions boundary conditions other than R or NS are
possible.

Each set of integers $m_i$ (each vector ${\bfmath\alpha}$) defines a {\it
sector} in the spectrum of string states. We have $\prod_i M_i$ such
sectors.
For example, the set of basis vectors always include the vector
[\Ref{KLT}]
$$ {\bf W}_0\ =\ \left( (\frac12)^{22} \vert
(\frac12) (\frac12\frac12\frac12)^3\right) \ ,
\efr
which describes the NS boundary condition for all fermions, and the
vector ${\bf W} = {\bf 0}$ which describes the R boundary
conditions for all fermions.

The string states in the sector specified by ${\bfmath\alpha}$ are
space-time bosons
(fermions) depending on whether the first right-moving component
$$ \alpha_1 = \sum_i m_i ({\bf W}_i)_{(1)} \equiv \sum_i m_i s_i \efr
(which specifies the boundary condition for the supercurrent) takes the
value $1/2$ $(0)$ mod $1$, and we will refer to it as a bosonic
(fermionic) sector.

In a bosonic (fermionic) sector, the set of all possible string states
are constructed from the  superghost vacuum with charge
$q' = -1$ $(q' = -1/2)$ [\Ref{FMS}],
and the set of {\it allowed\/} string states is specified by
the GSO projections, which in the Lorentz-covariant formulation assume
the form
$$
{\bf W}_i\cdot {\bf N}_{\modone{{\bfmath\alpha}}} - s_i
(N^{(0)}_{\modone{\alpha_{1}}}
 - N^{(\beta \gamma)}_{\modone{\alpha_{1}}}) \eqmodone
\sum_j k_{ij}m_j + s_i + k_{0i} - {\bf W}_i \cdot
\modone{{\bfmath\alpha}}  \ ,
\nfr{GSOp}
as shown in appendix C.

Here the inner-product of two vectors, such as ${\bf W}_i \cdot {\bf
N}$, includes a factor of $(-1)$ for
right-moving components. Also, for any real number $\alpha$ we define
$\modone\alpha\equiv\alpha - \Delta$, where $0\leq\modone\alpha< 1$ and
$\Delta \in {\Bbb Z}$.

${\bf N}_{\modone{\bfmath\alpha}}$ is the vector of fermion number operators
in the sector ${\bfmath\alpha}$, $N^{(0)}_{\modone{\alpha_1}}$ is the
number operator for the ``longitudinal'' complex fermion
$\psi_{(0)}$ and $N^{(\beta \gamma)}_{\modone{\alpha_1}}$ is the
superghost number operator.

If we introduce mode expansions
$$\eqalignno{
\psi_{(l)} (z) & = \ \sum_{q \in {\Bbb Z}}
\psi^{(l)}_{q-\modone{\alpha_l}} z^{-q+\modone{\alpha_l}-1/2} &
\nameali{cfive} \cr
\beta(z) & = \ \sum_{q \in {\Bbb Z}} \beta_{q-\modone{\alpha_1}}
z^{-q+\modone{\alpha_1}-3/2}
\cr
\gamma(z) & = \ \sum_{q \in {\Bbb Z}} \gamma_{q-\modone{\alpha_1}}
z^{-q+\modone{\alpha_1}+1/2} \ ,
\cr} $$
where
$$\eqalignno{
& \{ \psi^{(l)}_{q-\modone{\alpha_l}},
\psi^{(l')\, *}_{q'+\modone{\alpha_l}} \} =
\delta_{q+q'} \delta^{l,l'} \cr
& [ \gamma_{q-\modone{\alpha_1}}, \beta_{q'+\modone{\alpha_1}} ] =
\delta_{q+q'} \ ,
\cr} $$
we may write the $l$th fermion number operator as
$$ N^{(l)}_{\modone{\alpha_l}} = \sum_{q=1}^{\infty} \left[
n^{(l)}_{q+\modone{\alpha_l}-1} - n^{(l) *}_{q-\modone{\alpha_l}}
\right] \ , \nfr{fermnum}
with the fermion and anti-fermion mode occupation numbers defined by
$$
n^{(l)}_{q+\modone{\alpha_l}-1} = \psi^{(l)}_{-q-\modone{\alpha_l}+1}
\psi^{(l)*}_{q+\modone{\alpha_l}-1}\ ,\qquad
n^{(l)*}_{q-\modone{\alpha_l}} =
\psi^{(l)*}_{-q+\modone{\alpha_l}} \psi^{(l)}_{q-\modone{\alpha_l}}\ ;
\efr
and
$$ N^{(\beta \gamma)}_{\modone{\alpha_1}} = -\sum_{q=1}^{\infty} \left[
\beta_{-q+\modone{\alpha_1}} \gamma_{q-\modone{\alpha_1}} +
\gamma_{-q+1-\modone{\alpha_1}} \beta_{q-1+\modone{\alpha_1}} \right] \ .
\nfr{sghostnum}
Notice that $N^{(\beta \gamma)}_{\modone{\alpha_1}} = 0$ for states in the
superghost vacuum.
The GSO projections \GSOp\ are
parametrized by the quantities $k_{ij}$. As shown
in Ref.~[\Ref{KLT}] consistency at the $1$-loop level requires the
$k_{ij}$ and the ${\bf W}_i$ to satisfy the following conditions
$$\eqalignno{& k_{ij} +k_{ji} \eqmodone {\bf W}_i \cdot {\bf W}_j
&\nameali{kijeq}\cr
& M_j k_{ij} \eqmodone 0 \cr
& k_{ii} + k_{i0} + s_i -\frac12 {\bf W}_i \cdot {\bf W}_i \eqmodone
0 \ . \cr}
$$

On the torus the spin structure $\left[ {}^{\alpha_{l}}_{\beta_{l}}
\right]$
of the fermion $(l)$ is parametrized by two sets of integers, $m_i$ and
$n_i$, each taking values in $\{0,\ldots,M_i-1\}$:
$$\eqalignno{ & {\bfmath\alpha}  =
\sum_{i=0,1,\ldots} m_i {\bf W}_i  &
\nameali{spinstructures} \cr
 & {\bfmath\beta}  =
\sum_{i=0,1,\ldots} n_i {\bf W}_i  \ . \cr }
$$
The $m_i$ specify the sector of states being propagated in the loop. The
$n_i$ specify the boundary conditions when going around the time-like
direction of the torus. We sum over the spin structures by summing over
the $(\prod_i M_i)^2$
possible values of these integers.
The summation over the $n_i$ enforces the GSO projection on the states
propagating in the loop. Therefore the sum over spin structures may also
be viewed as a sum over the full spectrum of GSO projected states
circulating in the loop.

The $1$-loop partition function of the KLT model can be written as
$$\eqalignno{
{\cal Z}\ =\ &\sum_{m_i,n_j}  C^{{\bfmath\alpha}}_{{\bfmath\beta}} \int
{\di^2\tau\over
({\rm Im}\tau)^2 }  \left( \bar\eta(\bar\tau)\right) ^{-24}
\prod_{l=1}^{22}
\Tetab{\bar\alpha_l}{\bar\beta_l}(0\vert\bar\tau) \ \times &
\nameali{partf} \cr
&\qquad\qquad\qquad\qquad\quad\ \left( \eta(\tau)\right)^{-12}
\prod_{l=1}^{10}\Teta{\alpha_l}{\beta_l}(0\vert\tau)  {1 \over {\rm Im}
\tau} \ , \cr}
$$
where the summation coefficients are given by
$$C^{{\bfmath\alpha}}_{{\bfmath\beta}} = {1\over \prod_i M_i}
\exp\left\{- 2\pi
i\left[ \sum_i (n_i + \delta_{i,0})\left( \sum_j k_{ij} m_j
+s_i -k_{i0}\right) + \sum_i m_i s_i +\frac12 \right] \right\} \, .
\nfr{phases}
These coefficients are chosen so that all states in the GSO-projected
spectrum describing space-time bosons (fermions) contribute to the
partition function with weight $+1$ $(-1)$. Using the properties \kijeq\
it is straightforward to check that the partition function \partf\ is
modular invariant. Our expression \phases\ for the summation
coefficients
is somewhat simpler than that given in ref.~[\Ref{KLT}], thanks to
certain phases being absorbed into the definition of the $\Theta$
function (see appendix A for conventions).

To conclude this subsection, we recall the ``mass formula" [\Ref{KLT}].
We know that only states satisfying the level-matching condition $L_0 =
\bar{L}_0$ can propagate, and these have a mass given by
$$
{\alpha^{\prime} \over 4} M^2 = \bar{L}_0 -{\alpha^\prime\over 4} p^2 =
L_0 -
{\alpha^\prime\over 4} p^2 \ .
\efr
For states in the (super) ghost vacuum
$$\eqalignno{ L_0\ =\ & {\alpha^\prime\over 4} p^2 + \sum_{l=0}^{10}
\left\{ E_{\modone{\alpha_l}} + \sum_{q=1}^\infty \left(
( q+\modone{\alpha_l}-1) \, n^{(l)}_{q+\modone{\alpha_l}-1}
\right.\right.&\numali\cr
& \left.\left. + (q - \modone{\alpha_l}) \, n^{(l)*}_{q-\modone{\alpha_l}}
\right) \right\} + \sum_{q=1}^\infty
q a_{-q} \cdot a_q -1 + E^{(\beta \gamma)}_{\modone{\alpha_1}} \ , \cr}
$$
where
$a_q^{\mu}$ are the (right-moving) modes of $X^{\mu}(z,\bar{z})$,
$E^{(\beta \gamma)}_{\modone{\alpha_1}}$ is the superghost vacuum energy,
which equals $+1/2$ ($+3/8$) in a bosonic (fermionic) sector,
and the vacuum energy of the
$l$'th complex fermion (relative to the conformal vacuum) is
$$
E_{\modone{\alpha_l}}=\frac12\left(\modone{\alpha_l}-\frac12\right)^2\ .
\efr
The contribution of minus one represents the reparametrization ghost vacuum
energy.

The same formula holds for $\bar{L}_0$, without the superghost vacuum
energy, and with the left movers substituted
for the right movers.

In each sector, the vacuum energies of the left- and right-movers are
given by
$$
E_{\rm left} = \sum_{l=1}^{22} E_{\modone{\bar{\alpha}_l}} - 1 \ , \qquad
\quad E_{\rm right} = \sum_{l=0}^{10} E_{\modone{\alpha_l}} - 1 +
E^{(\beta \gamma)}_{\modone{\alpha_1}} \ .
\nfr{vacenergy}
If we restrict ourselves to vectors ${\bf W}_i$ where all components are
either $0$ or $1/2$, only NS and R boundary conditions are possible for
any given fermion. In the first case, the vacuum is the conformal one,
$\vert 0 \rangle$; in the
second, the vacuum is twofold degenerate and if we represent the zero
modes in terms of Pauli matrices
$$ \psi_0^{(l)} = \frac12 \left( \sigma_1^{(l)} + i \sigma_2^{(l)}
\right) \qquad {\rm and} \qquad
\psi_0^{(l) *} = \frac12 \left( \sigma_1^{(l)} - i \sigma_2^{(l)}
\right) \ ,
\efr
the vacua can be labelled $\vert a_l \rangle$, where $a_l = \pm 1/2$ is
the eigenvalue of $\frac12 \sigma_3^{(l)}$. The
fermion number operator \fermnum\ can then be written as
$$ N^{(l)}_{0} = \sum_{q=1}^{\infty} \left[ n^{(l)}_q - n^{(l)
*}_q \right] + \frac12 (1+ \sigma_3^{(l)} ) \  \efr
--- so the zero mode part
counts the state $\ket{-\frac12}$ with number zero, and $\ket{+\frac12}$
with number one.
\section{Our toy model}
The model we choose to work with has been already proposed in
ref.~[\Ref{KLT}]. It has two main advantages: The gauge group contains
a $U(1)$; and the spectrum can be made $N=1$ spacetime
supersymmetric by choosing appropriate values for the quantities
$k_{ij}$. This means that we can study at the same time spacetime
supersymmetric and non-supersymmetric models.

The model is specified by the following boundary vectors
$$\eqalignno{ & {\bf W}_0\ =\ \left( (\frac12)^{22} \vert
(\frac12) (\frac12\frac12\frac12)^3\right) &\nameali{vectors} \cr
& {\bf W}_1\ =\ \left( (\frac12)^{22} \vert
(0) (0\frac12\frac12)^3\right) \cr
& {\bf W}_2\ =\ \left( (\frac12)^{14} (0)^8 \vert
(0) (0\frac12\frac12) (\frac12 0 \frac12)^2\right) \cr
& {\bf W}_3\ =\ \left( (\frac12)^7 (0)^7 (\frac12)^3 (0)^5 \vert
(0) (\frac12 0 \frac12) (0\frac12\frac12) (\frac12 \frac12 0)\right)\cr
& {\bf W}_4\ =\ \left( (0)^7 (0)^7 (\frac12)^2 (0) (0)^5\vert
(0) (0\frac12\frac12) (\frac12\frac120) (\frac12\frac12 0)\right)\ .\cr}
$$
Since all entries are $0$ or $1/2$, $M_i=2$ and
$$ m_i,\, n_j \ =\  \{0,1\}
\efr
with $i,j=0,\dots , 4$. This implies that the on the torus we have a
total of $2^5
\times 2^5 = 1024$ spin structures.

We introduce the shorthand notation
$$ {\bfmath\alpha} = \sum_{i=0}^4 m_i {\bf W}_i \equiv {\bf W}_{\rm
subscript} \ , \efr
where ``subscript'' is the list of those $i$ for which $m_i = 1$.
For example the sector specified by $m_0=1$, $m_1=1$, $m_2=0$,
$m_3=0$ and $m_4=1$ is called ${\bf W}_{014}= {\bf W}_0 + {\bf W}_1
+ {\bf W}_4$. The only exception is the sector for
which all the $m_i$ are zero which we will just denote by ${\bfmath\alpha}
= {\bf 0}$.

The consistency conditions \kijeq\ are satisfied by any set of $k_{ij}$
satisfying the following matrix equation
$$
\pmatrix{k_{00} & k_{01} & k_{02} & k_{03} & k_{04} \cr
k_{10} & k_{11} & k_{12} & k_{13} & k_{14} \cr
k_{20} & k_{21} & k_{22} & k_{23} & k_{24} \cr
k_{30} & k_{31} & k_{32} & k_{33} & k_{34} \cr
k_{40} & k_{41} & k_{42} & k_{43} & k_{44} \cr}
\eqmodone \pmatrix{k_{00} & k_{01} & k_{02} & k_{03} & k_{04} \cr
k_{01} & k_{01} & k_{12} & k_{13} & k_{14} \cr
k_{02} & k_{12} + \frac12 & k_{02} & k_{23} & k_{24} \cr
k_{03} & k_{13} + \frac12 & k_{23} & k_{03}+\frac12 & k_{34} \cr
k_{04} & k_{14}+ \frac12 & k_{24} & k_{34} + \frac12 & k_{04}+\frac12
\cr} \ ,
\efr
where
$$ k_{ij} \ =\ \{0,\frac12\} \ .
\efr
Hence, the independent $k_{ij}$ can be chosen to be $k_{00}$ and
$k_{ij}$ with $i<j$.

{}From a quick glance at the left-moving part of the vectors \vectors,
it is obvious that the world-sheet fermions are grouped together
according to ${\bf W}_4$: For example the first seven
left-moving complex fermions always have the same spin structure; from
the corresponding $14$ real fermions we may build up the Ka\v{c}-Moody
algebra of $SO(14)$. It is therefore no surprise that the gauge group of
our model turns out to be
$$
SO(14)\otimes SO(14) \otimes SO(4) \otimes U(1) \otimes SO(10) \ ,
\nfr{group}
where the $U(1)$  is actually realized as an $SO(2)$. To prove it we
need to show that the gluons (massless spin $1$ states) existing in the
physical spectrum do indeed fill out the adjoint representation of the
group \group.
\section{The spectrum}
{\topinsert
$$\vbox{\offinterlineskip \halign{\strut#& \vrule# & #\hfill &
\vrule# & \hfill#\hfill & \hfill#\hfill & \hfill#\hfill &
\hfill#\hfill & \vrule# \cr
\noalign{\hrule}
&\ &\ sector\ &\ &\ $E_{\rm left}$\ &\ $E_{\rm right}$\
&\ $\modone{\alpha_1}$\ & $\ \modone{\bar{\alpha}_{17}}$\ &\cr
\noalign{\hrule}
&& \ {\bf 0}    && 7/4 & 3/4 & 0 & 0 &\cr
&& ${\bf W}_{4}$ && 3/2 & 0   & 0 & 0 &\cr
&& ${\bf W}_{3}$ && 1/2 & 0   & 0 & 1/2 &\cr
&& ${\bf W}_{34}$ && 3/4 & 1/4 & 0 & 1/2 &\cr
&& ${\bf W}_{2}$ && 0 & 0 & 0 & 0 &\cr
&& ${\bf W}_{24}$ && $-1/4$ & 1/4 & 0 & 0 &\cr
&& ${\bf W}_{23}$ && 1/2 & 0 & 0 & 1/2 &\cr
&& ${\bf W}_{234}$ && 3/4 & 1/4 & 0 & 1/2 &\cr
&& ${\bf W}_{1}$ && $-1$ & 0 & 0 & 1/2 &\cr
&& ${\bf W}_{14}$ && $-3/4$ & 1/4 & 0 & 1/2 &\cr
&& ${\bf W}_{13}$ && 1/4 & 1/4 & 0 & 0 &\cr
&& ${\bf W}_{134}$ && 0 & 0 & 0 & 0 &\cr
&& ${\bf W}_{12}$ && 3/4 & 1/4 & 0 & 1/2 &\cr
&& ${\bf W}_{124}$ && 1 & 1/2 & 0 & 1/2 &\cr
&& ${\bf W}_{123}$ && 1/4 & 1/4 & 0 & 0 &\cr
&& ${\bf W}_{1234}$ && 0 & 0 & 0 & 0 &\cr
&& ${\bf W}_{0}$ && $-1$ & $-1/2$ & 1/2 & 1/2 &\cr
&& ${\bf W}_{04}$ && $-3/4$ & 1/4 & 1/2 & 1/2 &\cr
&& ${\bf W}_{03}$ && 1/4 & 1/4 & 1/2 & 0 &\cr
&& ${\bf W}_{034}$ && 0 & 0 & 1/2 & 0 &\cr
&& ${\bf W}_{02}$ && 3/4 & 1/4 & 1/2 & 1/2 &\cr
&& ${\bf W}_{024}$ && 1 & 0 & 1/2 & 1/2 &\cr
&& ${\bf W}_{023}$ && 1/4 & 1/4 & 1/2 & 0 &\cr
&& ${\bf W}_{0234}$ && 0 & 0 & 1/2 & 0 &\cr
&& ${\bf W}_{01}$ && 7/4 & 1/4 & 1/2 & 0 &\cr
&& ${\bf W}_{014}$ && 3/2 & 0 & 1/2 & 0 &\cr
&& ${\bf W}_{013}$ && 1/2 & 0 & 1/2 & 1/2 &\cr
&& ${\bf W}_{0134}$ && 3/4 & 1/4 & 1/2 & 1/2 &\cr
&& ${\bf W}_{012}$ && 0 & 0 & 1/2 & 0 &\cr
&& ${\bf W}_{0124}$ && $-1/4$ & $-1/4$ & 1/2 & 0 &\cr
&& ${\bf W}_{0123}$ && 1/2 & 0 & 1/2 & 1/2 &\cr
&& ${\bf W}_{01234}$ && 3/4 & 1/4 & 1/2 & 1/2 &\cr
\noalign{\hrule}
}}$$
\centerline{{\bf Table 1:} Vacuum energies of the $32$ sectors.}
\sjump
\endinsert}
To compute the spectrum of our model we first need to know the vacuum
energies of all sectors characterized by the $\{m_i\}$ using eq.~\vacenergy.
These are summarized in table 1 for our toy model.
In this table we also list whether the sector is bosonic or fermionic
(whether $\modone{\alpha_1} = 1/2$ or $0$) and the value
$\modone{\bar{\alpha}_{17}}$
indicating whether the seventeenth left-moving fermion,
which carries the $U(1)$ charge, has R or NS boundary conditions.
For sectors with $\modone{\bar{\alpha}_{17}} = 0$ the vacuum state
carries $U(1)$ charge $\pm 1/2$.
{\topinsert
$$\vbox{\hbox{\ifdoublepage\hskip-12pt\fi\vbox{\offinterlineskip
\halign{\strut#& \vrule# & #\hfill &
\vrule# & \hfill#\hfill & \hfill#\hfill & \ #\hfill & #\hfill &
\vrule# \cr
\noalign{\hrule}
&\ &\ sector\ &\ &\ $\alpha' M^2$\ &\ Spin\ & \hfill\ State\ & &\cr
\noalign{\hrule}
&& && & & & & \cr
&& ${\bf W}_{0}$ && $-2$ & 0 & $\bar\psi^{\overline{m}}_{-1/2,(\bar{l})}
\ket{0}_L \otimes \ket{0}_R$ & $\bar{l}=1,\dots,22$ &\cr
&& ${\bf W}_{0124}$ && $-1$ & 0 & $\ket{\bar{a}_{17,18,
\dots,22}}_L\otimes\ket{a_{3,4}}_R$ & &\cr
&& ${\bf W}_{012}$ && 0 & 0 & $\ket{\bar{a}_{15,16,17,18,\dots,22}}_L
\otimes\ket{a_{5,6,8,9}}_R$ & &\cr
&& ${\bf W}_{2}$ && 0 & 1/2 & $\ket{\bar{a}_{15,16,17,18,\dots,22}}_L
\otimes\ket{\alpha,a_{2,6,9}}_R$ & &\cr
&& ${\bf W}_{034}$ && 0 & 0 & $\ket{\bar{a}_{1,\dots,7,
17}}_L\otimes\ket{a_{2,3,5,7}}_R$ & &\cr
&& ${\bf W}_{134}$ && 0 & 1/2 & $\ket{\bar{a}_{1,\dots,7,
17}}_L\otimes\ket{\alpha,a_{3,7,8}}_R$ & &\cr
&& ${\bf W}_{0234}$ && 0 & 0 & $\ket{\bar{a}_{8,\dots,14,
17}}_L\otimes\ket{a_{2,4,8,10}}_R$ & &\cr
&& ${\bf W}_{1234}$ && 0 & 1/2 & $\ket{\bar{a}_{8,\dots,14,
17}}_L\otimes\ket{\alpha,a_{4,5,10}}_R$ & &\cr
&& ${\bf W}_{0}$ && 0 & 0 & $\bar\psi_{-1/2,(\bar{l})}^{\overline{m}}
\bar{\psi}^{\overline{n}}_{-1/2,(\bar{k})} \ket{0}_L \otimes \psi_{-1/2,
(j)}^m \ket{0}_R$ & $j=2,\dots,10$ &\cr
&& ${\bf W}_{1}$ && 0 & 1/2 & $\bar\psi_{-1/2,(\bar{l})}^{\overline{m}}
\bar\psi^{\overline{n}}_{-1/2,(\bar{k})} \ket{0}_L \otimes
\ket{\alpha,a_{2,
5,8}}_R$ & &\cr
&& ${\bf W}_{0}$ && 0 & 1 & $\bar\psi_{-1/2,(\bar{l})}^{\overline{m}}
\bar\psi^{\overline{n}}_{-1/2,(\bar{k})} \ket{0}_L \otimes
\psi_{-1/2}^{\mu}
\ket{0}_R$ & &\cr
&& ${\bf W}_{1}$ && 0 & 3/2,1/2 & $\bar{a}_{-1}^{\mu} \ket{0}_L \otimes
\ket{\alpha,a_{2,5,8}}_R$ & &\cr
&& ${\bf W}_{0}$ && 0 & 2,0 & $\bar{a}_{-1}^{\mu} \ket{0}_L \otimes
\psi_{-1/2}^{\nu}\ket{0}_R$ & &\cr
&& ${\bf W}_{0}$ && 0 & 1 & $\bar{a}_{-1}^{\mu} \ket{0}_L \otimes
\psi_{-1/2, (j)}^{m}\ket{0}_R$ & $j=2,\dots,10$ &\cr
&& ${\bf W}_{04}$ && 1 & 0 & $\bar\psi_{-1/2,(\bar{l})}^{\overline{m}}
\bar\psi^{\overline{n}}_{-1/2,(\bar{k})} \ket{\bar{a}_{15,16}}_L
\otimes \ket{a_{3,4,5,6,8,9}}_R$ &
$\bar{l},\bar{k}\neq 15,16$ &\cr
&& ${\bf W}_{14}$ && 1 & 1/2 & $\bar\psi_{-1/2,(\bar{l})}^{\overline{m}}
\bar\psi^{\overline{n}}_{-1/2,(\bar{k})} \ket{\bar{a}_{15,16}}_L
\otimes \ket{\alpha,a_{2,3,4,6,9}}_R$ &
$\bar{l},\bar{k}\neq 15,16$ &\cr
&& ${\bf W}_{04}$ && 1 & 0 & $\bar\psi_{-1,(\bar{l})}^{\overline{m}}
\ket{\bar{a}_{15,16}}_L
\otimes \ket{a_{3,4,5,6,8,9}}_R$ &
$\bar{l} = 15,16$ &\cr
&& ${\bf W}_{14}$ && 1 & 1/2 & $\bar\psi_{-1,(\bar{l})}^{\overline{m}}
\ket{\bar{a}_{15,16}}_L
\otimes \ket{\alpha,a_{2,3,4,6,9}}_R$ &
$\bar{l} = 15,16$ &\cr
&& ${\bf W}_{04}$ && 1 & 1 & $\bar{a}_{-1}^{\mu} \ket{\bar{a}_{15,16}}_L
\otimes \ket{a_{3,4,5,6,8,9}}_R$ & &\cr
&& ${\bf W}_{14}$ && 1 & 3/2,1/2 & $\bar{a}_{-1}^{\mu}
\ket{\bar{a}_{15,16}}_L
\otimes \ket{\alpha,a_{2,3,4,6,9}}_R$ & &\cr
&& ${\bf W}_{03}$ && 1 & 0 & $\ket{\bar{a}_{1,\dots,7,
15,16,17}}_L\otimes
\ket{a_{2,4,6,7,8,9}}_R$ & &\cr
&& ${\bf W}_{13}$ && 1 & 1/2 & $\ket{\bar{a}_{1,\dots,7,
15,16,17}}_L\otimes
\ket{\alpha,a_{4,5,6,7,9}}_R$ & &\cr
&& ${\bf W}_{023}$ && 1 & 0 & $\ket{\bar{a}_{8,\dots,16,
17}}_L\otimes\ket{a_{2,3,5,6,9,10}}_R$ & &\cr
&& ${\bf W}_{123}$ && 1 & 1/2 & $\ket{\bar{a}_{8,\dots,16,
17}}_L\otimes\ket{\alpha,a_{3,6,8,9,10}}_R$ & &\cr
&& ${\bf W}_{0124}$ && 1 & 0 & $\bar\psi^{\overline{m}}_{-1/2,(\bar{l})}
\ket{\bar{a}_{17,18,\dots,22}}_L \otimes
\psi^{m}_{-1/2,(k)}\ket{a_{3,4}}_R$ & &\cr
&& && & & \hfill\hfill $\bar{l}=1,\dots,16$, $k=2,5,\dots,10$ & &\cr
&& ${\bf W}_{24}$ && 1 & 1/2 & $\bar\psi^{\overline{m}}_{-1/2,(\bar{l})}
\ket{\bar{a}_{17,18,\dots,22}}_L \otimes
\ket{\alpha,a_{2,3,4,5,8}}_R$ & &\cr
&& && & & \hfill\hfill $\bar{l}=1,\dots,16$ & &\cr
&& ${\bf W}_{0124}$ && 1 & 1 & $\bar\psi^{\overline{m}}_{-1/2,(\bar{l})}
\ket{\bar{a}_{17,18,\dots,22}}_L\otimes \psi_{-1/2}^{\mu} \ket{a_{3,4}}_R$
& &\cr
&& && & & & & \cr
\noalign{\hrule}
}}}}$$
\centerline{{\bf Table 2:} Lighter states in the spectrum, before
implementing the GSO projection.}
\sjump
\endinsert}
{}From this table it is simple, sector by sector, to construct the excited
states by
acting on the vacuum with the creation operators. Again we restrict
ourselves to states in the (super) ghost vacuum.

In table 2 we list all such states up to
mass level $\alpha' M^2 = 1$.
Obviously some of these states will be projected out of the
spectrum by the GSO projection.
We introduced a shorthand notation for a set of several R
vacua, for example
$$ \bar{a}_{17,18,\ldots,22} \equiv
\bar{a}_{17},\bar{a}_{18},\ldots,\bar{a}_{22} \ , \efr
where all $\bar{a}_l$ and $a_l$ take values $\pm 1/2$, and
$\alpha \equiv (a_0,a_1)$ is a space-time spinor index (not to be
confused with the spin structure, of course). Indices $\overline{m},
\overline{n},m$ take values $1,2$ and unless otherwise stated, indices
$\bar{l},\bar{k}$ take values $1,\dots,22$.

The sector ${\bf W}_0$ contains the standard charged tachyon, the
would-be gauge bosons, the graviton, dilaton and axion, as well as some
further spin 0 and spin 1 states. In the sector
${\bf W}_1$ we find gauginos and gravitinos. The number of space-time
supersymmetries is given by the number of gravitinos that survive the
GSO projection.
\section{The GSO projection conditions}
\setchap{sectGSO}
Now we turn our attention to the GSO projections \GSOp. We will
demonstrate how they reduce the spectrum of our toy model
by means of a few significant examples.

Let us first consider what happens to the states in the ${\bf
W}_0$-sector. In this sector the GSO projections \GSOp\ are reduced to
$$\eqalignno{ & \frac12 \left[
\sum_{l=1}^{22} \bar{N}^{(\bar{l})}_{\modone{\bar{\alpha}_l}}
- \sum_{l=0}^{10} N^{(l)}_{\modone{\alpha_l}} \right] \eqmodone \frac12
&\nameali{GSOzero} \cr
& \frac12 \left[
\sum_{l=1}^{22} \bar{N}^{(\bar{l})}_{\modone{\bar{\alpha}_l}}
- \sum_{l=3,4,6,7,9,10} N^{(l)}_{\modone{\alpha_l}} \right] \eqmodone 0
&\nameali{GSOone} \cr
& \frac12 \left[
\sum_{l=1}^{14} \bar{N}^{(\bar{l})}_{\modone{\bar{\alpha}_l}}
- \sum_{l=3,4,5,7,8,10} N^{(l)}_{\modone{\alpha_l}} \right] \eqmodone 0
&\nameali{GSOtwo} \cr
& \frac12 \left[
\sum_{l=1}^{7} \bar{N}^{(\bar{l})}_{\modone{\bar{\alpha}_l}} +
\sum_{l=15}^{17} \bar{N}^{(\bar{l})}_{\modone{\bar{\alpha}_l}}
- \sum_{l=2,4,6,7,8,9} N^{(l)}_{\modone{\alpha_l}} \right] \eqmodone 0
&\nameali{GSOthree} \cr
& \frac12 \left[
\sum_{l=15}^{16} \bar{N}^{(\bar{l})}_{\modone{\bar{\alpha}_l}}
- \sum_{l=3,4,5,6,8,9} N^{(l)}_{\modone{\alpha_l}} \right] \eqmodone 0 \ .
&\nameali{GSOfour}\cr}
$$
The tachyon has no excitations of the right-movers, and only a single
excitation of the left-movers, i.e. $\sum_{l=1}^{22}
\bar{N}^{(\bar{l})}_{\modone{\bar{\alpha}_l}} = 1$ and
$N^{(l)}_{\modone{\alpha_l}} = 0$ for
$l=0,\dots,10$. Thus, it fails to satisfy eq.~\GSOone\ and is projected
out.

Now consider the would-be gauge bosons. They have
$N^{(0)}_{\modone{\alpha_1}}
+ N^{(1)}_{\modone{\alpha_1}} = 1$ and $N^{(l)}_{\modone{\alpha_l}} = 0$ for
$l=2,\dots,10$. Equations \GSOone -\GSOfour\ then force both of the two
left-moving excitations to belong to the same group of fermions:
Either $\{\overline{1},\dots,\overline{7}\}$, $\{\overline{8},\dots,
\overline{14}\}$,
$\{\overline{15},\overline{16}\}$, $\{\overline{17}\}$ or
$\{\overline{18},\dots,\overline{22}\}$.
Accordingly, the gauge bosons fill out the adjoint
representation of the group \group. The ``extra'' massless
spin $1$ states, where
the vector index is carried by the oscillator $\bar{a}_{-1}^{\mu}$, are
all projected out. They have $N^{(l)}_{\modone{\alpha_l}} = \delta_{j,l}$ for
some $j=2,\dots,10$; by eq.~\GSOone\ this $j$ must be either $2$, $5$ or
$8$. By eq.~\GSOtwo\ it can only be $2$. But eq.~\GSOthree\ rules out
even
this possibility.

Next we consider the ${\bf W}_1$-sector, in order to see how many of the
eight gravitinos and gauginos survive the GSO projection. For the
gravitinos only zero mode excitations of the fermions $0,1,2,5,8$ are
allowed. Thus the five projection conditions \GSOp\ are reduced to
$$\eqalignno{  -\frac12 \left[N^{(0)}_0 +
N^{(1)}_0 + N^{(2)}_0 + N^{(5)}_0 + N^{(8)}_0\right] & \eqmodone \
k_{00} + k_{01} +\frac12 &\numali\cr
0 & \eqmodone \ 0\cr
-\frac12\left[N^{(5)}_0 + N^{(8)}_0
\right] & \eqmodone \ k_{02} + k_{12}\cr
-\frac12\left[N^{(2)}_0 + N^{(8)}_0
\right] & \eqmodone \ k_{03} + k_{13}\cr
-\frac12\left[N^{(5)}_0 + N^{(8)}_0
\right] & \eqmodone \ k_{04} + k_{14}\cr}
$$
from which it follows that a single gravitino survives in the physical
spectrum if and only if
$$
k_{02}+ k_{12} \eqmodone k_{04} + k_{14} \ .
\nfr{Susyc}
This is then the condition for the model to be $N=1$ supersymmetric.
The same analysis applies to the gauginos, leading
again to this condition for spacetime supersymmetry.

It is convenient to rewrite the GSO conditions for the Ramond zero modes
in term of Pauli matrices.
Consider the generic projection condition
$$ \frac12 \left(\sum_{l\in \bar{I}} \bar{N}^{(l)}_0
- \sum_{l \in I} N^{(l)}_0 \right) \eqmodone r \ ,
\efr
where $r \in \{0,1/2\}$ and the left-hand side involves a total of $m$
number operators. Then, since $N^{(l)}_0 = \frac12 (1+\sigma^{(l)}_3)$
for zero-mode excitations,
this projection condition can be rewritten as
$$\eqalignno{ & \bigotimes_{l\in \{\bar{I},I\}} \sigma^{(l)}_3 \ =\
\exp\left\{2\pi
i [ r +\frac12 ] \right\} \qquad {\rm for} \qquad
m\ {\rm odd} &\numali \cr
& \bigotimes_{l\in \{ \bar{I},I\}} \sigma^{(l)}_3 \ =\ \exp\left\{2\pi i
r \right\} \qquad\qquad {\rm for} \qquad
m\ {\rm even} \ . \cr}
$$
For example, the projection conditions for the gravitino can be
rewritten
as
$$\eqalignno{ & \bigotimes_{l=0,1,2,5,8} \sigma^{(l)}_3 \ =\ \exp
\left\{2\pi i
\left[ k_{00} + k_{01} \right] \right\} &\numali \cr
& \bigotimes_{l=5,8} \sigma^{(l)}_3 \ =\ \exp\left\{2\pi i
\left[ k_{02} + k_{12} \right] \right\}\cr
& \bigotimes_{l=2,8} \sigma^{(l)}_3 \ =\ \exp\left\{2\pi i
\left[ k_{03} + k_{13} \right] \right\}\cr
& \bigotimes_{l=5,8} \sigma^{(l)}_3 \ =\ \exp\left\{2\pi i
\left[ k_{04} + k_{14} \right] \right\}\ .\cr}
$$
Finally we list the GSO projection conditions for the $\alpha' M^2=1$
spacetime fermions existing in the ${\bf W}_{13}$ sector:
$$\eqalignno{&\Gamma^5 \otimes \sigma_3^{(5)} \ =\ \exp\left\{2\pi i
\left[k_{00} + k_{01} + k_{03}+ k_{13} +\frac12 \right]\right\} &
\nameali{gsospinor}\cr
&\sigma_3^{(\overline{17})}\otimes \sigma_3^{(4)} \ =\ \exp\left\{2\pi i
\left[k_{02} + k_{12} + k_{13}+ k_{23} +k_{04} + k_{14} + k_{34}
+\frac12 \right]\right\} \cr
&\Gamma_{SO(14)} \otimes \sigma_3^{(\overline{17})}
\otimes  \Gamma^5 \otimes\sigma_3^{(7)} \ =\ \exp\left\{2\pi i
\left[k_{00} + k_{01} + k_{03}+ k_{04} +k_{14} + k_{34}
+\frac12 \right]\right\} \cr
&\Gamma_{SO(4)} \otimes \sigma_3^{(\overline{17})}
\otimes \Gamma^5 \otimes\sigma^{(6)}_3
\otimes\sigma^{(9)}_3 = \exp\left\{2\pi i\left[k_{00} + k_{01} +
k_{02}+ k_{03} +k_{12} + k_{23} +\frac12 \right]\right\} ,\cr}
$$
where we introduced the space-time chirality operator $\Gamma^5 \equiv
\sigma_3^{(0)} \otimes \sigma_3^{(1)}$.  In a similar way, we introduce
the chirality matrices in the spinor representation of
the gauge groups $SO(14)$ and $SO(4)$, they are $\Gamma_{SO(14)} =
\bigotimes_{l=1}^7\sigma_3^{(\bar{l})}$ and $\Gamma_{SO(4)} =
\bigotimes_{l=15}^{16}\sigma^{(\bar{l})}_3$.

For any choice of the $k_{ij}$, the first equation tells us that the
internal ``spin'' in space $(5)$ is completely determined by the
spacetime chirality. Both chiralities are possible, as they should be
for a massive fermion. From the second equation we learn that
the ``spin'' in space $(4)$ is completely determined by the
$U(1)$ charge of the particle; and finally, $\sigma_3^{(7)}$ and
$\sigma_3^{(6)}$ are specified by the $SO(14)$ and $SO(4)$ chirality,
respectively. Thus, for any choice of $U(1)$ charge and $SO(14)$ and
$SO(4)$ chiralities, there exist two spin $1/2$ fermions labelled by
the eigenvalues of $\sigma_3^{(9)}$, $\pm 1$.

Analogously, one can check that the projection conditions for
the supersymmetric scalar partners of these fermions, in the
${\bf W}_{03}$ sector, leave only two free indices, the family
index in the ninth space and the index in the eighth space which
labels the two supersymmetric scalar partners of each massive
fermion.
\section{Supersymmetry of the spectrum}
\setchap{sectSUSY}
We are now in a position to understand how spacetime supersymmetry
manifests itself in the spectrum.

A necessary (but not sufficient) condition for a
generic KLT model to be spacetime supersymmetric is that,
among the vectors describing the possible boundary conditions, there
exists one, ${\bf W}_{\rm SUSY}$, where all components corresponding to
fermions carrying gauge charges are zero, and the first right-moving
component is $1/2$. It is not hard to see why: If the model is
supersymmetric, then for any state in some given sector ${\bfmath\alpha}$
there must exist an associated sector $\tilde{{\bfmath\alpha}}$
containing the superpartner state. If the original sector
is fermionic $(\alpha_1 = 0)$ the associated one is bosonic
$(\tilde{\alpha}_1 = 1/2)$ and vice versa;
furthermore, if the two states are to have the same charges, it is
necessary for the gauge charges of one sector to
run over the same set of values as those of the other sector, that is,
for all world-sheet fermions carrying gauge charges to have the
same boundary conditions in the two sectors. Thus,
${\bf W}_{\rm SUSY} = {\bfmath\alpha} - \tilde{{\bfmath\alpha}}$.

In our toy model ${\bf W}_{\rm SUSY}$ must have the form
$$
\left((0)^{22} \vert (\frac12)(***)(***)(***)\right)\ .
\efr
There is only one such vector, namely
$$
{\bf W}_{\rm SUSY} =
{\bf W}_0 + {\bf W}_1 \ =\ \left( (0)^{22} \vert (\frac12) (\frac12
00)^3
\right) \ .
\efr
In conclusion, our toy model is spacetime supersymmetric if and only if
equation \Susyc\ holds. And given a
state in the supersymmetric model in the sector $m{\bf W}$, the
superpartner resides in the sector ${\bf W}_0 + {\bf W}_1 + m{\bf W}$.

Notice that ${\bf W}_{\rm SUSY}$ also exchanges the boundary conditions
of
the internal world-sheet fermions $\psi_{(2)}$, $\psi_{(5)}$ and
$\psi_{(8)}$.
The associated degrees of freedom are not family indices for the states
and should be considered instead as enumerative indices for the elements
of the
spacetime supermultiplets.

\chapter{Amplitudes, Vertex Operators and Cocycles}
In this section we introduce the tools necessary for the computation of
arbitrary amplitudes in a KLT string model. We will restrict ourselves
to one-loop amplitudes but the generalization to higher loops is
straightforward. For convenience we will also adopt the operator
formalism, even if it is quite simple to express the following
formul\ae\ in terms of Polyakov path integrals.

We define the $T$-matrix element as the connected $S$-matrix element
with certain normalization factors removed
$$\eqalignno{
& { \langle \lambda_1, \dots , \lambda_{N_{\rm out}} \vert S_c \vert
\lambda_{N_{\rm out}+1}, \dots , \lambda_{N_{\rm out} + N_{\rm in}}
\rangle \over \prod_{i=1}^{N_{\rm tot}} \left( \langle \lambda_i \vert
\lambda_i \rangle \right)^{1/2} }  = & \nameali{Smatrix} \cr
& i (2\pi)^4 \delta^4 (p_1 + \dots p_{N_{\rm tot}}) \prod_{i=1}^{N_{\rm
tot}} (2 E_i V)^{-1/2} \ T(\lambda_1, \dots , \lambda_{N_{\rm out}} \vert
\lambda_{N_{\rm out}+1}, \dots , \lambda_{N_{\rm out} + N_{\rm in}} )
 \ , \cr } $$
where $N_{\rm tot} = N_{\rm in} + N_{\rm out}$ is the total number of
external states. All momenta are oriented inwards so that a string state
is considered ingoing (outgoing) if $p_i^0 > 0$ ($p_i^0 < 0$). $E_i =
|p_i^0|$ is the energy of the $i$'th string state and $V$ is the usual
volume-of-the-world factor.

Corresponding to each state $\vert \lambda \rangle$ we have a vertex
operator ${\cal V}_{\vert \lambda \rangle} (z,\bar{z})$ and
the 1-loop contribution to the $T$-matrix element, $T^{1-{\rm loop}}$,
is given by the operator formula
$$\eqalignno{ & T^{1-{\rm loop}}
(\lambda_1, \dots , \lambda_{N_{\rm out}} \vert
\lambda_{N_{\rm out}+1}, \dots , \lambda_{N_{\rm out} + N_{\rm in}} ) \ =
\ C_{g=1} \int \prod_{I=1}^{N_{\rm tot}}
\di^2 m^I\ \times & \nameali{Tmatrix} \cr
&  \sum_{m_i,n_j}  C^{{\bfmath\alpha}}_{{\bfmath\beta}}
\wew{\left| \prod_{I=1}^{N_{\rm tot}} (\eta_I \vert b)
\prod_{i=1}^{N_{\rm tot}} c(z_i) \right|^2
\prod_{A=1}^{N_B+N_{FP}}
\Pi (w_A)\ {\cal V}_{\langle \lambda_1 \vert} (z_1,\bar{z}_1) \dots
{\cal V}_{\vert \lambda_{N_{\rm tot}} \rangle} (z_{N_{\rm tot}},
\bar{z}_{N_{\rm tot}})}  \ . \cr} $$
Here the constant $C_{g=1}$ gives the correct normalization of the vacuum
amplitude. In $D=4$ it is given by [\Ref{Kaj}]
$$ C_{g=1} = \left( {1 \over 2\pi } \right)^2 (\alpha')^{-2} \ .
\nfr{vacuumnorm}
$m^I$ is a modular parameter, $\eta_I$ is the corresponding Beltrami
differential~[\Ref{Phong}], and the overlap $(\eta_I \vert b)$ with
the antighost field $b$ is given explicitly in ref.~[\Ref{Kaj2}].
The integral is over one
fundamental domain of $N$-punctured genus-one moduli space.
By definition the correlator $\wew{\dots}$ includes the partition
function (our conventions and normalizations for the partition function
can be found in appendices A and B).

In an amplitude involving $N_B$ space-time bosons and $2N_{FP}$
space-time fermions we have the insertion of $N_B + N_{FP}$
Picture Changing Operators (PCOs) $\Pi(w_A)$, given by eq.~\fdef{pco}
below, at
arbitrary points $w_A$ on the Riemann surface. In practical calculations
we will always choose to insert one PCO at each of the vertex operators
describing the space-time bosons. This leaves $N_{FP}$ PCOs at
arbitrary points.

In order to introduce explicitly the vertex operators it is convenient
to bosonize all complex fermions according to
$$\eqalignno{ &
\psi_{(l)}(z) = e^{\phi_{(l)}(z)} c_{(l)} \qquad\qquad\qquad
\psi_{(l)}^*(z) = e^{-\phi_{(l)}(z)} c^*_{(l)} &
\nameali{bosonization} \cr
& \bar{\psi}_{(\bar{l})}(\bar{z}) = e^{\bar{\phi}_{(\bar{l})}(\bar{z})}
c_{(\bar{l})} \qquad\qquad\qquad
\bar{\psi}_{(\bar{l})}^*(\bar{z}) = e^{-\bar{\phi}_{(\bar{l})}(\bar{z})}
c^*_{(\bar{l})} \ , \cr }  $$
where
the scalar field $\phi_{(l)}$ has operator product expansion (OPE)
$$ \phi_{(l)}(z) \phi_{(k)}(w) = +\, \delta_{l,k} \log (z-w) + \ldots \ .
\efr
The cocycle factors $c_{(l)}$ guarantee the correct anti-commutation
relations between different fermions. We will return to them in the next
subsection.

The ground state in the sector specified by $\alpha_l$ is created from
the conformal vacuum by the spin field operator
$$ S^{(l)}_{a_l}(z) = e^{a_l \phi_{(l)}(z) } (c_{(l)})^{a_l} \ ,
\nfr{bosspinf}
with $a_l \in [-\frac12 ; \frac12 ]$ given by $\frac12 - \alpha_l
\ {\rm mod} \ 1$. Notice that  the R case ($\alpha_l=0$)
is unique in having two vacua, corresponding to
$a_l = \pm 1/2$. We will sometimes use the
abbreviation $S_{\pm} \equiv S_{\pm 1/2}$. An expression similar to
\bosspinf\ holds for the left-movers.

The scalar field is related to the fermion number current by
$$ \partial \phi_{(l)} = - \psi_{(l)}^{*} \psi_{(l)} = - i \psi_{(l)}^1
\psi_{(l)}^2 \ , \nfr{current}
and the corresponding number operators
$$ J_0^{(l)} = \oint_0 {dz \over 2\pi i} \partial \phi_{(l)} (z)
\nfr{unumb}
satisfy
$$ [J_0^{(l)}, \phi_{(k)}] = \delta^l_k \nfr{phinumb}
and differ from the fermion number operators \fermnum\ only by a
constant term:
$$ J_0^{(l)} = N^{(l)}_{\modone{\alpha_l}} + \modone{1-\alpha_l} -
\frac12 \ . \efr

We also ``bosonize'' the superghosts in the standard way
$$\eqalignno{ & \beta = \partial \xi \, e^{-\phi} (c_{(11)})^{-1} &
\nameali{bossupgh} \cr
& \gamma = e^{\phi} c_{(11)} \, \eta \ , \cr } $$
where
the scalar field $\phi$ has the ``wrong'' metric
$$ \phi(z) \phi(w) = - \log (z-w) + \ldots \ , \nfr{supghOPE}
and $c_{(11)}$ is another cocycle factor.
The PCO now assumes the form
$$\Pi = 2c\del \xi + 2 e^\phi c_{(11)}T^{[X,\psi]}_F -
\frac12 \del(e^{2\phi} (c_{(11)})^2 \eta b)
-\frac12 e^{2\phi}(c_{(11)})^2 (\del\eta) b \ .
\nfr{pco}
If we define $\phi_{(11)}
\equiv \phi$ the superghost part of any physical state vertex operator is
given by eq.~\bosspinf\ for $l=11$, with
$$ a_{11} = -\frac12 - \modone{\alpha_1} = \left\{ \matrix{ -1 & {\rm in \
bosonic \ sector} \cr -1/2 & {\rm in \ fermionic \ sector} \cr }
\right. \ . \nfr{supghcharge}
In any given sector ${\bfmath\alpha}$ the vertex operator describing the
ground state of momentum $p$ now assumes the form
$$ {\cal V} = {\cal N} \cdot \prod_{\bar{l} = \bar{1}}^{\overline{22}}
\bar{S}^{(\bar{l})}_{\bar{a}_l}
\prod_{l=0}^{11}  S^{(l)}_{a_l} \cdot  e^{i k\cdot X}
\equiv {\cal N} \cdot S_{\Bbb A} \cdot e^{i k\cdot X} \ ,
\nfr{vertop}
where ${\Bbb A} \equiv (A;a_{11}) \equiv
(\bar{a}_1,\ldots,\bar{a}_{22};a_0,a_1,\ldots;a_{11})$ and
we introduced the dimensionless momentum
$k_\mu \equiv \sqrt{{\alpha^\prime \over 2}} \, p_\mu$.
The normalization constant ${\cal N}$ may be found using the method of
ref.~[\Ref{Kaj}] (see also Appendix \fdef{Appnormvert}).

Vertex operators describing excited states are constructed using the
standard connection between mode operators and field operators.
Physical external states are described by vertex operators ${\cal V}$
such that $\bar{c} c {\cal V}$ is BRST invariant.
The BRST currents are given by
$$\eqalignno{ & j_{\rm BRST} = c T_B^{[X,\psi,\beta,\gamma]} - c b
\partial c - T_F^{[X,\psi]} e^{\phi} c_{(11)} \, \eta - {1 \over 4}
e^{2\phi} (c_{(11)})^2 \, \eta (\partial \eta) b \cr
& \barj_{\rm BRST} = \bar{c} \bar{T}_B^{[\bar{X},\bar{\psi}]} - \bar{c}
\bar{b} \bar{\partial} \bar{c} \ , & \nameali{BRST} \cr } $$
where $T_B$ and $\bar{T}_B$ are the energy-momentum tensors.
The first-order pole in the OPE of $\barj_{\rm BRST}$ with $\bar{c} c
{\cal V}$, as well as the first order pole in the OPE of the first two
terms of $j_{\rm BRST}$ with $\bar{c} c {\cal V}$, vanish merely by
imposing that the vertex operator ${\cal V}$ is a primary
conformal field of dimension one. In particular this
implies that the string states satisfy the mass-shell
condition $\bar{L}_0 = L_0 = 0$.

The last term in $j_{\rm BRST}$ has a non-singular OPE with
$\bar{c} c {\cal V}$ for any  operator ${\cal V}$ whose
superghost part is given by $e^{-\phi}$ or $e^{-\phi/2}$. Therefore the
BRST-invariance is reduced to the requirement that the first-order pole
in the OPE
$e^{\phi(w)} c_{(11)} T_F^{[X,\psi]}(w) \ c\bar{c} {\cal V}(z,\bar{z})$
should vanish.
For a gauge boson, this equation becomes the transversality condition
$$
\epsilon\cdot k \ =\ 0\ ,
\efr
whereas for a spacetime fermion it becomes the ``Dirac equation'', as
it will be discussed in subsection \S\fdef{vertoy}.

\section{Choosing cocycles}
In this subsection we consider in detail how to define the cocycle
operators introduced by the bosonization \bosonization.

The simplest example is the case of just two complex fermions, where we
would define
$$ c_{(1)} = {\bf 1} \qquad {\rm and} \qquad c_{(2)} = e^{\pm i \pi
J_0^{(1)} } \ . \efr
Clearly $\psi_{(1)}(z_1) \psi_{(2)} (z_2) = - \psi_{(2)} (z_2)
\psi_{(1)}(z_1)$ regardless of which sign is chosen in the definition of
$c_{(2)}$. But in the presence of spin fields the two choices of sign
are no longer equivalent. Moving, say, $\psi_{(1)}$ through the spin
field operator $S^{(2)}_{a_2}$ we pick up a phase $e^{\mp i \pi a_2}$.
In general, when more than two fermions are involved, the cocycles
involve a choice of many signs. But
the various signs are not all independent: They have to be chosen in
such a way that the left- and right-moving BRST currents have well-defined
statistics with respect to all vertex operators, i.e. they are only
allowed to pick up a possible {\it overall} phase when moved through a vertex
operator. The relative signs between different terms should not change.
Otherwise, a product of BRST invariant vertex operators would not
necessarily be BRST invariant. Likewise, we must require that all
Ka\v{c}-Moody currents satisfy Bose statistics with respect to all
vertex operators; otherwise, a product of vertex operators ${\cal
V}_i$ transforming in various representations $D_i$ of the gauge group
would not necessarily transform in the tensor representation $\otimes_i
D_i$. We also require that the PCO \pco\ should obey Bose statistics
with respect to all vertex operators.
In the present subsection we discuss how to make a consistent
choice of cocycles, and we present an explicit solution in the case of
our toy model. The discussion generalizes that of ref.~[\Ref{Koste}].

We write the cocycle operators as follows
$$\eqalignno{
& c_{(\bar{l})} = c_{\rm gh}^{(\bar{l})} \cdot
\exp \left\{ i \pi \sum_{j=1}^{l-1} Y_{\bar{l} \barj}
\bar{J}^{(\barj)}_0  \right\}   \qquad {\rm for}
\qquad \bar{l} = \bar{1},\dots,\overline{22} & \nameali{cocycles} \cr
& c_{(l)} = c_{\rm gh}^{(l)} \cdot
\exp \left\{ i \pi \left( \sum_{j=1}^{22} Y_{l \barj}
\bar{J}^{(\barj)}_0 + \sum_{j=0}^{l-1} Y_{lj} J^{(j)}_0 \right)  \right\}
\qquad {\rm for}
\qquad l = 0,1,\dots,10,11 \ , \cr } $$
with
$$ \eqalignno{
& c_{\rm gh}^{(l)} \equiv \exp\{ -i \pi \varepsilon^{(l)} N_{(\eta,\xi)} \}
\exp \{ i \pi \varepsilon^{(l)} (N_{(b,c)} - N_{(\bar{b},\bar{c})}) \} &
\nameali{cgh} \cr
& c_{\rm gh}^{(\bar{l})} \equiv \exp\{ -i \pi \varepsilon^{(\bar{l})}
N_{(\eta,\xi)} \}
\exp \{ i \pi \varepsilon^{(\bar{l})} (N_{(b,c)} - N_{(\bar{b},\bar{c})})
\} \ . } $$
Here all the parameters $Y$, as well as the $\varepsilon$,
take values either $+1$ or $-1$,
and $N_{(b,c)}$, $N_{(\bar{b},\bar{c})}$ and $N_{(\eta,\xi)}$ are the
number operators of the $(b,c)$, $(\bar{b},\bar{c})$ and $(\eta,\xi)$
systems respectively. The form chosen for $c_{\rm gh}^{(l)}$ and
$c_{\rm gh}^{(\bar{l})}$ is one of convenience: It ensures that the
first term, $2c\partial \xi$, in the PCO, as well as operators like
$\bar{b} b$ and $\bar{c} c$, commute with any spin field operator
\bosspinf.

It is convenient to introduce a more compact notation: Let capital
indices $K$ and $L$ run over the set of values $\{ \bar{1}, \dots,
\overline{22}; 0,1, \dots, 10,11\}$. Define
$$ \Phi_{(L)} =  \left\{
\matrix{  \bar{\phi}_{(\bar{l})} & {\rm
for} &  L=\bar{l} = \bar{1},\dots,\overline{22} \cr
\phi_{(l)} & {\rm
for} &  \ L=l = 0,1,\dots,10,11  \cr } \right. \ . \nfr{newphi}
We may then recast the definitions \cocycles\ on the form
$$ C_{(L)} = C_{\rm gh}^{(L)} \cdot e^{i \pi e_{(L)} \cdot Y \cdot J_0 }
\ ,
\efr
where the $(\overline{22} \vert 12) \times (\overline{22} \vert 12 )$ matrix
$$ Y_{L L'} =  \left[ \matrix{ Y_{\bar{l} \bar{l}'} & 0 \cr
Y_{l \bar{l}'} & Y_{l l'} \cr } \right]   \efr
is lower triangular, $J_0$ is the $(\overline{22} \vert 12)$ vector of number
operators \unumb\ and $e_{(L)}$ is the unit vector with components $(e_{(L)})_K
= \delta_{L,K}$.

{}From the definitions \bosonization , \bosspinf\ and \cocycles\
one finds for $K \neq L$
$$ \Psi_{(L)}(z,\bar{z}) S^{(K)}_{{\Bbb A}_K}(w,\bar{w}) =
S^{(K)}_{{\Bbb A}_K}(w,\bar{w})
\Psi_{(L)}(z,\bar{z}) e^{-i \pi Y_{KL}
{\Bbb A}_L + i \pi Y_{LK} {\Bbb A}_K } \ , \nfr{cone}
where we introduced the rather obvious notation
$$ \Psi_{(L)} = e^{\Phi_{(L)}} C_{(L)} \qquad {\rm and} \qquad
S^{(L)}_{{\Bbb A}_L} = e^{{\Bbb A}_L \Phi_{(L)}} (C_{(L)})^{{\Bbb A}_L}
\ . \efr
In order to generalize eq.~\cone\ to the case $L=K$ we study the branch
cut behaviour present in the OPE:
$$\eqalignno{ & \psi_{(l)} (z) S^{(l)}_{a_l} (w) = (z-w)^{a_l}
e^{(1+a_l) \phi^{(l)} (w) } + \dots \qquad {\rm for} \qquad
l=0,1,\dots,10  \cr
& \bar{\psi}_{(\bar{l})} (\bar{z}) \bar{S}^{(\bar{l})}_{\bar{a}_l}
(\bar{w}) = (\bar{z}-\bar{w})^{\bar{a}_l}
e^{(1+\bar{a}_l) \bar{\phi}^{(\bar{l})} (\bar{w}) } + \dots \qquad
{\rm for} \qquad \bar{l}=\bar{1},\dots,\overline{22} \cr
& e^{\phi(z)} e^{a_{11} \phi(w)} \ \ = (z-w)^{-a_{11}}
e^{(1+a_{11}) \phi (w) } + \dots  \ . & \nameali{ctwo} \cr } $$
If we make the phase choice
$$ \left( {z-w \over w-z} \right) = e^{\varepsilon i \pi} \qquad ,
\qquad \varepsilon
= \pm 1 \ , \efr
eqs.~\cone\ and \ctwo\ may be summarized in a single equation
$$ \Psi_{(L)}(z,\bar{z}) S^{(K)}_{{\Bbb A}_K}(w,\bar{w}) =
S^{(K)}_{{\Bbb A}_K}(w,\bar{w}) \Psi_{(L)}(z,\bar{z}) e^{i \pi
\tilde{Y}_{LK} {\Bbb A}_K } \ , \nfr{cthree}
where we introduced the matrix $\tilde{Y}$ obtained by anti-symmetrizing
the lower-triangular matrix, $Y$, and adding the diagonal elements
$$ \tilde{Y}_{LL} = \left\{ \matrix{ -\varepsilon & {\rm for} & L =
\bar{l} =
\bar{1}, \dots, \overline{22} \cr
\varepsilon & {\rm for} & L = l = 0,1,\dots,10 \cr
-\varepsilon & {\rm for} & L=11 \cr } \right. \ .  \nfr{ytilde}
In order to ensure that all spin field operators commute with $\beta$
and $\gamma$ we take
$$ \varepsilon^{(L)} = \tilde{Y}_{11,L} \ .  \efr
For a generic product of spin fields, as defined in eq.~\vertop\ we find
by repeated use of \cthree\
$$ \Psi_{(L)} (z,\bar{z}) S_{\Bbb A} (w,\bar{w}) = S_{\Bbb A} (w,\bar{w})
\Psi_{(L)} (z,\bar{z}) e^{i \pi \varphi_L [{\Bbb A}]} \ , \efr
with
$$\varphi_L [{\Bbb A}] \equiv \sum_K \tilde{Y}_{LK} {\Bbb A}_K
\ {\rm mod} \ 2 \ , \efr
while $\Psi_{(L)}^{*}(z,\bar{z})$ picks up the complex conjugate phase.

We are now in a position to investigate the constraints on the matrix
$\tilde{Y}$ imposed by the requirement that the BRST current should
have well-defined statistics with respect to all vertex operators.
Since all raising
(and lowering) operators do have well-defined statistics
it is sufficient to consider just the ground state vertex
operators. These are given by eq.~\vertop\ with
$$ {\Bbb A}_L = \frac12 - \sum_i m_i ({\bf V}_i)_{(L)} + {\Bbb N}_L \ ,
\nfr{partchoice}
where ${\bf V}_i$ is the $(\overline{22}\vert 12)$ vector obtained from
${\bf W}_i$ by adding the components $({\bf V}_i)_{(0)}$ and
$({\bf V}_i)_{(11)}$ given in the obvious way by
$$ ({\bf V}_i)_{(0)} = ({\bf V}_i)_{(11)} = ({\bf V}_i)_{(1)} =
({\bf W}_i)_{(1)} = s_i \ , \efr
and ${\Bbb N}_L$ is a set of appropriate integers.
Now we see that whereas the left-moving BRST current has well-defined
statistics with respect to $S_{\Bbb A}$,
$$ \barj_{BRST}(\bar{z}) S_{\Bbb A}(w,\bar{w}) = S_{\Bbb A} (w,\bar{w})
\barj_{BRST} (\bar{z}) e^{i\pi \varphi_{11}[{\Bbb A}]} \ , \efr
the different terms in the right-moving BRST current will in general not
pick up the same phase. Particularly non-trivial is the requirement that
all terms in the supercurrent \supcur\ pick up the same phase.

First of all, in order for the real fermions $\psi_{(l)}^m,
l=0,1,\dots,10$ to have well-defined statistics we need to
have
$$ \varphi_l [{\Bbb A}] = {\rm integer} \qquad {\rm for} \quad l=0,1,\dots,10
\ . \nfr{constone}
Since by the constraint \supcurconst\ the number of right-moving
fermions having R boundary conditions (and hence half-integer $a_l$) is
even, eq.~\constone\ is actually equivalent to
$$ \sum_{k=1}^{22} \tilde{Y}_{l \bar{k}} \bar{a}_k = {\rm integer}
\qquad {\rm for} \quad l=0,1,\dots,10 \ . \nfr{constonee}
If we further require
$$ \varphi_0 [{\Bbb A}] \eqmodtwo \varphi_1 [{\Bbb A}] \eqmodtwo \sum_{k=2,3,4}
\varphi_k [{\Bbb A}] \eqmodtwo  \sum_{k=5,6,7} \varphi_k [{\Bbb A}] \eqmodtwo
\sum_{k=8,9,10} \varphi_k [{\Bbb A}] \ ,
\nfr{consttwo}
we find
$$ T_F^{[X,\psi]}(z) S_{\Bbb A} (w,\bar{w}) =
S_{\Bbb A} (w,\bar{w})T_F^{[X,\psi]}(z)
e^{i\pi \varphi_1[{\Bbb A}]} \ . \efr
If we also require
$$ \varphi_{11}[{\Bbb A}] \eqmodtwo \varphi_1[{\Bbb A}] \ , \nfr{constthree}
the entire BRST current $j_{BRST}$ will satisfy
$$ j_{BRST}(z) S_{{\Bbb A}}(w,\bar{w}) = S_{{\Bbb A}}(w,\bar{w}) j_{BRST}(z)
e^{i \pi \varphi_1 [{\Bbb A}]} = \pm S_{{\Bbb A}}(w,\bar{w})
j_{BRST}(z) \ , \efr
and the PCO \pco\ will pick up no phase at all
$$ \Pi (z) S_{{\Bbb A}} (w,\bar{w}) = S_{\Bbb A}(w,\bar{w}) \Pi(z) \ . \efr
The constraints \constone, \consttwo\ and \constthree\
should be satisfied for all
sectors. If we insert the value \partchoice\
and define the $(\overline{22} \vert 12)$ vectors
$$ (\tilde{\bf V}_i)_{(L)} \equiv \frac12 \sum_K \tilde{Y}_{LK} ( {\bf
V}_i )_{(K)} \ , \efr
the constraints \constone, \consttwo\ and \constthree\ are seen to be
equivalent to
$$ \eqalignno{ & (\tilde{\bf V}_i)_{(0)} \eqmodone (\tilde{\bf V}_i)_{(1)}
\eqmodone & \nameali{vone} \cr
& \sum_{k=2,3,4} (\tilde{\bf V}_i)_{(k)} \eqmodone
\sum_{k=5,6,7} (\tilde{\bf V}_i)_{(k)} \eqmodone
\sum_{k=8,9,10} (\tilde{\bf V}_i)_{(k)} \eqmodone
(\tilde{\bf V}_i)_{(11)} \cr } $$
and
$$ 2 (\tilde{\bf V}_i)_{(1)} \eqmodone 0 \ , \efr
regardless of the values of the integers ${\Bbb N}_L$.
That is, the right-moving components of the vectors $\tilde{{\bf V}}_i$
should satisfy exactly the same properties as the right-moving
components of the vectors ${\bf V}_i$.

The requirement that all Ka\v{c}-Moody currents should have Bose statistics
with respect to $S_{\Bbb A}$
places further constraints on the matrix $\tilde{Y}$;
these constraints will
also involve the left-moving components of $\tilde{\bf V}_i$.
To be more explicit we consider our toy model.

\section{Choosing cocycles in the toy model}

In the toy model described in subsection 1.2 the Ka\v{c}-Moody currents
corresponding to the gauge group \group\ are given by
$\bar{\psi}^m_{(\bar{l})} \bar{\psi}^n_{(\bar{k})} (\bar{z})$, where
$m,n = 1,2$ and $\bar{l}$ and $\bar{k}$ {\it both} belong to {\it one}
of the five subsets $\{\bar{1},\dots,\bar{7}\}$,
$\{\bar{8},\dots,\overline{14}\}$, $\{ \overline{15},\overline{16}\}$,
$\{ \overline{17}\}$, $\{\overline{18},\dots,\overline{22}\}$.

In order for these currents to have Bose statistics with respect to
any operator $S_{\Bbb A}$ we need the phases
$\varphi_{\bar{l}}[{\Bbb A}]$ to be integer, and to assume always
the same value (mod 2) for any value of $\bar{l}$ inside one of the
above subsets. This translates into the requirement that the left-moving
components of the vectors $\tilde{\bf V}_i$ should be either integer or
half-integer and satisfy
$$\eqalignno{ & (\tilde{\bf V}_i)_{(\bar{1})} \eqmodone
(\tilde{\bf V}_i)_{(\bar{2})} \eqmodone \dots \eqmodone
(\tilde{\bf V}_i)_{(\bar{7})} & \nameali{vtwo} \cr
&  (\tilde{\bf V}_i)_{(\bar{8})} \eqmodone \dots \eqmodone
(\tilde{\bf V}_i)_{(\overline{14})}  \cr
&  (\tilde{\bf V}_i)_{(\overline{15})} \eqmodone
(\tilde{\bf V}_i)_{(\overline{16})}  \cr
&  (\tilde{\bf V}_i)_{(\overline{18})} \eqmodone \dots \eqmodone
(\tilde{\bf V}_i)_{(\overline{22})} \ . \cr } $$
In any model based on vectors ${\bf W}_i$ which have only $0$ and $1/2$
entries (so that all $M_i=2$) the phases $\varphi_K[{\Bbb A}]$ are guaranteed
to be integer, since all ${\Bbb A}_K$ are either integer or half-integer and
the number of ${\Bbb A}_K$ that are half-integer is always even. The latter
statement follows from the conditions \kijeq\ which imply that ${\bf
W}_i \cdot {\bf W}_i = 2k_{ii}$ is an integer, so that the vector ${\bf
W}_i$ contains an even number of components that are $1/2$ (and hence
also an even number that are $0$). This property is inherited by any
vector $\sum_i m_i {\bf W}_i$.

The constraints \vone\ and \vtwo\ allow many choices for the cocycle
matrix $Y_{LL'}$ since the number of free variables far exceeds the
number of constraints. A convenient choice is to take~\note{For these
matrices we adopt the convention that where there is a missing entry,
one should put zero.}
$$ Y = \left[ \matrix{ Y_{7,7} & {} & {} & {} & {}  &
{} & {} & {} & {} & {}
\cr
{\bf 1}_{7,7} & Y_{7,7} &  {} & {} & {}  & {} & {} & {} &
{} & {}
\cr
{\bf 1}_{2,7} & {\bf 1}_{2,7} & {\bf 1}^{\rm LT}_{2,2} & {} & {} &
 {} & {} & {} & {} & {}
\cr
{\bf 1}_{1,7} & {\bf 1}_{1,7} & {\bf 1}_{1,2} & 0 & {}  & {}
& {} & {} & {} & {}
\cr
{\bf 1}_{5,7} & {\bf 1}_{5,7} & {\bf 1}_{5,2} & {\bf 1}_{5,1} & Y_{5,5}
 & {} & {} & {} & {} & {}
\cr
{\bf 1}_{2,7} & {\bf 1}_{2,7} & {\bf 1}_{2,2} & {\bf 1}_{2,1} &
{\bf 1}_{2,5}  & {\bf 1}^{\rm LT}_{2,2} & {} & {} & {} & {}
\cr
{\bf 1}_{3,7} & {\bf 1}_{3,7} & {\bf 1}_{3,2} & {\bf 1}_{3,1} &
{\bf 1}_{3,5}  & {\bf 1}_{3,2} & Y_{3,3} & {} & {} & {}
\cr
{\bf 1}_{3,7} & {\bf 1}_{3,7} & Y_{3,2} & {\bf 1}_{3,1} & {\bf 1}_{3,5}
 & {\bf 1}_{3,2} & {\bf 1}_{3,3} & {\bf 1}^{\rm LT}_{3,3} & {}
& {}
\cr
{\bf 1}_{3,7} & {\bf 1}_{3,7} & {\bf 1}_{3,2} & {\bf 1}_{3,1} &
{\bf 1}_{3,5}  & {\bf 1}_{3,2} & {\bf 1}_{3,3} & {\bf 1}_{3,3}
& Y_{3,3} & {}
\cr
-{\bf 1}_{1,7} & -{\bf 1}_{1,7} & -{\bf 1}_{1,2} & -1 & -{\bf 1}_{1,5}
 & 1 \ -1 & -{\bf 1}_{1,3} & -{\bf 1}_{1,3} & -{\bf 1}_{1,3} & 0
\cr}
\right] \ ,  \nfr{ccocycle}
where ${\bf 1}_{m,n}$ is the $m \times n$ matrix with all elements equal
to 1; ${\bf 1}^{\rm LT}_{m,m}$ is the $m\times m$ matrix which has all
elements $1$ in the lower triangle and the rest equal to zero; and
$$ Y_{7,7} = \left( \matrix{ 0 & {} & {} & {} & {} & {} & {} \cr
1 & 0 & {} & {} & {} & {} & {} \cr
1 & 1 & 0 & {} & {} & {} & {} \cr
1 & 1 & 1 & 0 & {} & {} & {} \cr
1 & 1 & 1 & 1 & 0 & {} & {} \cr
1 & 1 & 1 & 1 & 1 & 0 & {} \cr
-1 & 1 & -1 & 1 & -1 & 1 & 0 \cr } \right) \efr
$$ Y_{5,5} = \left( \matrix{ 0 & {} & {} & {} & {}  \cr
1 & 0 & {} & {} & {}  \cr
-1 & -1 & 0 & {} & {}  \cr
1 & 1 & -1 & 0 & {}  \cr
-1 & 1 & -1 & 1 & 0  \cr } \right) \efr
$$ Y_{3,3} = \left( \matrix{ 0 & {} & {}   \cr
1 & 0 & {}  \cr
-1 & 1 & 0  \cr } \right) \efr
$$ Y_{3,2} = \left( \matrix{ 1 & 1    \cr
-1 & 1   \cr
1 & 1   \cr } \right) \ . \efr
We define the set of 4-dimensional gamma matrices by means of the OPE
between the real space-time fermions $\psi^{\mu}$ and the space-time spin
field $S_{\alpha} \equiv S^{(0)}_{a_0} S^{(1)}_{a_1}$:
$$ \psi^{\mu} (z) S_{\alpha} (w) \eqope {1 \over \sqrt{2} }
(\Gamma^{\mu})_{\alpha}^{\ \beta} {S_{\beta} (w) \over \sqrt{z-w} } \ .
\nfr{gammadef}
Then, corresponding to each of the two possible choices for $Y_{10}$ we
have an explicit representation of the gamma matrices. The cocycle
choice \ccocycle\ (which has $Y_{10} = +1$) gives rise to
$$\eqalignno{ & \Gamma^0 = \sigma_1^{(0)} \otimes \sigma_0^{(1)} &
\nameali{gammamatrix} \cr
& \Gamma^1 = - \sigma_2^{(0)} \otimes \sigma_0^{(1)} \cr
& \Gamma^2 = \sigma_3^{(0)} \otimes \sigma_2^{(1)} \cr
& \Gamma^3 = \sigma_3^{(0)} \otimes \sigma_1^{(1)} \ , \cr } $$
where $\sigma_0$ denotes the $2 \times 2$ unit matrix.
Choosing instead $Y_{10} = -1$ would change the sign of $\Gamma^2$ and
$\Gamma^3$.
Notice that this way of defining the gamma matrices makes no reference
to the choice of model or the particular string state we happen to
consider. However, it has its own drawbacks.
Indeed, in computing amplitudes involving vertex operators like \vertop,
the gamma matrices arise from an operator product expansion like
$\psi^\mu(z) S_{\Bbb A}(w,\bar{w})$.
Obviously, in moving $\psi^{\mu} (z)$ across all the left-moving spin
fields, using eq.~\cthree, one will acquire the phase factor
$$ \exp \{ i \pi \sum_{l=1}^{22} \tilde{Y}_{s\bar{l}} \bar{a}_l \} \ , \efr
where $s=0$ for $\mu=0,1$ and $s=1$ for $\mu=2,3$. By eq.~\constonee\
this phase factor is just a sign, but it could still differ in the two
cases $s=0$ and $s=1$. In this case, to
have a Lorentz covariant formulation, one would need to redefine, say,
the gamma matrices $\Gamma^2$ and $\Gamma^3$ by a sign as compared to
\gammamatrix. To avoid this, one can make a
cocycles' choice such that
$$\sum_{l=1}^{22} (Y_{0\bar{l}} - Y_{1\bar{l}})
\bar{a}_{l}\eqmodtwo 0 \ .
\nfr{nonewgamma}
This condition is trivially satisfied by our cocycles' choice
eq.~\ccocycle.

\section{Vertex operators in the toy model}
\setchap{vertoy}

In this subsection we will introduce the vertex operators necessary for the
computation of the amplitude we have chosen to consider: The one-loop
three-point amplitude of a ``photon'' (that is, the $U(1)$ gauge boson)
and two massive charged fermions. We choose to consider the
$\alpha' M^2 = 1$ spacetime fermions that form the ground states in
the ${\bf W}_{13}$ sector. They have nonzero $U(1)$ charge and
belong to a $(\frac12, 0)$ multiplet when the model is spacetime
supersymmetric.
We call them ``electrons'' (``positrons'') depending on whether the
$U(1)$ charge is negative (positive). Obviously these names should not
be taken too literally.

The vertex operator for the photon is given by [\Ref{FMS},\Ref{Kaj}]:
$$
{\cal V}^{(-1)}_{{\rm photon}} (z,\bar{z};k;\epsilon) = {\kappa
\over \pi} \bar\psi_{(
\overline{17})}\bar\psi_{(\overline{17})}^* (\bar{z}) \
\epsilon \cdot \psi (z) \, e^{-\phi(z)} \, (c_{(11)})^{-1} e^{ik\cdot
X(z,\bar{z})}\ ,
\nfr{photonn}
where $\epsilon \cdot \epsilon = 1$ and the gravitational coupling
$\kappa$ is related to Newton's constant by $\kappa^2 = 8 \pi G_N$.
Here the label $(-1)$ specifies the superghost charge of the vertex
(i.e. the ``picture''). For future convenience we also give the once
picture-changed version of this vertex
$$ \eqalignno{ &
{\cal V}^{(0)}_{{\rm photon}} (z,\bar{z};k;\epsilon) =
\lim_{w \rightarrow z} \Pi (w) \, {\cal V}^{(-1)}_{{\rm photon}}
(z,\bar{z};k;\epsilon) = & \nameali{photon} \cr
 &\qquad\qquad  -i {\kappa \over \pi} \bar\psi_{(
\overline{17})}\bar\psi_{(\overline{17})}^* (\bar{z}) \
\left[
\epsilon \cdot \partial_z X(z) - i k\cdot \psi (z) \epsilon \cdot \psi (z)
\right] \,  e^{ik\cdot X(z,\bar{z})} \ , \cr } $$
where $k^2 = \epsilon \cdot k = 0$.

The vertex operator for the electron/positron is given by
$$ {\cal V}^{(-1/2)} (z,\bar{z};k;{\Bbb V}) \ = \
 N_f \ \overline{\bf V}^{\bar{a}}
\bar{S}_{\bar{a}} (\bar{z}) \times  {\bf V}^a   \ S_a (z) \
e^{-\frac12\phi(z)} (c_{(11)})^{-1/2}\ e^{ik\cdot X(z,\bar{z})} \ ,
\nfr{fermion}
where
$$ \bar{S}_{\bar{a}}(\bar{z}) \equiv
\prod_{l=1}^7 \prod_{l=15}^{17} \bar{S}^{(\bar{l})}_{\bar{a}_l} (\bar{z})
\qquad {\rm and} \qquad
S_a(z) \equiv \prod_{l=0,1,4,5,6,7,9} S^{(l)}_{a_l}(z) \ . \efr
The normalization constant $N_f$
is computed in appendix \fdef{Appnormvert}.
The left-moving spinor indices $\bar{a} =
\{\bar{a}_1,\dots,\bar{a}_7; \bar{a}_{15},\bar{a}_{16};\bar{a}_{17}\}$
all take values $\pm 1/2$ and indicate that the fermion transforms
in the spinor representation of the first $SO(14)$ and of the $SO(4)$,
and $\bar{a}_{17} = \pm \frac12$ is the $U(1)$ charge.
The right-moving spinor index $a = (\alpha; a_{int}) =
(a_0,a_1;a_4,a_5,a_6,a_7,a_9)$ also takes values $\pm 1/2$ and
consists of the $4$-dimensional
space-time spinor index $\alpha$ as well as family and enumerative
indices.

The spinor decomposes accordingly
$$ {\Bbb V}^A = \overline{\bf V}^{\bar{a}} \ {\bf V}^a \ , \efr
where
$$ \overline{\bf V}^{\bar{a}} =
\bar{V}_{SO(14)}^{\bar{a}_1,\dots,\bar{a}_7} \
\bar{V}_{SO(4)}^{\bar{a}_{15},\bar{a}_{16}} \
\bar{V}^{\bar{a}_{17}}_{U(1)} \ , \efr
and
$$ {\bf V}^a = V^{\alpha} v^{a_{int}} = V^{\alpha} \prod_{l=4,5,6,7,9}
v_{(l)}^{a_l} \efr
is the product of the space-time spinor $V^{\alpha}$ and the
two-dimensional ``internal'' spinors $v_{(l)}^{a_l}$.
The right-moving
spinor ${\bf V}$ satisfies a ``Dirac equation'', which (as explained
at the beginning of section \S 2) is obtained
from the requirement that the single pole in the OPE of $e^{\phi}
T_F^{[X,\psi]}$ with the vertex operator \fermion\ should vanish.
One finds
$$\eqalignno{&
{\bf V}^T (k) (\slashchar{k} + {\bf M})  \ =\ 0 &\nameali{dirac} \cr
& {\bf M} \equiv - {1 \over 2} \Gamma^5 \otimes
\sigma_3^{(4)} \otimes \left(\sigma_1^{(5)} \otimes\sigma_1^{(6)}
\otimes\sigma_1^{(7)} + \sigma_2^{(5)}\otimes\sigma_2^{(6)}
\otimes\sigma_2^{(7)}\right)\otimes
\sigma^{(9)}_0 \ . \cr}
$$
In this formula $\sigma^{(l)}_m$ for $m=1,2,3$ are the Pauli matrices
acting in the $(l)$ space whereas $\sigma^{(l)}_0$ is the two
dimensional identity matrix acting on the $(l)$ space.
The overall sign of the ``mass operator'' ${\bf M}$ depends on the
cocycle choice. The sign quoted in \dirac\ corresponds to the choice
\ccocycle.

For a generic ground state \vertop\ it is convenient to define
a ``generalized charge conjugation
matrix" ${\Bbb C}$ by
$$S_{\Bbb A} (z,\bar{z}) S_{\Bbb B} (w,\bar{w}) \eqope
{\Bbb C}_{A B} \delta_{a_{11},b_{11}} \
{1\over (z-w)^p} {1 \over (\bar{z} - \bar{w})^{\bar{p}} }
\ , \efr
where $p=\sum_{l=0}^{11} (a_l)^2$ and
$\bar{p}=\sum_{l=1}^{22} (\bar{a}_l)^2$.
This matrix is related to the choice of cocycles by
$$ {\Bbb C}_{A B} = e^{i \pi {\Bbb A} \cdot Y \cdot {\Bbb B} }
\delta_{A + B}  \ ,  \efr
where it is understood that $a_{11} = b_{11}$ is given by \supghcharge.
In the case of the electron/positron \fermion\ the cocycle choice
\ccocycle\ leads to~\note{We hope that the reader will not be confused
by ${\rm C}$ the conjugation matrices, $C_{g=1}$ the normalization of the
amplitude as in eq.~\vacuumnorm, and $C_{{\bfmath\beta}}^{{\bfmath\alpha}}$
the phase coefficients of the sum over the spin structures given by
eq.~\phases.}
$$ {\Bbb C}_{AB} = i e^{i \pi \varphi_{\rm c}}
\left( {\rm C}_{SO(14)} \otimes {\rm C}_{SO(4)} \otimes
\sigma_1^{(\overline{17})} \right)_{\bar{a} \bar{b}} \
{\bf C}_{ab} \ , \nfr{bigcmatrix}
where
$$ {\rm C}_{SO(14)} = \sigma_2^{(\bar{1})} \otimes \sigma_1^{(\bar{2})}
\otimes \sigma_2^{(\bar{3})} \otimes \sigma_1^{(\bar{4})} \otimes
\sigma_2^{(\bar{5})} \otimes \sigma_1^{(\bar{6})} \otimes
\sigma_2^{(\bar{7})} \efr
and
$$
{\rm C}_{SO(4)} = \sigma_2^{(\overline{15})} \otimes
\sigma_1^{(\overline{16})}
\efr
are standard spinor metrics; and
$$ {\bf C}_{ab} =
{\rm C} \otimes \sigma_1^{(4)} \otimes \sigma_1^{(5)} \otimes \sigma_2^{(6)}
\otimes \sigma_1^{(7)} \otimes \sigma_1^{(9)} \ , \nfr{cmatrix}
where
$$ {\rm C} = \, -i\, \sigma_2^{(0)} \otimes \sigma_1^{(1)} \efr
is the standard charge conjugation matrix which satisfies
$$ \Gamma^\mu {\rm C} \ =\ - {\rm C} \left( \Gamma^\mu \right)^T
\qquad \qquad C^T = C^{-1} = - C
\ . \efr
The phase appearing in \bigcmatrix\ depends on whether
$\bar{b}_1 + \dots +\bar{b}_7 + \bar{b}_{15} + \bar{b}_{16} +
\bar{b}_{17} + b_0 + b_1 + b_4 + \dots + b_7 + b_9 - 1/2$ is an even or
an odd integer. When ${\Bbb C}$ acts on the spinor ${\Bbb V}$ this is in
turn determined by the GSO projections
\gsospinor. One finds
$$ \varphi_{\rm c} = + 3/4 + \modone{k_{00} + k_{01} + k_{03}} \ . \efr
Notice that the charge conjugation matrix satisfies
$$
{\bf M} {\bf C} \ =\ {\bf C} {\bf M}^T \ .
\efr
For a generic choice of cocycles, the precise form of ${\Bbb C}$ can
change, but one may verify that any choice of cocycles consistent with
the constraints \consttwo\ leads to a matrix ${\bf C}$ such that
$$
\Gamma^\mu {\rm C} \ =\ -\eta\, {\rm C} \left( \Gamma^\mu \right)^T
\qquad\qquad {\bf M} {\bf C} \ =\ \eta\,{\bf C} {\bf M}^T \ ,
\nfr{MCsign}
with $\eta=\pm 1$, so that the Dirac equation \dirac\ is equivalent to
$$ (\slashchar{k} - {\bf M} ) {\bf C} {\bf V} (k) = 0 \ . \nfr{newdirac}
The choice $\eta=+1$ is preferable since only then is C the standard
charge-conjugation matrix, but in what follows we will only need
eq.~\newdirac.
\chapter{A Sample Calculation: The Anomalous Magnetic Moment at 1 Loop}
In this subsection we perform the explicit computation of a 1-loop
amplitude with
external space-time fermions.
It is convenient to demonstrate how the machinery works in
a simple example, the procedure
for any other 1-loop amplitude being completely analogous.

For the reasons explained in the introduction, we have chosen to
compute the three point amplitude of one photon \photonn\ and two
``electrons/positrons'' \fermion, i.e. we consider, say, the process
$$ \ e^{\pm} \ \rightarrow \ e^{\pm} \ + \ \gamma  \ . \efr
The $1$-loop $T$-matrix element for this process is given by
eq.~\Tmatrix:
$$\eqalignno{ & T^{1-{\rm loop}}
(\ e^{\pm} \ \rightarrow \ e^{\pm} \ + \ \gamma ) \ = \ &
\nameali{fundfora} \cr
& C_{g=1} \int {\rm d}^2\tau {\rm d}^2z_1 {\rm d}^2z_2 \
\sum_{m_i,n_j} C^{{\bfmath\alpha}}_{{\bfmath\beta}} \
\langle \langle \,
\vert (\eta_{\tau}\vert b) (\eta_{z_1}\vert b) (\eta_{z_2}\vert b)
\, c(z)c(z_1)c(z_2)\vert^2 \ \times \cr
&\Pi(w_1)\ \Pi (w_2) \ {\cal V}^{(-1)}_{\rm photon} (z,\bar{z};k;\epsilon)
\ {\cal V}^{(-1/2)} (z_1,\bar{z}_1;k_1;{\Bbb V}_1) \
{\cal V}^{(-1/2)} (z_2,\bar{z}_2;k_2;{\Bbb V}_2) \, \rangle \rangle \ , \cr }
$$
where we used translational invariance of the torus to fix the position
$z$ of the photon vertex operator at an arbitrary value.

We would like also to stress that the computation will be done without
explicitly using the cocycles' choice eq.~\ccocycle, we will only need
eq.~\newdirac\ which follows from
the general properties of the cocycles as
discussed in the previous section.

Before turning to the actual computation of the correlation functions
appearing in \fundfora\ we would like to make a few observations.
\section{Decomposition in Lorentz structures}
As it is obvious from Lorentz covariance and from the spacetime
structure of the
expression \fundfora\ the amplitude must have an on-shell
Lorentz decomposition which can be written as follows
$$ \eqalignno{ & T^{1-{\rm loop}}
(\ e^{\pm} \ \rightarrow \ e^{\pm} \ + \ \gamma ) \ =
\ \epsilon_\mu {\bf V}_1^T {\bf M}^2\, \Gamma^\mu {\bf C\, V}_2\,
T^{1-{\rm loop}}_{{\rm REN}}
+ & \nameali{decomp} \cr
&\qquad \epsilon_\mu  k_\nu {\bf V}^T_1 {\bf M}\, \Gamma^{\mu\nu} {\bf C
\, V}_2 \, T^{1-{\rm loop}}_{{\rm AMM}}
+  \epsilon_\mu k_\nu {\bf V}_1^T {\bf M}\, \Gamma^{\mu\nu}
\Gamma^5 {\bf C\, V}_2\,T^{1-{\rm loop}}_{{\rm PEDM}}\ . \cr}
$$
Here the first term has the same structure as the tree-level amplitude
and will be absorbed by a combination of vertex, wave-function and mass
renormalization. The second and third terms contribute to
the Anomalous Magnetic Moment and the Pseudo-Electric Dipole
moment respectively. Our aim is to compute these
two contributions, $T^{1-{\rm loop}}_{{\rm AMM}}$ and
$T^{1-{\rm loop}}_{{\rm PEDM}}$, and discuss when they vanish.

To arrive at the decomposition \decomp\ is actually quite non-trivial in
our bosonized approach. As is clear from the OPE \gammadef\ each factor
$\psi^\mu$ appearing in \fundfora\ should give rise to a gamma matrix.
However, the gamma matrices only appear after all the cocycle algebra
has been performed. Furthermore, Lorentz covariance requires that the
quantities $T^{1-{\rm loop}}_{{\rm REN}}$, $T^{1-{\rm loop}}_{{\rm
AMM}}$ and $T^{1-{\rm loop}}_{{\rm PEDM}}$ do not depend on the values of
the Lorentz vector indices. This only turns out to be the case by means
of some non-trivial identities in theta functions.
\section{Dependence on the point of insertion of the PCOs}
Before proceeding, it is convenient to make a quick analysis of the
dependence of the world-sheet integrand appearing in \fundfora\
on the PCO insertion points  $w_1$ and $w_2$.

Suppose we take the derivative of the amplitude with respect to
$w_1$. We know that the result must be zero because the amplitude
should not depend on the point of insertion of the PCO operator. In
general this comes about only {\it after} integrating over the moduli
--- the differentiation with respect to $w_1$ gives rise to a total
derivative in the integrand. However, in the present case things are
more simple. Indeed,
substituting  $\Pi(w_1)= 2 \{Q_{BRST}, \xi(w_1)\}$ in the amplitude and
then moving the BRST commutator onto the other operators we find
$$\eqalignno{ & \del_{w_1} T^{1-{\rm loop}}
(\ e^{\pm} \ \rightarrow \ e^{\pm} \ + \ \gamma ) \ = \ & \nameali{totder} \cr
& -2 C_{g=1} \int {\rm d}^2\tau {\rm d}^2z_1 {\rm d}^2z_2 \
\sum_{m_i,n_j} C^{{\bfmath\alpha}}_{{\bfmath\beta}} \ \sum_{m^I =
\tau,z_1,z_2} {\partial \over \partial m^I} \,  \cr
& \langle \langle \, (\bar{\eta}_{\bar{\tau}}\vert \bar{b})
(\bar{\eta}_{\bar{z}_1}\vert \bar{b}) (\bar{\eta}_{\bar{z}_2}\vert
\bar{b}) {\del \over \del (\eta_{m^I} \vert b ) } \left\{
(\eta_{\tau}\vert b) (\eta_{z_1}\vert b) (\eta_{z_2}\vert b) \right\}
\, \vert c(z)c(z_1)c(z_2)\vert^2 \ \times \cr
& \del_{w_1}\xi(w_1)\ \Pi(w_2) \ {\cal V}^{(-1)}_{{\rm photon}}
(z,\bar{z};k;\epsilon) \ {\cal V}^{(-1/2)}
(z_1,\bar{z}_1;k_1;{\Bbb V}_1) \
{\cal V}^{(-1/2)} (z_2,\bar{z}_2;k_2;{\Bbb V}_2) \, \rangle
\rangle
\ , \cr }
$$
where we used that $\Pi$, as well as $\bar{c}c{\cal V}$, are BRST
invariant and that~[\Ref{Martinec}]
$$ \wew{\{Q_{{\rm BRST}}, (\eta_I\vert
b)\} \dots } = \wew{ (\eta_I \vert T_B ) \dots } = {\del  \over \del
m^I} \wew{\dots} \ . \efr
Now, by superghost charge conservation, only the part with superghost
number two in $\Pi(w_2)$ can give a non zero contribution to the
integrand. But this part of the PCO (the last two terms in eq.~\pco)
is made up only of ghosts and superghosts and thus the only $\psi^{\mu}$
appearing is the one residing in the superghost charge $(-1)$ photon
vertex operator \photonn. The Lorentz structure of the total derivative
\totder\ is therefore seen to contain only a single gamma-matrix,
contracted with the photon polarization $\epsilon$, that is, the total
derivative contributes only to the renormalization part of the
amplitude, $T^{1-{\rm loop}}_{{\rm REN}}$.

This is very fortunate, because it means that the {\it integrands\/}
appearing in
the expression for $T_{{\rm AMM}}^{1-{\rm loop}}$ and $T^{1-{\rm
loop}}_{{\rm PEDM}}$ are {\it independent} of $w_1$ and $w_2$. In
particular, the vanishing of these quantities are not obscured by the
presence of any total derivative.
Let us also note that $T^{1-{\rm loop}}_{{\rm REN}}$ is ill-defined
on-shell, since the modular integral contains divergencies in the
corners of moduli space where the loop is isolated on an external leg,
corresponding to the pinching limits $z_1 \rightarrow z_2$, $z
\rightarrow z_1$ and $z \rightarrow z_2$, as well as in the limits
corresponding to tadpole diagrams ($|z_1-z| \ll |z_2-z| \rightarrow 0$,
$|z_2-z| \ll |z_1-z| \rightarrow 0$ and $|z_1-z_2| \ll |z_1-z|
\rightarrow 0$). Some regularization and renormalization procedure is
needed to properly treat this part of the amplitude. On the other hand,
the integrands appearing in $T^{1-{\rm loop}}_{{\rm AMM}}$ and
$T^{1-{\rm loop}}_{{\rm PEDM}}$ are completely well-behaved in all these
pinching limits.

In performing the actual calculation of the amplitude it is convenient
to take the limit $w_2
\rightarrow z$ so to represent the photon by the zero superghost number
vertex operator \photon.
The other PCO we retain at an arbitrary point, $w \equiv w_1$.
By taking the limit $w_2 \rightarrow z$ in eq.~\totder\ it is easy to
see that superghost charge conservation now forces the total derivative
to vanish altogether, meaning that the integrand must be
explicitly independent of $w \equiv w_1$.

One might think that it would be
advantageous to take also the limit, say, $w \rightarrow z_1$, so as to
picture-change one of the ``electron/positron'' vertex operators; but retaining
$w$ at an arbitrary point actually leads to simpler calculations, even
though we have to deal with four rather than three vertex insertions.
A similar observation was made in ref.~[\Ref{Atick}]. Furthermore, the
eventual independence of $w$ provides a powerful check of the result.

The form of the amplitude from which we start is then
$$\eqalignno{ & T^{1-{\rm loop}}
(\ e^{\pm} \ \rightarrow \ e^{\pm} \ + \ \gamma ) \ = \ &
\nameali{fundforb} \cr
&\qquad C_{g=1} \int {\rm d}^2\tau {\rm d}^2z_1 {\rm d}^2z_2 \
\sum_{m_i,n_j} C^{{\bfmath\alpha}}_{{\bfmath\beta}}   \
\langle \langle \vert (\eta_{\tau}\vert b) (\eta_{z_1}\vert b)
(\eta_{z_2}\vert b)
\, c(z)c(z_1)c(z_2)\vert^2 \ \times \cr
&\qquad\Pi(w) \ {\cal V}^{(0)}_{\rm photon} (z,\bar{z};k;\epsilon)
\ {\cal V}^{(-1/2)} (z_1,\bar{z}_1;k_1;{\Bbb V}_1) \
{\cal V}^{(-1/2)} (z_2,\bar{z}_2;k_2;{\Bbb V}_2) \rangle
\rangle \ . \cr }
$$
\section{Computation of correlators}
\setchap{sectccorr}
If we substitute the explicit form of the vertex operators eqs.
\photon\ and \fermion\ in eq.~\fundforb, we obtain after some
rearranging of operators
$$\eqalignno{& T^{1-{\rm loop}} (\ e^{\pm} \ \rightarrow
\ e^{\pm} \ + \ \gamma ) \ = \ & \nameali{fundforc} \cr
&\ \ - C_{g=1} {\kappa\over \pi} (N_f)^2 \ {\Bbb V}_1^{A} {\Bbb V}_2^B
\sum_{m_i,n_j} C^{{\bfmath\alpha}}_{{\bfmath\beta}}  \
e^{i\pi {\Bbb A} \cdot Y\cdot {\Bbb B}}\int {{\rm d}^2 k \over \bar{k}^2k^2}
{{\rm d}^2z_1 {\rm d}^2z_2 \over \bar\omega(\bar{z}) \omega(z)}
\prod_{n=1}^\infty \vert 1 - k^n \vert^4 \times T_{L} \times T_R
\cr}
$$
where
$$ T_{L} \ =\  \wew{
\prod_{l=1}^{7} \prod_{l=15}^{16}
\left(\bar{S}_{\bar{a}_l}^{(\bar{l})} (\bar{z}_1)
\bar{S}_{\bar{b}_l}^{(\bar{l})} (\bar{z}_2)\right) \,
 \bar\del\bar\phi_{(\overline{17})}(\bar{z})
\bar{S}_{\bar{a}_{17}}^{(\overline{17})} (\bar{z}_1)
\bar{S}_{\bar{b}_{17}}^{(\overline{17})} (\bar{z}_2)} \nfr{LmP}
and
$$\eqalignno{ T_R\ =\ & \langle \langle \,
\left(\epsilon\cdot \del X(z) -i k\cdot \psi(z)
\,\epsilon\cdot \psi(z)\right)\ \times  &\nameali{RmP}\cr
& \left(\del X \cdot \psi(w) +  \sum_{m=1}^2 \psi_{(5)}^m \psi_{(6)}^m
\psi_{(7)}^m (w)\right) \prod_{l=0,1,4,5,6,7,9} \left(S_{a_l}^{(l)}
(z_1) S_{b_l}^{(l)}(z_2) \right)\ \times  \cr
& e^{ik\cdot X(z,\bar{z})} e^{ik_1\cdot X(z_1,\bar{z}_1)}
e^{ik_2\cdot X(z_2,\bar{z}_2)} e^{\phi(w)} e^{-\frac12\phi(z_1)}
e^{-\frac12\phi(z_2)} \, \rangle \rangle \ .\cr}
$$

Here we already integrated out the reparametrization ghosts by means
of the formula given in appendix \fdef{Appghosts}. The multiplier $k=
\exp\{2\pi i \tau\}$  should of course not be confused with the photon
momentum. Fermion number
conservation implies that only the two terms in the supercurrent
displayed give rise to a non-zero correlation function.

We now turn our attention to the computation of all correlators appearing
in eqs. \LmP\  and \RmP. We will not discuss in detail the correlators
involving the $X^\mu$ fields which are rather trivial and can be easily
reconstructed using the Wick theorem, with the contraction given by the
bosonic Green function (see Appendix \fdef{Appconv} for conventions).

For each bosonized complex fermion it is convenient to define a
correlation function $\vev{\dots}$ where the non-zero mode part of the
partition function has been removed
$$
\wew{ {\cal O}_1(z_1) \dots {\cal O}_N(z_N)}_{(l)}\ =\
\prod_{n=1}^\infty (1-k^n)^{-1} \vev{{\cal O}_1(z_1) \dots {\cal O}_N
(z_N) }_{(l)} \ .\nfr{gencorr}
The subscript $(l)$ is there to remind us that the correlator depends on
the spin structure.
The fundamental correlator $\vev{\prod_{i=1}^N e^{q_i\phi(z_i)}}$ is
given in Appendix \fdef{Appconv} and correlators involving $\del\phi$
can be obtained from this by differentiation.  Notice that
$$ \vev{ {\bf 1} }_{(l)} = \Theta \left[ {}^{\alpha_l}_{\beta_l}
\right] (0 \vert \tau) \ , \efr
which vanishes when the spin structure is odd.

The fundamental spin field correlator is
$$ \vev{S_{a_l}^{(l)} (z_1) S_{b_l}^{(l)} (z_2) }_{(l)} \ =\
\left( (\sigma_3^{(l)})^{S_l} \sigma_1^{(l)}\right)_{a_l,b_l}
\vev{S_+(z_1) S_-(z_2)}_{(l)} \ ,
\nfr{spincorr}
where we introduced $$S_l\equiv (1-2\alpha_l)(1+2\beta_l) \ , \efr
which is $0$ ($1$) mod 2 depending on whether the spin structure
$\left[{}^{\alpha_l}_{\beta_l}\right]$ is even (odd). Notice that the
correlator \spincorr\ develops a dependence on the sign of the charge
$a_l$ whenever the spin structure is odd. The correlator $\vev{ S_+
(z_1) S_- (z_2)}$ is given explicitly in appendix \fdef{Appconv}.

The other correlators that we need are
$$\eqalignno{
&\vev{\bar\del\bar\phi_{(\overline{17})}(\bar{z})
\bar{S}_{\bar{a}_{17}}^{(\overline{17})}(\bar{z}_1)
\bar{S}_{\bar{b}_{17}}^{(\overline{17})}(\bar{z}_2)} \ =
 & \nameali{correls} \cr
& \qquad\left( (\sigma_3^{(\overline{17})})^{1+\overline{S}_{17}}
\sigma_1^{(\overline{17})}\right)_{\bar{a}_{17}\bar{b}_{17}} \,
\vev{\bar{S}_+(\bar{z}_1)\bar{S}_-(\bar{z}_2)}_{(\overline{17})} \,
\IIb{\bar\alpha_{17}}{\bar\beta_{17}}(\bar{z},\bar{z}_1,\bar{z}_2)\ ,\cr
&\vev{\psi^\rho (w)\, S_{a_0}^{(0)} (z_1) \, S_{b_0}^{(0)}(z_2)\,
S_{a_1}^{(1)}(z_1)\, S_{b_1}^{(1)} (z_2) } \ = \cr
&\qquad {1\over \sqrt{2} } e^{-i\pi a_1Y_{10}b_0}(\Gamma^\rho
(\Gamma^5)^{S_1} \tilde{{\rm C}})_{\alpha\beta} \,
\left(\vev{S_+(z_1)S_-(z_2)}_{(1)} \right)^2\,
\Ical{\alpha_1}{\beta_1}{(w,z_1,z_2)}\ ,\cr
&\vev{(\sum_{m=1}^2 \psi^m_{(5)}\psi^m_{(6)}\psi^m_{(7)})(w)\,
S_{a_5}^{(5)} (z_1) \, S_{b_5}^{(5)}(z_2)\,
S_{a_6}^{(6)}(z_1)\, S_{b_6}^{(6)} (z_2)\, S_{a_7}^{(7)}(z_1)
\, S_{b_7}^{(7)} (z_2) }=\cr
&\qquad -{1\over 2 \sqrt{2}}
\left( \tilde{{\bf M}}  \,((\sigma_3^{(5)})^{S_5} \sigma_1^{(5)}
\otimes (\sigma_3^{(6)})^{S_6} \sigma_1^{(6)} \otimes
(\sigma_3^{(7)})^{S_7} \sigma_1^{(7)})\right)_{a_5a_6a_7,b_5b_6b_7}
\ \times\cr  &\qquad
\prod_{l=5,6,7} \vev{S_+(z_1)S_-(z_2)}_{(l)} \,
\prod_{l=5,6,7} \Ical{\alpha_l}{\beta_l}(w,z_1,z_2)\ ,  \cr
&\vev{\psi^\mu\psi^\nu(z) \, \psi^\rho (w)\,
S_{a_0}^{(0)} (z_1) \, S_{b_0}^{(0)}(z_2)\,
S_{a_1}^{(1)}(z_1)\, S_{b_1}^{(1)} (z_2)}=\ -{1\over \sqrt{2}}
e^{-i\pi a_1 Y_{10} b_0} \ \times\cr
&\qquad\qquad \left\{ (\Gamma^{\mu\nu\rho}(\Gamma_{5})^{S_1}
\tilde{{\rm C}})_{\alpha\beta}
\,\GGm{\alpha_1}{\beta_1}(z,w;z_1,z_2) \ +\right.\cr
&\qquad\qquad\qquad \left.\left((g^{\mu\rho}\Gamma^\nu
-g^{\nu\rho}\Gamma^\mu)(\Gamma_{5})^{S_1}\tilde{{\rm C}}\right)_{\alpha\beta}
\GGp{\alpha_1}{\beta_1}(z,w;z_1,z_2)\right\}\ \times\cr
&\qquad\qquad  \left(\vev{S_+(z_1)S_-(z_2)}_{(1)} \right)^2
\, \Ical{\alpha_1}{\beta_1}(z,z_1,z_2)\ , \cr
&\vev{\psi^\mu\psi^\nu(z) \, S_{a_0}^{(0)} (z_1) \,
S_{b_0}^{(0)}(z_2)\, S_{a_1}^{(1)}(z_1)\, S_{b_1}^{(1)} (z_2)}=
\  -{1\over 2}e^{-i\pi a_1 Y_{10} b_0} \ \times\cr
&\qquad\qquad \left( \Gamma^{\mu\nu}
(\Gamma^5)^{S_1} \tilde{{\rm C}}\right)_{\alpha\beta}\,
\left(\vev{S_+(z_1)S_-(z_2)}_{(1)}\right)^2  \,
\II{\alpha_1}{\beta_1}(z,z_1,z_2)  \ . \cr}
$$
Here $\Gamma^{\mu \nu}$ and $\Gamma^{\mu \nu \rho}$ are products of
gamma matrices antisymmetrized with unit weight; we also introduced the
abbreviations
$$\eqalignno{ & \tilde{\rm C}_{\alpha \beta} \equiv \delta_{a_0+b_0}
\delta_{a_1+b_1} e^{i \pi a_1 Y_{10} b_0} & \nameali{abbr} \cr
& \tilde{\bf M} \equiv \sigma_1^{(5)} \sigma_1^{(6)} \sigma_1^{(7)} -
\sigma_2^{(5)} \sigma_2^{(6)} \sigma_2^{(7)} \cr } $$
and defined the following functions of the world-sheet coordinates
$$\eqalignno{
& \Ical\alpha\beta(z,z_1,z_2) = \sqrt{ E(z_1,z_2) \over E(z,z_1)
E(z,z_2) } {\Theta \left[ {}^{\alpha}_{\beta} \right] (\mu_z \vert \tau)
\over \Theta \left[ {}^{\alpha}_{\beta} \right] (\frac12 \nu_{12} \vert
\tau)} \ , & \nameali{eIIGG} \cr
& \II\alpha\beta(z,z_1,z_2) = \del_z \log {E(z,z_1)\over
E(z,z_2)} + 2 {\omega(z)\over 2\pi i} \del_\nu \log \Teta\alpha\beta
(\nu\vert\tau)\vert_{\nu=\frac12\nu_{12}}&\cr
&\qquad = \left( \Ical\alpha\beta(z,z_1,z_2) \right)^2 \ , \cr
& \GGp\alpha\beta(z,w;z_1,z_2)={1\over 2 E(z,w)} \left\{
{\Teta\alpha\beta (\rho_{z,w}\vert\tau)\over
\Teta\alpha\beta(\frac12\nu_{12}\vert\tau)}
\sqrt{{E(z,z_1) E(w,z_2)\over E(w,z_1) E(z,z_2)}} \ + \right. \cr
&\left. \qquad\qquad\qquad\qquad\qquad
{\Teta\alpha\beta (\rho_{w,z}\vert\tau)\over
\Teta\alpha\beta(\frac12\nu_{12}\vert\tau)}
\sqrt{{E(w,z_1) E(z,z_2)\over E(z,z_1) E(w,z_2)}} \right\}\cr
&\qquad = {\Ical\alpha\beta(w,z_1,z_2)\over \Ical\alpha\beta
(z,z_1,z_2)} \left\{\del_z\log{E(z,w)\over\sqrt{E(z,z_1)E(z,z_2)}}
\ +\right. \cr&\left. \qquad\qquad\qquad\qquad\qquad
{\omega(z)\over 2\pi i} \del_\nu\log\Teta\alpha\beta(\nu\vert\tau)
\vert_{\nu=\mu_w} \right\} \ , \cr
& \GGm\alpha\beta(z,w;z_1,z_2)={1\over 2 E(z,w)} \left\{
{\Teta\alpha\beta (\rho_{z,w}\vert\tau)\over
\Teta\alpha\beta(\frac12\nu_{12}\vert\tau)}
\sqrt{{E(z,z_1) E(w,z_2)\over E(w,z_1) E(z,z_2)}} \ - \right. \cr
&\left. \qquad\qquad\qquad\qquad\qquad
{\Teta\alpha\beta (\rho_{w,z}\vert\tau)\over
\Teta\alpha\beta(\frac12\nu_{12}\vert\tau)}
\sqrt{{E(w,z_1) E(z,z_2)\over E(z,z_1) E(w,z_2)}} \right\}\cr
&\qquad = {1\over 2}\Ical\alpha\beta(z,z_1,z_2)\Ical\alpha\beta
(w,z_1,z_2) \ , \cr } $$
where
$$\eqalignno{
& \nu_{12} \equiv \int^{z_1}_{z_2} \frac\omega{2\pi i} &\numali\cr
& \mu_z \equiv \int^z \frac\omega{2\pi i} -\frac12 \int^{z_1}
\frac\omega{2\pi i} -\frac12 \int^{z_2}\frac\omega{2\pi i}  \cr
& \rho_{z,w} \equiv \int^z_w \frac\omega{2\pi i} +
\frac12 \int^{z_1}_{z_2}\frac\omega{2\pi i} \ . \cr }
$$
To arrive at the correlators \correls\ is quite tedious. In the next
subsection we give an explicit example. Notice that in \eIIGG\
we give two
different expressions for the functions $\II\alpha\beta$ and
$G^{\pm}\left[ {}^{\alpha}_{\beta} \right]$. These two expressions
appear when we compute the correlators \correls\ for different values of
the Lorentz vector indices. Lorentz covariance implies that the two
expressions are identical. This may also be proved directly. In appendix
\fdef{AppIIfunct}\ we sketch the proof of the equivalence of the two
forms given for the function $\II\alpha\beta$.

Finally, the correlator involving the superghosts is computed in Appendix
\fdef{Appghosts} and is given by
$$ \eqalignno{ &
\wew{ e^{\phi(w)}
e^{-\frac12\phi(z_1)} e^{-\frac12\phi(z_2)} }  \ =  &
\nameali{sghostcorr} \cr
& (-1)^{S_1} k^{1/2} \prod_{n=1}^{\infty} (1-k^n) {(\omega(z_1)
\omega(z_2))^{1/2} \over \omega(w) }
{ 1 \over \vev{S_+(z_1)
S_-(z_2)}_{(0)} \Ical{\alpha_1}{\beta_1}(w,z_1,z_2) } \ . \cr } $$
\section{The explicit computation of a correlator}
In this subsection we outline the computation of the
correlator
$$
\vev{ \psi^\mu\psi^\nu(w) S_{a_0}^{(0)}(z_1) S_{b_0}^{(0)}(z_2)
S_{a_1}^{(1)}(z_1) S_{b_1}^{(1)}(z_2)} \ .
\nfr{protcorr}
The other correlators in \correls\ can be obtained in a similar way.
We will compute \protcorr\ in two cases: For $\mu=0$, $\nu=1$ and for
$\mu=0$, $\nu=2$. The other cases can be worked out similarly.

Let us consider first  $\mu=0$, $\nu=1$. Since $\psi^0\psi^1(w) =
i\del \phi_{(0)}(w)$ we get
$$\eqalignno{
&\vev{ \psi^0\psi^1(w) S_{a_0}^{(0)}(z_1) S_{b_0}^{(0)}(z_2)
S_{a_1}^{(1)}(z_1) S_{b_1}^{(1)}(z_2)} \ =\ & \nameali{exampone}  \cr
&\qquad\qquad\qquad\qquad \vev{i\del \phi (w)S_{a_0}(z_1) S_{b_0}
(z_2)}_{(0)}\, \vev{S_{a_1}(z_1) S_{b_1}(z_2)}_{(1)}\ .\cr}
$$
Bosonizing the spin fields and using the formul\ae\ given in Appendix
\fdef{Appconv}, we get
$$\eqalignno{
&\vev{ \psi^0\psi^1(w) S_{a_0}^{(0)}(z_1) S_{b_0}^{(0)}(z_2)
S_{a_1}^{(1)}(z_1) S_{b_1}^{(1)}(z_2)} \ =\
i a_0 \left( (\sigma_3)^{S_0} \sigma_1 \right)_{a_0 b_0} \left(
(\sigma_3)^{S_1} \sigma_1 \right)_{a_1 b_1} \ \times\cr
&\qquad\qquad \left(\del_w \log {E(w,z_1)\over E(w,z_2)} +2
{\omega(w)\over 2\pi i}\del_\nu\log\Teta{\alpha_0}{\beta_0}
(\nu\vert\tau)\vert_{\nu=\frac12\nu_{12}}\right)\ \times\cr
&\qquad\qquad \vev{S_+(z_1) S_-(z_2)}_{(0)} \vev{S_+(z_1)
S_-(z_2)}_{(1)}\ .&\numali \cr}
$$
Since by the definitions \abbr\ and \gammamatrix\
$$ i a_0 \left( (\sigma_3)^{S_0} \sigma_1 \right)_{a_0 b_0} \left(
(\sigma_3)^{S_1} \sigma_1 \right)_{a_1 b_1} =
-\frac12 \left( \Gamma^{01} \left( \Gamma^5 \right)^{S_1} \tilde{\rm C}
\right)_{\alpha \beta} e^{-i \pi a_1 Y_{10} b_0 } \ , \efr
eq.~\exampone\ yields the result quoted in \correls, with
$\II\alpha\beta$ given by the first expression appearing in \eIIGG.

We now consider the case $\mu=0$, $\nu=2$, where we have
$$\psi^0\psi^2(w) = \frac12
\left( e^{\phi^{(0)}(w)} c_{(0)} + e^{-\phi^{(0)}(w)} (c_{(0)})^{-1}
\right)\,
\left( e^{\phi^{(1)}(w)} c_{(1)} + e^{-\phi^{(1)}(w)} (c_{(1)})^{-1}
\right) \ , \efr
and then
$$\eqalignno{
&\vev{ \psi^0\psi^2(w) S_{a_0}^{(0)}(z_1) S_{b_0}^{(0)}(z_2)
S_{a_1}^{(1)}(z_1) S_{b_1}^{(1)}(z_2)} \ =\  &\numali\cr
&\qquad\qquad
-\frac12 \left( \delta_{1+a_0+b_0} + (-1)^{S_0} \delta_{-1+a_0+b_0}
\right) \left( \delta_{1+a_1+b_1} + (-1)^{S_1} \delta_{-1+a_1+b_1}
\right)\ \times \cr
& \qquad\qquad
{E(z_1,z_2)\over E(w,z_1)E(w,z_2)} \left( {\Teta{\alpha_1}{\beta_1}
(\mu_w\vert\tau) \over \Teta{\alpha_1}{\beta_1}(\frac12\nu_{12}\vert\tau)}
\right)^2\ \times\cr &\qquad\qquad
\vev{S_+(z_1) S_-(z_2)}_{(0)} \vev{S_+(z_1)
S_-(z_2)}_{(1)} \ . \cr}
$$
It is straightforward to check that
$$ \eqalignno{&\left( \delta_{1+a_0+b_0} + (-1)^{S_0} \delta_{-1+a_0+b_0}
\right) \left( \delta_{1+a_1+b_1} + (-1)^{S_1} \delta_{-1+a_1+b_1}
\right)\  = \cr
&\qquad\qquad\qquad\qquad
\left( \Gamma^{02} \left( \Gamma^5 \right)^{S_1} \tilde{\rm
C} \right)_{\alpha \beta} e^{-i \pi a_1 Y_{10} b_0} \ , &\numali\cr}$$
so that we obtain again the result quoted in \correls\ with
$\II\alpha\beta$ now given by the second expression in \eIIGG.
Thus, once the equality of the two expressions for $\II\alpha\beta$
is proven (see Appendix \fdef{AppIIfunct}), one obtains a Lorentz
covariant formula for the correlator \protcorr:
$$\eqalignno{
&\vev{ \psi^\mu\psi^\nu(w) S_{a_0}^{(0)}(z_1) S_{b_0}^{(0)}(z_2)
S_{a_1}^{(1)}(z_1) S_{b_1}^{(1)}(z_2)} \ =\ &\numali\cr
&\qquad\quad -{1\over 2}e^{-i\pi a_1Y_{10}b_0}
\left(\Gamma_{\mu\nu} (\Gamma^5)^{S_1} \tilde{\rm C} \right)_{\alpha\beta}
\II{\alpha_1}{\beta_1}(w,z_1,z_2)\,
\left(\vev{S_+(z_1)S_-(z_2)}_{(1)}\right)^2 .\cr}
$$
In the same way one derives all the formul\ae\ in eq.~\correls.
\section{Using the GSO projections and Dirac equation}
To arrive at the final form of the amplitude, we have yet to make
various simplifications. Due to lack of space and the obvious unnecessity
of entering into too many details, we will just indicate the
main steps.

First of all, notice that substituting the correlators \correls\ into
eq.~\fundforc, we do not reconstruct directly the charge conjugation
matrix ${\Bbb C}$ or the mass matrix ${\bf M}$.
Indeed, even if in eq.~\fundforc\ there is an
overall phase factor $e^{i\pi {\Bbb A} \cdot Y\cdot {\Bbb B}}$ there also
appears a factor of the form $ \left(\sigma^{(l)}_n
(\sigma_3^{(l)})^{S_l} \sigma_1 \right)_{a_l,b_l} = \left(\sigma^{(l)}_n
(\sigma_3^{(l)})^{S_l}\right)_{a_l,b_l'}(\sigma_1)_{b_l',b_l}$
($n=0,1,2,3$) for each left and right moving fermion with R boundary
conditions. What we need to do is to rewrite
$e^{i\pi {\Bbb A} \cdot Y \cdot {\Bbb B}} = e^{i \pi {\Bbb B}' \cdot Y
\cdot {\Bbb B}} e^{i \pi ({\Bbb A} - {\Bbb B}') \cdot Y \cdot {\Bbb B} }$
where the first factor is what we need to reconstruct the ${\Bbb C}$
matrix and the second can be rewritten as a product of $\sigma_3$
matrices acting directly on the spinor ${\Bbb V}_2$. In this way one
obtains the following relations
$$\eqalignno{ & e^{i \pi {\Bbb A} \cdot Y \cdot {\Bbb B} } \prod_{l=1}^7
\prod_{l=15}^{16} \left( (\sigma_3^{(\bar{l})})^{\bar{S}_l}
\sigma_1^{(\bar{l})} \right)_{\bar{a}_l \bar{b}_l}
\left( (\sigma_3^{(\overline{17})})^{1+\bar{S}_{17}}
\sigma_1^{(\overline{17})} \right)_{\bar{a}_{17} \bar{b}_{17}}\ \times\cr
&\qquad\qquad\left(\Gamma_{*} \left( \Gamma^5 \right)^{S_1} \tilde{\rm C}
\right)_{\alpha \beta} \prod_{l=4,5,6,7,9}
\left( (\sigma_3^{(l)})^{S_l}
\sigma_1^{(l)} \right)_{a_l b_l} e^{-i \pi a_1 Y_{10} b_0} \ =  \cr
& - \left( \left( \Gamma_{SO(14)} \right)^{1+\bar{S}_1}
\left( \Gamma_{SO(4)} \right)^{1+\bar{S}_{15}}
\left( \sigma_3^{(\overline{17})}  \right)^{\bar{S}_{17}}
\Gamma_* {\bf \Gamma}_{\bf S} {\Bbb C} \right)_{AB} \ ,
&\nameali{cocgyma}\cr } $$
where $\Gamma_*$ denotes either $\Gamma^{\rho}$ or $\Gamma^{\mu \nu
\rho}$; and
$$\eqalignno{ & e^{i \pi {\Bbb A} \cdot Y \cdot {\Bbb B} } \prod_{l=1}^7
\prod_{l=15}^{16} \left( (\sigma_3^{(\bar{l})})^{\bar{S}_l}
\sigma_1^{(\bar{l})} \right)_{\bar{a}_l \bar{b}_l}
\left( (\sigma_3^{(\overline{17})})^{1+\bar{S}_{17}}
\sigma_1^{(\overline{17})} \right)_{\bar{a}_{17} \bar{b}_{17}}
\left(\Gamma_{*} \left( \Gamma^5 \right)^{S_1} \tilde{\rm C}
\right)_{\alpha \beta} \ \times\cr
&\qquad\qquad\qquad \left( \tilde{\bf M} \prod_{l=4,5,6,7,9}
\left( (\sigma_3^{(l)})^{S_l}
\sigma_1^{(l)} \right)_{a_4 a_5 a_6 a_7 a_9, b_4 b_5 b_6 b_7 b_9}
e^{-i \pi a_1 Y_{10} b_0} \right)\ =  \cr
& - 2 i \left( \left( \Gamma_{SO(14)} \right)^{1+\bar{S}_1}
\left( \Gamma_{SO(4)} \right)^{1+\bar{S}_{15}}
\left( \sigma_3^{(\overline{17})}  \right)^{\bar{S}_{17}}
\Gamma_* {\bf M} {\bf \Gamma}_{\bf S} {\Bbb C} \right)_{AB} \ ,
&\nameali{cocgymb}\cr}$$
with $\Gamma_*$ now denoting either $1$ or $\Gamma^{\mu \nu}$.
Here we defined
$$ {\bf \Gamma}_{\bf S} \equiv \left( \Gamma^5 \right)^{S_1}
\otimes_{l=4,5,6,7,9} \left( \sigma_3^{(l)} \right)^{S_l}\ .\efr
In equations \cocgyma\ and \cocgymb\ we used the
cocycles' choice \ccocycle\ since the general expression turns out to
be quite long and in this subsection we are interested only in indicating
to the reader the various steps needed to arrive at the final form
of the amplitude. In any case the reader can easily obtain the corresponding
expressions depending explicitly on the cocycles. The form of these
equations, when one leaves unspecified the choice of cocycles, differs
from equations \cocgyma\ and \cocgymb\ only for some signs appearing in the
definition of the gamma matrices and of the mass matrix ${\bf M}$.
As it will be immediately obvious, the fact that in eqs.  \cocgyma\
and \cocgymb\ we used the cocycles' choice \ccocycle\ has no consequences
on the generality of what follows.

Each term in our amplitude now has the following general
structure for what concerns the dependence on the external
gauge and Lorentz spinor indices:
$$
(\Gamma_{SO(14)})^{1+\bar{S}_1}
(\Gamma_{SO(4)})^{1+\bar{S}_{15}}
\left( (\sigma_3^{(\overline{17})})^{\bar{S}_{17}} \right)
\otimes {\cal O}\, {\bf\Gamma}_{\bf S} {\Bbb C} \ ,
\nfr{generalform}
where ${\cal O}$ denotes either $\Gamma^{\rho}$, $\Gamma^{\mu \nu
\rho}$, ${\bf M}$ or $\Gamma^{\mu \nu} {\bf M}$.
Remembering that these structures are sandwiched between the
`1' and `2' spinors, we can use the GSO projection conditions
\gsospinor\ to rewrite \generalform\ on the form
$$ \Gamma_{SO(14)} \Gamma_{SO(4)}
(\sigma_3^{(\overline{17})})^{\bar{S}_{17}+S_4+S_6+S_7}
\otimes {\cal O} \left( \Gamma^5 \right)^{S_1+S_5+S_6+S_7} {\Bbb C}
\ \exp\{2\pi i [K_{GSO}]\}  \ , \efr
where
$$\eqalignno{
K_{GSO}\ =\ &(k_{00}+k_{01}+k_{03}+k_{13})S_5\ + &\numali\cr
& (\frac12+k_{02}+k_{12}+k_{13}+k_{23}+k_{04} +k_{14}+k_{34})S_4\ +\cr
& (k_{00}+k_{01}+k_{03}+k_{04}+k_{14}+k_{34})S_7\ +\cr
& (k_{00}+k_{01}+k_{02}+k_{03}+k_{12}+k_{23})S_{6}\ , \cr}
$$
and we used the fact (following directly from the form of the ${\bf
W}$-vectors \vectors) that $\bar{S}_1 = S_7$ and $\bar{S}_{15} = S_6 =
S_9$. It is straightforward to verify that
$$\bar{S}_{17} + S_4 \eqmodtwo S_1 + S_5 + \bar{S}_{18} + S_2 + S_3 +
S_8 \efr
and, since the amplitude contains the overall factor
$$
\prod_{l=8}^{14} \Tetab{\bar\alpha_l}{\bar\beta_l}(0\vert\bar\tau)
\prod_{l=18}^{22}
\Tetab{\bar\alpha_l}{\bar\beta_l}(0\vert\bar\tau)
\prod_{l=2,3,8,10} \Teta{\alpha_l}{\beta_l}(0\vert\tau)\ ,
\nfr{zeroteta}
which vanishes whenever $\bar{S}_{8}$, $\bar{S}_{18}$, $S_2$, $S_3$,
$S_8$ or $S_{10}$ equals 1 (mod 2), it is legitimate to use the
following identity
$$
\bar{S}_{17}+S_4\ \eqmodtwo \ S_1+S_5 \ .
\nfr{effectident}
Then it is convenient to introduce
$$
S\equiv S_1+S_5+S_6+S_7 \eqmodtwo n_2m_4+n_4m_2 + \bar{S}_{18} + S_2 +
S_3 + S_8
\efr
so that the general term \generalform\ becomes
$$ \Gamma_{SO(14)} \Gamma_{SO(4)}
\left( \sigma_3^{(\overline{17})} \right)^{S} \otimes
{\cal O}\left( \Gamma^5 \right)^{S} {\Bbb C}
\ \exp\{2\pi i [K_{GSO}]\} \ . \efr
The phase factor $\exp\{2\pi i K_{GSO}\}$ combines with the summation
coefficient $C^{{\bfmath\alpha}}_{{\bfmath\beta}}$, given by
eq.~\phases, which in the particular
case of our toy model can be written as
$$\eqalignno{ C^{{\bfmath\alpha}}_{{\bfmath\beta}}\ =\ &\exp\{2\pi i
[ \frac12(m_0+n_0)
+ \frac12n_0(m_2+m_3+m_4) +\frac12 n_1^\prime (m_2+m_3+m_4)\ + \cr
&\frac12 m_3n_3 + \frac12 m_4n_4 +\frac12 m_3n_4 +
 S_4(k_{13}+k_{23}+k_{34}+\frac12) \ + \cr
& S_1 (k_{00}+k_{01}+k_{02}+k_{03}+k_{04}+k_{12}+k_{23}+k_{24}+\frac12)
\ +\cr
& S_5(k_{02}+k_{04}+k_{12}+k_{13}+k_{23}+k_{24}) \ +\cr
& S_6(k_{04}+k_{24}) + S_7(k_{04}+k_{23}+k_{24}+k_{34}) \ + \cr
& (k_{04}+k_{14}-k_{02}-k_{12})(m_4n_0 + n_4 m_0 + m_1'n_4+m_4n_1') ] \}
\ ,
&\numali\cr}
$$
where we put $\bar{S}_{8}, \bar{S}_{18}, S_2,S_3,S_8$ and $S_{10}$ equal
to zero throughout, made use of the identity
$$
S_1 + \bar{S}_{18} + S_4 + S_8 + m_1 (n_2+n_3+n_4) + n_1 (m_2+m_3+m_4)
\eqmodtwo 0 \efr
as well as \effectident\ and introduced $m_1^\prime=m_0+m_1$
and $n_1^\prime=n_0+n_1$ for future convenience.

Putting together all the phases (including the $(-1)^{S_1}$ which comes
from the superghost correlator \sghostcorr) we arrive at the overall
phase
$$\eqalignno{ K^{{\bfmath\alpha}}_{{\bfmath\beta}}\ \equiv\ &
(-1)^{S_1} e^{2\pi i K_{GSO} } C^{{\bfmath\alpha}}_{{\bfmath \beta}}\ =
&\nameali{sumcoeff} \cr
=\ & \exp\left\{2\pi i\left[ \frac12(m_0+n_0) +
\frac12 n_0(m_2+m_3+m_4) \ + \right.\right. \cr
&\frac12 n_1^\prime (m_2+m_3+m_4) +
\frac12 m_3n_3 + \frac12 m_4n_4 + \frac12 m_3n_4\ +  \cr
& \left( k_{00} + k_{01} + k_{02} + k_{03} + k_{04} + k_{12}
+ k_{23} + k_{24} \right)\, S\ +\cr
&\left.\left.(k_{04}+k_{14}-k_{02}-k_{12})(
S_4 + S_7 + m_4n_0 +m_0n_4+m_1^\prime n_4+m_4n_1^\prime)\right]\right\}
\ .
\cr}
$$
It is obvious
from eq.~\Susyc\ that for spacetime supersymmetric models, the last
term in $K_{{\bfmath\beta}}^{{\bfmath\alpha}}$ vanishes.
This will be the key point in the proof of the vanishing of the
Anomalous Magnetic Moment for spacetime supersymmetric models.

Finally, to rewrite all Lorentz structures appearing in the amplitude on
the form of eq.~\decomp\ one has to use the on-shell conditions $k^2=0$,
$k_1^2=k_2^2 = - \frac12$, momentum conservation $k+k_1+k_2 = 0$, and
the Dirac equations
$$ {\bf V}_1^T (\slashchar{k}_1 + {\bf M} ) = 0 = (\slashchar{k}_2 -
{\bf M} ) {\bf C} {\bf V}_2 \efr
from which one may derive the following useful identities
$$\eqalignno{ & {\bf V}_1^T (\Gamma^5)^S \slashchar{k} {\bf C} {\bf V}_2
\epsilon \cdot k_1 = \frac12 (1-(-1)^S) \epsilon_{\mu} k_{\nu} {\bf
V}_1^T {\bf M} \Gamma^5 \Gamma^{\mu \nu} {\bf C} {\bf V}_2 &
\nameali{usefulident} \cr
& {\bf V}_1^T \frac12 ( \slashchar{k}_1 - \slashchar{k}_2) (\Gamma^5)^S
{\bf C} {\bf V}_2 \epsilon \cdot k_1 = \frac12 (1+ (-1)^S) \left\{ -
{\bf V}_1^T {\bf M}^2 \slashchar{\epsilon} {\bf C} {\bf V}_2 + \frac12
{\bf V}_1^T {\bf M} \Gamma^{\mu \nu} {\bf C} {\bf V}_2 \epsilon_{\mu}
k_{\nu} \right\} \ . \cr} $$
\section{The final form of the amplitude}
In this subsection we display the final form of the amplitude that we
arrive at following the steps described in the previous subsections.
It is equivalent to the form of a field theory amplitude where the internal
momenta are already integrated away
whereas the integrals over the Schwinger proper-times are still to be done.
In other words, all Lorentz algebra
is already done and what is left to do is  an adimensional integral.

We present the partial amplitudes $T^{1-{\rm loop}}_{{\rm REN,AMM,PEDM}}$
as defined in eq.~\decomp, as follows
$$\eqalignno{
T^{1-{\rm loop}}_{\rm REN,AMM,PEDM} & =  C_{g=1} {\kappa \over \pi} (N_f)^2
i e^{i \pi \varphi_{\rm c}} {1 \over 4 \sqrt{2}}\
\left( \bar{V}_{SO(14),1}^T \Gamma_{SO(14)} {\rm C}_{SO(14)}
\bar{V}_{SO(14),2} \right)\ \times\cr
&\left( \bar{V}^T_{SO(4),1} \Gamma_{SO(4)}
{\rm C}_{SO(4)} \bar{V}_{SO(4),2} \right)\
\left( \bar{V}^T_{U(1),1} (\sigma_3^{(\overline{17})})^{S}
\sigma_1^{(\overline{17})} \bar{V}_{U(1),2} \right)\ \times\cr
&\sum_{n_i,m_i} K^{{\bfmath \alpha}}_{{\bfmath\beta}} \,
\int {{\rm d}^2 \tau\over ({\rm Im}\tau)^2 } {{\rm d}^2z_1
{\rm d}^2z_2 \over \bar{\omega} (\bar{z}) \omega(z)} \, \exp\left\{2\pi i
(\bar\tau -\frac12\tau)\right\}\ \times & \nameali{result} \cr
&\exp\left\{ \frac12 G_B (z_1,\bar{z}_1,z_2,\bar{z}_2)\right\}\ \times\
 \bar{{\cal Z}}_L\ \times\ {\cal Z}_R \ \times\ {\cal I}_{\rm REN,AMM,PEDM}
\cr}
$$
with
$$\eqalignno{
\bar{{\cal Z}}_L = &(\bar{\eta} (\bar{\tau}) )^{-24}
\prod_{l=1}^7 \prod_{l=15}^{17} \bar{\Theta} \left[
{}^{\bar{\alpha}_{l}}_{\bar{\beta}_{l}} \right] (\frac12
\bar{\nu}_{12} \vert \bar{\tau} )\ \times & \nameali{zl} \cr
 & \prod_{l=8}^{14} \prod_{l=18}^{22} \bar{\Theta} \left[
{}^{\bar{\alpha}_{l}}_{\bar{\beta}_{l}} \right] ( 0 \vert \bar{\tau} )
\times (\bar{E} (\bar{z}_1,\bar{z}_2))^{-5/2}
\bar{I} \left[{}^{\bar{\alpha}_{17}}_{\bar{\beta}_{17}} \right]
(\bar{z},\bar{z}_1,\bar{z}_2) \ ,
\cr } $$
$$ \eqalignno{
{\cal Z}_R = &(\eta(\tau))^{-12} {\sqrt{\omega(z_1)
\omega(z_2)}\over \omega (w)} \prod_{l=1,4,5,6,7,9}
\Teta{\alpha_l}{\beta_l} (\frac12 \nu_{12} \vert \tau)\ \times &
\nameali{zr} \cr
& \prod_{l=2,3,8,10} \Teta{\alpha_l}{\beta_l}
(0 \vert \tau)\times (E(z_1,z_2))^{-3/2} \ , \cr } $$
and finally
$$ \eqalignno{
{\cal I}_{\rm REN} = & (1+ (-1)^S) \left\{ \del_zG_B(z,z_1) -
\del_z G_B (z,z_2)  \right\} \times & \nameali{ivr}\cr
& \left\{ \del_wG_B(w,z_1) - \del_w G_B(w,z_2) -
{\prod_{l=5,6,7}\Ical{\alpha_l}{\beta_l}(w,z_1,z_2) \over
\Ical{\alpha_1}{\beta_1}(w,z_1,z_2)} \right\} \ , \cr } $$
$$ \eqalignno{
{\cal I}_{\rm AMM} = &- \frac12 (1+ (-1)^S) \times & \nameali{iamm} \cr
& \left\{ \del_zG_B(z,z_1) -
\del_z G_B (z,z_2) - \II{\alpha_1}{\beta_1}(z,z_1,z_2)
\right\} \times & \cr
& \left\{ \del_wG_B(w,z_1) - \del_w G_B(w,z_2) -
{\prod_{l=5,6,7}\Ical{\alpha_l}{\beta_l}(w,z_1,z_2) \over
\Ical{\alpha_1}{\beta_1}(w,z_1,z_2)} \right\} \ , \cr } $$
$$ \eqalignno{
{\cal I}_{\rm PEDM}& = (1-(-1)^S)\ \Biggl\{ (\del_zG_B(z,z_1) -
\del_z G_B (z,z_2))\ \times & \nameali{ipedm} \cr
& \left( \del_wG_B(w,z) - \frac12\del_w G_B (w,z_1) - \frac12\del_w
G_B (w,z_2)\  + \right. \cr
&\qquad\qquad \left.
\frac12 {\prod_{l=5,6,7}\Ical{\alpha_l}{\beta_l}(w,z_1,z_2)
\over \Ical{\alpha_1}{\beta_1}(w,z_1,z_2)}\right) +\cr
&\  (\del_w G_B(w,z_1) -
\del_w G_B(w,z_2))\, \GGp{\alpha_1}{\beta_1}(z,w;z_1,z_2)
{\Ical{\alpha_1}{\beta_1}(z,z_1,z_2)\over
\Ical{\alpha_1}{\beta_1}(w,z_1,z_2)} \ -\cr
&\ \left. \frac12 \II{\alpha_1}{\beta_1}(z,z_1,z_2)
{\prod_{l=5,6,7}\Ical{\alpha_l}{\beta_l}(w,z_1,z_2) \over
\Ical{\alpha_1}{\beta_1}(w,z_1,z_2)} \right\}\ . \cr } $$
Notice that spin structures with $S= 0$ mod 2 contribute to $T_{\rm
REN}^{1-{\rm loop}}$ and $T_{\rm AMM}^{1-{\rm loop}}$ and those with
$S=1$ mod 2 contribute to $T_{\rm PEDM}^{1-{\rm loop}}$.
\section{$w$ independence}
As a first check on the correctness of our computation, we verify that
eq.~\result\ is independent of $w$, the point of insertion of the PCO
operator.

For ${\cal I}_{\rm REN}$ and ${\cal I}_{\rm AMM}$
this is simple enough. One may verify that for all spin structures
contributing to the REN and AMM partial amplitudes (i.e. for which $S=0$
mod 2 and the factor \zeroteta\ is nonzero) it is possible to write
$$
{\prod_{l=5,6,7}\Ical{\alpha_l}{\beta_l}(w,z_1,z_2) \over
\Ical{\alpha_1}{\beta_1}{(w,z_1,z_2)}}=\II{\alpha_L}{\beta_L}(w,z_1,z_2)
\ ,
\efr
where $L=6$ if $(m_4=m_2, n_4=n_2)$, $L=5$ otherwise.
Thus the term which depends on $w$ in
$T^{1-{\rm loop}}_{\rm REN}$ and $T^{1-{\rm loop}}_{\rm AMM}$ becomes
$$\eqalignno{ {1\over \omega(w)} & \left\{\del_wG_B(w,z_1)-\del_w
G_B(w,z_2) - \II{\alpha_L}{\beta_L}(w,z_1,z_2) \right\}\ =&\numali\cr
&\qquad {1\over\omega(w)} \left\{ {\omega(w)\over \log\vert k\vert}
\log\left\vert{z_2\over z_1}\right\vert  - 2 {\omega(w)\over 2\pi i}
\del_\nu\log\Teta{\alpha_L}{\beta_L}(\nu\vert\tau)
\vert_{\nu=\frac12\nu_{12}}\right\}\ , \cr } $$
which is explicitly independent of $w$.

For the PEDM things are more complicated. We have been able to prove the
$w$ independence only implicitly, i.e. we have checked that the quantity
${1\over \omega(w)}{\cal I}_{\rm PEDM}(w)$, which is a meromorphic function
of $w$ on the torus, does not have zeros and that the residues at all poles
vanish. Thus it is a constant (as a function of $w$) and hence
independent of $w$. We do not reproduce the details of this proof here
since we will not need it in what follows.
\chapter{The Vanishing of the AMM and PEDM}
We will now briefly discuss the properties of the
$T^{1-{\rm loop}}_{\rm AMM}$ and $T^{1-{\rm loop}}_{\rm PEDM}$.
We will not discuss the $T^{1-{\rm loop}}_{\rm REN}$
term since this requires a regularization and renormalization procedure,
as pointed out in the introduction and in section \S 3.2.
\section{The vanishing of the PEDM}
\setchap{PEDMsect}
Let us consider first the PEDM. Since only spin structures with $S=1$
mod 2 contribute to $T^{1-{\rm loop}}_{\rm PEDM}$, it is clear from
eq.~\result\ that this part of the amplitude has an anomalous dependence
on the sign of the $U(1)$ charge, as compared to the tree-level
amplitude and the AMM-part of the 1-loop amplitude, because of the
presence of an extra factor $(\sigma_3^{(\overline{17})})$.~\note{We
remind the reader that the $U(1)$ factor of the gauge group is associated
with the $\overline{17}$ world-sheet fermion.}
Hence, if we normalize our spinors ${\Bbb V}_1$ and ${\Bbb V}_2$ (see
appendix D for details) such that
$$ T^{\rm tree} (\ e^{+} \ \rightarrow \ e^{+} \ + \ \gamma )
= - T^{\rm tree} (\ e^{-} \ \rightarrow \ e^{-} \ + \ \gamma ) \ , \efr
which is the behaviour consistent with charge conjugation invariance of
the $S$-matrix (since the photon is a charge conjugation eigenstate with
eigenvalue $-1$), it is clear that we will then find
$$ T^{1-{\rm loop}}_{\rm PEDM} (\ e^{+} \ \rightarrow \ e^{+} \ + \ \gamma )
= + T^{1-{\rm loop}}_{\rm PEDM}
(\ e^{-} \ \rightarrow \ e^{-} \ + \ \gamma ) \ , \efr
i.e. the PEDM, besides violating P and T, also violates C, and then CPT.
But as it was pointed out in the introduction, the KLT models should not
violate CPT perturbatively (see refs.
[\Ref{Sonoda},\Ref{Pott},\Ref{Us}]). Then $T^{1-{\rm loop}}_{\rm PEDM}$
must vanish.

It is possible to show that for spacetime supersymmetric models, the sum
over the spin structures implementing the spacetime supersymmetry makes
$T^{1-{\rm loop}}_{\rm PEDM}$ vanish point by point in moduli space.
We will not show this in details, but it works in a similar,
although more complicated, manner
as for the AMM which we will consider in the next subsection. Anyway,
for the non spacetime supersymmetric models the sum over the spin
structures does not vanish point by point in moduli space.~\note{We checked
this also numerically.}

Indeed the reason for the vanishing of $T^{1-{\rm loop}}_{\rm PEDM}$
is more general. One can show that the following identity holds:
$$
{\prod_{l=5,6,7}\Ical{\alpha_l}{\beta_l}(w,z_1,z_2) \over
\Ical{\alpha_1}{\beta_1}(w,z_1,z_2)}\ =\ (-1)^{1+S}
{\prod_{l=5,6,7}\Ical{\alpha_l}{\beta_l}(w,z_2,z_1) \over
\Ical{\alpha_1}{\beta_1}(w,z_2,z_1)}\ .
\efr
Using this identity it is possible to prove that
$$\eqalignno{
&T_{\rm REN}(z_1,z_2)\ =\ +\ T_{\rm REN}(z_2,z_1)&\numali\cr
&T_{\rm AMM}(z_1,z_2)\ =\ +\ T_{\rm AMM}(z_2,z_1)\cr
&T_{\rm PEDM}(z_1,z_2)\ =\ -\ T_{\rm PEDM}(z_2,z_1) \ , \cr}
$$
where $T_{\rm REN,AMM,PEDM}(z_1,z_2)$ are the same expressions as
for $T^{1-{\rm loop}}_{\rm REN,AMM,PEDM}$ in eq.~\result\ but
stripped of the integrals $\int {\rm d}^2z_1 {\rm d}^2z_2$.
Thus for the PEDM the integrand is odd under the exchange of
$z_1\leftrightarrow z_2$, and then it vanishes.
\section{The vanishing of the AMM for SUSY models}
Finally, we show that the AMM vanishes for spacetime supersymmetric
models according to the Ferrara-Porrati sum rules [\Ref{FP}].

The vanishing of the AMM when the model has spacetime supersymmetry
must be due to the contributions from the superpartners circulating in the
loop cancelling one another.
Therefore we expect to get zero
by summing, for any given sector $m{\bf W}$, only over the sectors
$m{\bf W}$ and $m{\bf W} + {\bf W}_0 + {\bf W}_1$. We also have to
implement the GSO projections by summing over the $n_i$. However, since
the $m_i$ and $n_i$ are equivalent by modular invariance, we again
expect to get zero only by summing over boundary conditions $n{\bf W}$ and
$n{\bf W} + {\bf W}_0 + {\bf W}_1$. The explicit calculations vindicate
this belief. This means that it is sufficient to study the right-moving
part of the amplitude (see subsection \S \sectSUSY). It is
then convenient to change basis for the ${\bf W}$-vectors, or
equivalently, for the $(m_i,n_i)$. We introduce then
$(m^{\prime}_i, n^{\prime}_i)$ by
$$\eqalignno{ & m^{\prime}_1 = m_0 + m_1 \qquad\qquad\qquad
m^{\prime}_i = m_i \quad i\neq 1 &\nameali{newnm} \cr
& n^{\prime}_1 = n_0 + n_1 \qquad\qquad\qquad\ \  n^{\prime}_i = n_i
\quad i\neq 1 \ , \cr} $$
since
$$\eqalignno{ m_0{\bf W}_0 + m_1 {\bf W}_1 =\ & m_0 ({\bf W}_0+
{\bf W}_1) + (m_0+m_1) {\bf W}_1 = &\numali \cr
& m^{\prime}_0 ({\bf W}_0+{\bf W}_1) + m^{\prime}_1 {\bf W}_1\
({\rm mod~1}) \ .\cr} $$
In the rest of this section we will
work in the new basis and we will drop the primes.

The spin structures depending on $(m_0,n_0)$ are explicitly given by
$$\eqalignno{&
\alpha_1=\frac12 m_0\ ,\quad \alpha_2 = \frac12(m_0+m_3)\ ,\quad
\alpha_5 = \frac12(m_0+m_2+m_4) &\nameali{snomo} \cr
& \alpha_8 = \frac12 (m_0+m_2+m_3+m_4) \cr}
$$
and similarly for the $\beta_i$ in terms of the $n_i$.

In a model with spacetime supersymmetry $k_{02}+k_{12}=k_{04}+k_{14}$
mod~1 and the last term in the summation coefficient \sumcoeff \
drops out so that the dependence on
$(m_0,n_0)$ is given by
$$\eqalignno{
K^{{\bfmath\alpha}}_{{\bfmath\beta}}=\exp&\left\{2\pi i\left[
\frac12(m_0+n_0) + \frac12
n_0(m_2+m_3+m_4)\ \right.\right. & \nameali{summ} \cr
&\quad\left.\left. +\ \hbox{\rm terms not depending on }
(m_0,n_0)\right]
\right\} \ .\cr}
$$
Since we want to sum over $(m_0,n_0)$ keeping fixed all the other spin
structures, it is convenient to use eq.~\fdef{spinsshift}\ to reexpress
$\Theta \left[ {}^{\alpha_l}_{\beta_l} \right]$ for $l=2,5,8$ in terms
of $\Theta \left[ {}^{\alpha_1}_{\beta_1} \right]$. In doing so, the
argument of these theta functions is shifted by $(\beta_l - \beta_1)
- \tau (\alpha_l - \alpha_1)$
and we get also an overall phase $ \exp \left\{ 2\pi i [ \frac12 n_0
(m_2+m_3+m_4)] \right\} $ cancelling the one appearing in
\summ ,
so that effectively the summation coefficients for the sum over
$(m_0,n_0)$ are reduced to
$$
\exp\left\{2\pi i\left[ \frac12(m_0+n_0) + \ \hbox{\rm
terms not depending on } (m_0,n_0)\right]\right\}\ .
\efr
Thus we expect to prove the vanishing of the AMM in the
supersymmetric case using the standard Riemann identity
equation \fdef{acfor}.
Notice that this is not possible in the non-supersymmetric case
since there is an extra dependence on $(m_0,n_0)$ in the phase
of the coefficient \sumcoeff \ given by
$$
(k_{04}+k_{14}-k_{02}-k_{12})(m_4n_0+m_0n_4) = \frac12
(m_4n_0+m_0n_4)\ ({\rm mod}~1)\ .
\efr

To show that the AMM vanishes when the model is supersymmetric, it
is convenient to extract all factors depending on $(n_0,m_0)$ from
eqs. \result, \zr\ and \iamm, arriving at the quantity
$$\eqalignno{\sum_{m_0,n_0} K^{{\bfmath\alpha}}_{{\bfmath\beta}}
&\left( {1\over \log\vert k\vert}
\log \left\vert{z_1\over z_2}
\right\vert + {2\over 2\pi i} \left.\del_\nu \log
\Teta{\alpha_1}{\beta_1}(\nu\vert\tau)\right\vert_{\nu=\frac12 \nu_{12}}
\right)\ \times \cr
& \left( {1\over \log\vert k\vert} \log\left\vert {z_1\over z_2}
\right\vert + {2\over 2\pi i }\left. \del_\nu \log
\Teta{\alpha_L}{\beta_L}(\nu\vert \tau)\right\vert_{\nu=\frac12
\nu_{12}} \right)\ \times\cr
& \prod_{l=1,5}\Teta{\alpha_l}{\beta_l}(\frac12 \nu_{12}\vert\tau)
\prod_{l=2,8} \Teta{\alpha_l}{\beta_l} (0\vert\tau)\ .&\nameali{bfor}\cr}
$$
Using the Riemann identity eq.~\fdef{acfor}, one can prove that (in the
supersymmetric case)
$$\eqalignno{&\sum_{m_0,n_0} K^{{\bfmath\alpha}}_{{\bfmath\beta}}
\prod_{l=1,5}
\Teta{\alpha_l}{\beta_l} (\frac12 \nu_{12}\vert\tau) \prod_{l=2,8}
\Teta{\alpha_l}{\beta_l} (0\vert\tau) \ =\ 0 \ , &\numali\cr
&\sum_{m_0,n_0}K^{{\bfmath\alpha}}_{{\bfmath\beta}} \left( \left.
\del_\nu\log \Teta{\alpha_1}{\beta_1} (\nu\vert\tau)
\right\vert_{\nu=\frac12 \nu_{12}}
+ \left.\del_\nu\log\Teta{\alpha_5}{\beta_5}(\nu\vert\tau)
\right\vert_{\nu=\frac12 \nu_{12}}\right)\ \times\cr
&\qquad\qquad \prod_{l=1,5}\Teta{\alpha_l}{\beta_l}(\frac12
\nu_{12}\vert\tau)
\prod_{l=2,8}\Teta{\alpha_l}{\beta_l} (0\vert\tau)\ =\ 0 \ , \cr
&\sum_{m_0,n_0} K^{{\bfmath\alpha}}_{{\bfmath\beta}}
\left.\del_\nu\log\Teta{\alpha_1}{\beta_1}
(\nu\vert\tau)\right\vert_{\nu=\frac12 \nu_{12}} \ \left.\del_\nu\log
\Teta{\alpha_5}{\beta_5}(\nu\vert\tau)\right\vert_{\nu=\frac12
\nu_{12}}\
\times\cr
&\qquad\qquad\prod_{l=1,5}\Teta{\alpha_l}{\beta_l}(\frac12
\nu_{12}\vert\tau)
\prod_{l=2,8}\Teta{\alpha_l}{\beta_l}(0\vert\tau)\ =\ 0\ .\cr}
$$
Using these identities it is straightforward to show that the quantity
\bfor\ is zero,
and thus the AMM vanishes due to the spacetime supersymmetry, as it
follows from the sum rules [\Ref{FP}].

In the non-supersymmetric case the quantity \bfor\
is nonzero. Thus to compute the value of the AMM  one
also needs to sum over the other spin structures. This is
particularly difficult since the sum over all the other spin
structures involves more than 4 theta functions and both the
right- and left-movers. Obviously it is always possible to compute
numerically the AMM since the expansion in powers of
$\exp[-2\pi\, {\rm Im} \tau]$ converges very rapidly.

\acknowledgements
We would like to thank Paolo di Vecchia and Ed Witten for very
interesting discussions and suggestions.
\appendix{Notations, Conventions and Useful Formul\ae}
\setchap{Appconv}
In this appendix we will give the notations and conventions we have
adopted for the basic (correlation) functions on the torus. We begin
by giving our conventions for the Dedekind $\eta$-function and for the
theta functions.

The Dedekind $\eta$-function is given by
$$
\eta(\tau) = k^{1/24} \prod_{n=1}^{\infty} (1-k^n)\ \, , \qquad
k = e^{2\pi i \tau} \ ,
\nfr{adedekind}
and our conventions for the theta functions are
$$\eqalignno{
& \Teta{\alpha}{\beta}(\nu\vert\tau) = e^{i\pi(\frac12-\alpha)^2\tau}
e^{2\pi i(\frac12 +\beta)(\frac12-\alpha)} e^{2\pi i(\frac12
-\alpha)\nu}\ \times & \nameali{atheta}\cr
& \qquad \qquad\qquad \prod_{n=1}^\infty
(1-k^n)(1-k^{n+\alpha-1}e^{-2\pi i (\beta+\nu)}) (1-k^{n-\alpha}
e^{2\pi i (\beta+\nu)})\cr
& \qquad\qquad\quad =\ \sum_{r\in{\Bbb Z}} e^{\pi i (r+\frac12-\alpha)^2\tau
+2\pi i (r+\frac12-\alpha)(\nu +\beta+\frac12)} \cr
& \Theta_1\equiv\Teta00\
,\quad \Theta_2\equiv \Teta{\ 0}{1/2}\ ,\quad
\Theta_3\equiv\Teta{1/2}{1/2} \ ,\quad
\Theta_4 \equiv \Teta{1/2}{\ 0} \ .  \cr} $$
The theta functions satisfy the following relations
$$\eqalignno{&\Teta{\alpha+\Delta\alpha}{\beta+\Delta\beta}(\nu\vert
\tau) = \exp\left[2\pi i\left\{ \frac12 (\Delta\alpha)^2\tau -
\Delta\alpha(\nu+\beta+\Delta\beta +\frac12)\right\}\right]\ \times \cr
& \qquad\qquad\qquad\qquad \Teta\alpha\beta (\nu-\Delta\alpha\,\tau +
\Delta\beta\vert \tau)&\nameali{spinsshift} \cr
&\Teta\alpha\beta(\nu\vert\tau) = \exp\left[2\pi i\left\{ \frac12
\alpha^2\tau -\alpha(\nu +\beta+\frac12)\right\}\right]
\Theta_1 (\nu - \alpha\tau +\beta\vert \tau)&\nameali{thetazero}\cr}
$$
and
$$\eqalignno{&\Teta{\alpha+m}{\beta+n}(\nu\vert\tau) = \exp\left[
2\pi i (\frac12-\alpha)n\right] \Teta\alpha\beta(\nu\vert\tau)
&\numali\cr
&\Teta\alpha\beta(\nu + m\tau +n\vert\tau) = &\nameali{thetashift}\cr
&\qquad\quad
\exp\left[2\pi i\left\{-\frac12 m^2\tau +(\frac12-\alpha)n -m\nu
-m(\beta+\frac12)\right\}\right]\Teta\alpha\beta (\nu\vert\tau)
\ , \cr} $$
where $m,n$ are integer numbers.

The bosonic Green function on the torus is given by
$$
G_B(z_1,\bar{z}_1;z_2, \bar{z}_2) =
2\left[ \log \vert
E(z_1,z_2)\vert - \frac12 {\rm Re}\left( \int_{z_2}^ {z_1}\omega
\right)^2
{1\over 2\pi {\rm Im}\tau} \right]
\nfr{agb}
and the prime form is
$$
E(z_1,z_2) = {2\pi i \Theta_1(\nu_{12}
\vert \tau) \over \sqrt{\omega(z_1)\omega(z_2)}
\Theta^\prime_1 (0\vert \tau) } \ \, , \qquad \nu_{12} =
\int_{z_2}^{z_1} {\omega \over 2\pi i } \ ,
\nfr{aprimef}
where $\omega(z)$ is the holomorphic 1-form on the torus, normalized to
have period $2\pi i$ around the $a$-cycle. In the parametrization where
$\omega(z)=1/z$ the prime form \aprimef\ becomes
$$
E(z_1,z_2) = (z_1-z_2) \prod_{n=1}^\infty {(1-{z_1\over z_2} k^n )
(1-{z_2\over z_1}k^n) \over (1-k^n)^2 } \ .
\efr
The standard Riemann identity is ($\alpha,\beta\, = \{0,\frac12\}$)
$$
\sum_{\alpha,\beta} e^{2\pi i(\alpha+\beta)} \prod_{i=1}^4
\Teta\alpha\beta(x_i\vert\tau) \ =\ 0 \ ,
\nfr{acfor}
where one of the following equations must hold
$$\eqalignno{&x_1+x_2+x_3+x_4 = 0 \qquad\qquad\quad
x_1-x_2-x_3+x_4=0 &\numali\cr
&x_1-x_2+x_3-x_4 = 0 \qquad\qquad\quad x_1+x_2-x_3-x_4 = 0\ .\cr}
$$

\sjump
Having given our notations for the basic functions on the torus, we
can now turn to the correlators.

The spacetime coordinate fields $X^\mu$ satisfy the OPE
$$
X^\mu (z,\bar{z}) X^\nu (w,\bar{w})\ \buildchar{=}{OPE}{ } \ -\delta^{\mu\nu}
\left(\log(z-w) + \log (\bar{z}-\bar{w}) \right)
+\ \cdots \ .
\efr
Their one-loop partition function is given by
$$ Z_{X} = \prod_{n=1}^{\infty} \vert 1 - k^n \vert^{-8} (2\pi {\rm Im}
\tau)^{-2} \ , \efr
and the genus one correlator is
$$ \wew{X^\mu(z,\bar{z}) X^\nu(w,\bar{w})} \ =\ -\delta^{\mu\nu} G_B
(z,\bar{z};w,\bar{w}) \ Z_{X}\ . \efr
(notice that this correlator does not decompose in the product of a
holomorphic times an anti-holomorphic part, only $\del_w\del_z
\vev{X^\mu(z,\bar{z})X^\nu(w,\bar{w})}$ is holomorphic).

For the world-sheet fermions we have the following normalization
$$\eqalignno{ &\psi^\mu(z) \psi^\nu(w) \ \buildchar{=}{OPE}{ }\
{\delta^{\mu \nu} \over z-w}\ +\ \cdots &\numali\cr
&\psi^m_{(l)}(z) \psi^n_{(k)}(w) \ \buildchar{=}{OPE}{ }\
{\delta^{m,n}\delta_{l,k}\over z-w} \ +\ \cdots\ .\cr}
$$
They are
bosonized according to eq.~\bosonization\ and correlation functions are
defined as in eq.~\gencorr. The fundamental genus one correlator [\Ref{PDV1}]
is
$$\vev{\prod_{i=1}^N e^{q_i \phi(z_i)} }
\left[{}^\alpha_\beta\right]\ =\
\delta_{\sum_{i=1}^N q_i,0} \prod_{i<j} \left[ E(z_i,z_j)
\right]^{q_iq_j} \, \Teta\alpha\beta \left(\sum_{i=1}^N q_i \int^{z_i}
\frac\omega{2\pi i}
\vert \tau \right) \ ,  \nfr{aboscorr}
where we have
explicitly displayed the spin structure dependence of the
correlator, whereas in the paper we often adopt the following short-hand
notation
$$
\vev{\prod_{i=1}^N e^{q_i \phi(z_i)} }_{(l)}\ =\
\vev{\prod_{i=1}^N e^{q_i \phi(z_i)} } \left[{}^{\alpha_l}_{\beta_l}
\right]\ .
\efr
We use also the following notation
$$
\vev{S_+(z_1) S_-(z_2)}=\vev{e^{\frac12\phi(z_1)}e^{-\frac12\phi(z_2)}}=
 \left( E(z_1,z_2)\right)^{-\frac14}
\Teta\alpha\beta(\frac12\nu_{12} \vert\tau)\ ,
\efr
and we define the integer
$$
S_l \equiv (1-2\alpha_{l})(1+2\beta_{l}) \ ,
\nfr{sdef}
which is even (odd) whenever the $(l)$th spin structure is even (odd).

\appendix{Ghost and Superghost Correlators}
\setchap{Appghosts}

Since we choose to work in the Lorentz-covariant formulation, all
$1$-loop computations involve also the calculation of certain ghost and
superghost correlators on the torus.

The ghost correlator relevant for our $1$-loop scattering amplitudes
involving $N$ physical external states is
$$\eqalignno{
& {\rm d}^2 k \prod_{i=1}^{N-1} {\rm d}^2 z_i \
\wew{ \left| (\eta_k \vert b) \prod_{i=1}^{N-1}
(\eta_{z_i} \vert b) \prod_{i=1}^N c(z_i) \right|^2} & \nameali{bone} \cr
&\qquad\qquad = \ {{\rm d}^2 k \over \bar{k}^2 k^2} \prod_{i=1}^{N-1}
{\rm d}^2 z_i \left| {1 \over \omega(z_N)} \right|^2 \prod_{n=1}^{\infty}
\left| 1 - k^n \right|^4 \ . \cr}$$
Here $\eta_k, \eta_{z_i}$ are the Beltrami-differentials dual to the moduli
$k\ (=e^{2\pi i \tau})$ and $z_i$, and the point $z_N$ has been fixed using
the translational
invariance of the torus. $\omega$ is the holomorphic one-form,
normalized to have period $2\pi i$ around the $a$-cycle.
If the Picture Changing Operators are inserted at arbitrary points on
the torus other ghost correlators besides \bone\ will in general be
needed~[\Ref{Kaj2}].
However, for the 3-point calculations we consider in this paper
the correlator \bone\ will suffice.

The superghost correlators are most conveniently calculated in the
``bosonized'' formalism [\Ref{FMS}]. They have been given up to overall
numerical factors in ref.~[\Ref{Verlinde},\Ref{Verlinde2}].
However, such overall factors are
not unimportant since they may in general depend on the spin structure.
Therefore, we compute all superghost correlators using the $N$-point
$g$-loop vertex for the bosonized
$(\beta,\gamma)$-system [\Ref{PDV2}] which has been
obtained by the sewing technique [\Ref{PDV3}] and therefore automatically
includes all phase factors required by factorization.

As an example, consider the correlation function
$$\eqalignno{ &
\wew{ \prod_{i=1}^N e^{q_i
\phi(z_i)} \, \xi(z_{N+1}) } \ = & \nameali{btwo} \cr
&\qquad\qquad \prod_{n=1}^{\infty} (1-k^n) \prod_{i=1}^N
(\sigma(z_i))^{-2q_i}
\prod_{i < j} (E(z_i,z_j))^{-q_i q_j}\ \times \cr
&\qquad\qquad \left[ \Teta{\alpha}{\beta} \left( - \sum_{j=1}^N q_j
\int_{z_0}^{z_j}
{\omega \over 2\pi i} + 2 \Delta^{z_0} \vert \tau \right) \right]^{-1} \
, \cr} $$
where $q_1 + \ldots + q_N = 0$ and the insertion of the operator
$\xi(z_{N+1})$ is needed to saturate the integration over the $\xi$ zero
mode, a degree of freedom not present in the original $(\beta,\gamma)$
system. The result \btwo\ is obtained by saturating
the vertex given by eq.~(6.10) of ref.~[\Ref{PDV2}] with the following $N+1$
highest weight states~\note{Assuming $V_i'(0)=1$ for simplicity.}
$$ e^{q_i \phi^{(i)} (0)} \vert 0 \rangle_i \ \ \ {\rm for} \ \ \ i = 1,
\ldots , N  \qquad \qquad {\rm and} \qquad \qquad
e^{\chi^{(N+1)} (0)} \vert 0 \rangle_{N+1} \ . \efr
The result \btwo\ agrees with eq.~(36) of ref.~[\Ref{Verlinde}]
{\it except\/} for an
overall minus sign whenever the spin structure is odd. Since the parity
of the spin structure is modular invariant such a sign can never be
fixed by modular invariance, only by factorization.

At genus one
$$ \Delta^{z_0} = - {1 \over 4\pi i } \log k = - {1 \over 2} \tau \ ,
\nfr{bfive}
and $\sigma$ is a multivalued $1/2$-differential which reduces to
$$ \sigma(z) = 1 \nfr{bsix}
if we choose coordinates such that $\omega(z) = 1/z$.

Using eq.~\thetashift\ for the $\Theta$-functions
we arrive at our final expression
for the correlator \btwo
$$\eqalignno{ &
\wew{ \prod_{i=1}^N e^{q_i
\phi(z_i)} \ \xi(z_{N+1}) } \ = & \nameali{bseven} \cr
&\qquad\qquad (-1)^S k^{1/2}
\prod_{n=1}^{\infty} (1-k^n) \prod_{i=1}^N (\omega(z_i))^{-q_i}
\prod_{i < j} (E(z_i,z_j))^{-q_i q_j}\ \times \cr
&\qquad\qquad \left[ \Teta{\alpha}{\beta} \left( \sum_{j=1}^N q_j
\int_{z_0}^{z_j}
{\omega \over 2\pi i } \vert \tau \right) \right]^{-1} \ . \cr}$$
Here we dropped a phase factor $\exp \{ 2\pi i (1/2 + \beta) \}$ which
is already included in the KLT summation coefficient \phases. The remaining
phase factor $(-1)^S$, where $S = (1-2\alpha)(1+2\beta)$ is even (odd)
whenever the spin structure is even (odd), is crucial in order to obtain
a vanishing Anomalous Magnetic Moment in the supersymmetric case, as
discussed in section \S 4.
\appendix{Covariant Formulation of KLT Formalism}
\setchap{Applightcone}
In this appendix we outline the modifications encountered when
rephrasing the Kawai-Lewellen-Tye $4$-dimensional string models in a
Lorentz-covariant way. In particular, we obtain the covariant form of
the GSO projections.

The Kawai-Lewellen-Tye [\Ref{KLT}] construction of $4$-dimensional string
theories is performed in the light-cone gauge. We have $22$ left-moving
and $10$ right-moving complex fermions, whose $1$-loop partition
function is given by
$$ Z_{\rm fermion} = \sum_{n_i,m_j}
\tilde{C}^{{\bfmath\alpha}}_{{\bfmath\beta}}
Z^{{\bfmath\alpha}}_{{\bfmath\beta}} \ ,
\nfr{cone}
where
$$ Z^{{\bfmath\alpha}}_{{\bfmath\beta}} = {\rm Tr} \left[
\left( \prod_{l=1}^{22}
\bar{k}^{\overline{H}^{(\bar{l})}_{\modone{\bar{\alpha}_l}}} \right)
\left( \prod_{l=1}^{10}
k^{H^{(l)}_{\modone{\alpha_l}}} \right)
e^{2\pi i ({\bf W}_0
+ {\bfmath\beta} ) \cdot {\bf N}_{\modone{{\bfmath\alpha}}} }
\right] \nfr{ctwo}
is the partition function corresponding to a given set of spin
structures, specified as in eq.~\spinstructures\
by the integers $m_i$ and $n_i$. The summation
coefficients
$$ \tilde{C}^{{\bfmath\alpha}}_{{\bfmath\beta}} = {1 \over \prod_i M_i}
\exp \{ -2\pi i [ \sum_i (n_i + \delta_{i,0}) (\sum_j k_{ij} m_j + s_i +
k_{0i} - {\bf W}_i \cdot \modone{{\bfmath\alpha}} ) + \sum_i m_i s_i +
\frac12 ] \} \nfr{KLTcoeff}
are carefully constructed to
ensure modular invariance of $Z_{\rm fermion}$,
as well as the correct spin-statistics
relation (i.e. space-time bosons (fermions) contribute with weight $+1$
$(-1)$ to the partition function). $N_{\modone{{\bfmath\alpha}}}$ is the
vector of fermion number operators, given by eq.~\fermnum\ and the
Hamiltonians are given by
$$ H^{(l)}_{\modone{\alpha_l}} = \tilde{H}^{(l)}_{\modone{\alpha_l}}
+ \frac12((\alpha_l)^2 - \alpha_l + {1 \over 6}) \ , \efr
where
$$ \tilde{H}^{(l)}_{\modone{\alpha_l}} = \sum_{q=1}^{\infty}
((q+\modone{\alpha_l} -1) n^{(l)}_{q+\modone{\alpha_l}-1} +
(q-\modone{\alpha_l}) n^{(l)*}_{q-\modone{\alpha_l}} ) \ . \efr
We may now add the longitudinal complex fermion, $\psi_{(0)}$, and the
superghosts $\beta$ and $\gamma$. By world-sheet supersymmetry, both
carry the same spin structure as the transverse complex fermion,
$\psi_{(1)}$.
The corresponding partition functions are
$$\eqalignno{
& Z_{\psi_{(0)}} \ = \ {\rm Tr} \left[
k^{\tilde{H}^{(0)}_{\modone{\alpha_1}}} e^{-2\pi i
(1/2 + \beta_1) N^{(0)}_{\modone{\alpha_1}}} \right] & \nameali{cthree} \cr
& Z_{(\beta \gamma)} \ = \ {\rm Tr} \left[ k^{\tilde{H}^{(\beta
\gamma)}_{\modone{\alpha_1}}} e^{2\pi i \beta_1 N^{(\beta
\gamma)}_{\modone{\alpha_1}}}
\right] \ , & \nameali{cfour} \cr}$$
where $N^{(\beta\gamma)}_{\modone{\alpha_1}}$ is given by
eq.~\sghostnum\ and
$$ \tilde{H}^{(\beta \gamma)}_{\modone{\alpha_1}} = \sum_{q=1}^{\infty}
\left[ (q-1+\modone{\alpha_1})
\beta_{-q+1-\modone{\alpha_1}} \gamma_{q-1+\modone{\alpha_1}}
- (q-\modone{\alpha_1}) \gamma_{-q+\modone{\alpha_1}}
\beta_{q-\modone{\alpha_1}} \right]  \efr
corresponding to the choice of superghost vacuum
$\vert q' = -1/2 - \modone{\alpha_1} \rangle $, i.e. the superghost part of the
ground state vertex operator is $e^{-\phi}$ in a bosonic  sector
and $e^{-\phi /2}$ in a fermionic sector.

We have carefully chosen the definitions \cthree\ and \cfour\ to ensure that
$$ Z_{\psi_{(0)}} Z_{(\beta \gamma)} = 1 \nfr{ceight}
--- if the two factors did not cancel each other completely, their
inclusion would alter the already correct result \cone.

Using the expression \KLTcoeff\ for the summation coefficients,
the partition function \cone\ can now be written as
$$\eqalignno{
Z_{\rm fermion} = &{1 \over \prod_i M_i} \sum_{m_i,n_i} e^{-
2\pi i [ \sum_i (n_i +
\delta_{i,0}) (\sum_j k_{ij} m_j + s_i + k_{0i} - {\bf W}_i \cdot
\modone{{\bfmath\alpha}} ) + \sum_i m_i s_i + 1/2 ]}  \cr
& {\rm Tr} \left[ \left(\prod_{l=1}^{22}
\bar{k}^{\overline{H}^{(\bar{l})}_{\modone{\bar{\alpha}_l}}} \right)
\left( \prod_{l=1}^{10} k^{H^{(l)}_{\modone{\alpha_l}}} \right)
k^{\tilde{H}^{(0)}_{\modone{\alpha_1}}} k^{\tilde{H}^{(\beta
\gamma)}_{\modone{\alpha_1}}} e^{2\pi i
\sum_i (n_i + \delta_{i,0} ) {\bf W}_i \cdot {\bf
N}_{\modone{{\bfmath\alpha}}} } \right.
\cr & \left. e^{-2\pi i (1/2 + \sum_i n_i s_i )
N^{(0)}_{\modone{\alpha_1}} }
e^{2\pi i \sum_i n_i s_i N^{(\beta \gamma)}_{\modone{\alpha_1}} } \right]
\ . &\nameali{cnine} \cr}$$
Summing over the $n_i$ enforces the GSO projections in the loop. We
therefore arrive at the covariant form of the GSO projection conditions
$$ {\bf W}_i \cdot {\bf N}_{\modone{{\bfmath\alpha}}} - s_i
(N^{(0)}_{\modone{\alpha_1}} - N^{(\beta
\gamma)}_{\modone{\alpha_1}} ) \eqmodone \sum_j k_{ij} m_j + s_i + k_{0i}
- {\bf W}_i \cdot \modone{{\bfmath\alpha}}  \ . \nfr{cten}
The physical external states are in the superghost vacuum (with charge
$-1$ or $-1/2$) and therefore have $N^{(\beta
\gamma)}_{\modone{\alpha_1}} = 0$.
\appendix{Normalization of the ``Electron/Positron'' Vertex Operator}
\setchap{Appnormvert}
Factorization dictates that the problem of normalizing string amplitudes
can be separated into two independent problems: One, to fix the
normalization constant $C_g$ of the vacuum amplitude at genus $g$. The
other, to fix the normalization of each vertex operator in the theory.

The constant $C_g$ as well as the overall normalization of the photon
vertex operator has been derived in ref.~[\Ref{Kaj}] and are given by
eqs.~\vacuumnorm\ (for $g=1$) and \photonn\ respectively. In this appendix we
consider the normalization of the fermion vertex operator \fermion\
$$ {\cal V}^{-1/2}(z,\bar{z};k;{\Bbb V}) = N_f \overline{\bf V}^{\bar{a}}
\bar{S}_{\bar{a}} (\bar{z}) {\bf V}^a S_a (z) e^{-\phi(z)/2}
(c_{(11)})^{-1/2} e^{i k \cdot X(z,\bar{z})} \ . \nfr{electron}
Obviously, the quantity $N_f$ is not well defined until we have
specified our conventions for the spinors $\overline{\bf V}^{\bar{a}}$ and
${\bf V}^a$. This we will do in the following way: First we define the
spinors in the case of any incoming string state \electron\
with negative $U(1)$ charge (any incoming ``electron''). Next, we
define the anti-particle state (``positron'') corresponding to each of
these incoming ``electron'' states. Finally we define the out-going
state corresponding to any incoming state.

Having carefully normalized the spinors we may determine $N_f$ using the
method of ref.~[\Ref{Kaj}]: We consider the elastic scattering of a
photon and an
``electron'' at very high center-of-mass energies, where the
interactions are dominated by gravity, and require that the
tree-level amplitude for this process should reproduce the standard one
dictated by the principle of equivalence. This will yield an expression
for $N_f$ in terms of the gravitational coupling $\kappa$. Finally, to
relate $\kappa$ to the $U(1)$ charge, $e$, we consider the tree-level
amplitude of two ``electrons'' and a photon and require that we
reproduce the standard Yukawa coupling.

We write the spinor pertaining to an incoming ``electron'' as follows:
$$ {\Bbb V}^{-}_{\rm in} (k,s;\{\bar{q}_l\},f) \equiv
\overline{\bf V}^{-} (\{ \bar{q}_l \}) \otimes
{\bf P} V^{-}_{\rm in}(k,s) \otimes v_{(4)}^{-} \otimes
v_{(567)}^{-}(\{\bar{q}_l\},f) \otimes v_{(9)}^f \ . \nfr{inelectron}
Here we imagine
$\overline{\bf V}^{-}(\{ \bar{q}_l \})$
to be an eigenvector of the various ``charge''
operators $\sigma_3^{(\bar{l})}$, with eigenvalues $\bar{q}_l$,
$l=1,\dots,7;15,16$ and with $U(1)$-charge $\sigma_3^{(\overline{17})} =
-1$, normalized such that
$$ \left( \overline{\bf V}^{-}(\{ \bar{q}_l \}) \right)^{*} =
\overline{\bf V}^- (\{ \bar{q}_l \}) \qquad {\rm and} \qquad
\left( \overline{\bf V}^{-}(\{ \bar{q}_l \}) \right)^{\dagger}
\overline{\bf V}^{-}(\{\bar{q}_l\}) = 1 \ . \nfr{normone}
The projection operator
$$ {\bf P} = \frac12 ( {\bf 1} - \exp \{ 2\pi i [ k_{00} + k_{01} +
k_{03} + k_{13} ] \} \Gamma^5 \sigma_3^{(5)} ) \efr
enforces the first of the four GSO conditions given by eqs.~\gsospinor .
The space-time spinor $V^{-}_{\rm in}(k,s)$ satisfies the equation
$$ (\slashchar{k}^T - i \, ({\rm sign}) {1 \over \sqrt{2}} ) V^{-}_{\rm
in} (k,s) = 0
\nfr{almostdirac}
with~\note{In this appendix we adopt the cocycle choice \ccocycle\
throughout.}
$$ ({\rm sign}) = - \exp\{ 2\pi i [ k_{00} + k_{01} + k_{02} + k_{03} +
k_{04} + k_{12} + k_{14} + k_{23} + k_{34} ] \} \ . \efr
If we define new gamma matrices by
$$ \gamma^{\mu} \equiv -i (\Gamma^{\mu})^T \ , \efr
eq.~\almostdirac\ becomes the ordinary Dirac equation (in the notation
of ref.~[\Ref{IZ}]) and we may identify $V^-_{\rm in} (k,s)$ with the
standard $u$ spinors as follows
$$ V_{\rm in}^- (k,s) = \left\{  \matrix{u(p,s) & {\rm if} & ({\rm
sign}) = +1 \cr
\Gamma^5 u(p,s) & {\rm if} & ({\rm sign}) = -1 \cr } \right. \ ,
\efr
where $p = \sqrt{\alpha' \over 2} k$ is the dimensionful momentum and $s
= \pm 1/2$ is the spin in the rest frame.

The second GSO condition in \gsospinor\ specifies
$\sigma_3^{(4)}$ in terms of $\sigma_3^{(\overline{17})}=-1$,
and hence $v_{(4)}^{-}$ up to an overall constant.
Likewise, the ``family label'' $f = \pm 1$,
defined as the eigenvalue of $\sigma_3^{(9)}$, specifies $v_{(9)}^f$
up to a normalization constant.

We may normalize $v_{(4)}^-$ and $v_{(9)}^f$ in analogy with \normone
$$  (v_{(4)}^{-})^* = v_{(4)}^-  \qquad {\rm and} \qquad
(v_{(4)}^{-})^{\dagger} v_{(4)}^{-} = 1 \efr
$$ (v_{(9)}^{f})^* = v_{(9)}^f \qquad {\rm and} \qquad
(v_{(9)}^{f})^{\dagger} v_{(9)}^{f} = 1 \ . \efr
Finally, the spinor in the
spaces $(5)$, $(6)$ and $(7)$, $v_{(567)}^{-} (\{ \bar{q}_l \}; f)$,
satisfies the ``mass eigenvalue equation''
$$ -\frac12( \sigma_2^{(5)} \sigma_1^{(6)} \sigma_1^{(7)} +
\sigma_1^{(5)} \sigma_2^{(6)} \sigma_2^{(7)} ) \ v_{(567)}^{-} (\{
\bar{q}_l \}; f) = {1 \over
\sqrt{2}} \, v_{(567)}^{-} (\{ \bar{q}_l \}; f) \nfr{masseigen}
and is in fact specified up to an overall constant by the third and
fourth GSO projections in \gsospinor , once all charges and the ``family
label'' $f$ have been specified. It would be inconsistent
with eq.~\masseigen\ to take $v_{(567)}^- (\{ \bar{q}_l \}; f)$
to be real. Instead it is consistent to impose the Majorana-like
condition
$$ \sigma_3^{(5)} C_{(567)} v_{(567)}^- = (v_{(567)}^-)^* \efr
and
$$ (v_{(567)}^-)^{\dagger} v_{(567)}^- = 1 \ . \efr
Here $C_{(567)} \equiv \sigma_1^{(5)} \sigma_2^{(6)} \sigma_1^{(7)}$. Notice
that by the antisymmetry of $C_{(567)}$
$$ (v_{(567)}^-)^T C_{(567)} v_{(567)}^- = 0 \ . \efr

Given the spinor \inelectron , describing an incoming ``electron''
string state with $SO(14)$ and $SO(4)$ charges $\{ \bar{q}_l \}$, family
label $f$, momentum $p$ and spin $s$, we define the corresponding
incoming ``positron'' string state by
$$ {\Bbb V}^+_{\rm in} (k,s;\{ - \bar{q}_l \}; -f) = {{\bfmath \Sigma}}
\,
{\Bbb V}^-_{\rm in} (k,s;\{ \bar{q}_l \};f) \ , \efr
where the operator
$$ {\bfmath \Sigma} \equiv ({\rm sign}) \ C_{SO(14)} \Gamma_{SO(14)}
\otimes C_{SO(4)} \Gamma_{SO(4)} \otimes \sigma_1^{(\overline{17})}
\sigma_3^{(\overline{17})} \otimes \sigma_1^{(4)} \otimes \sigma_3^{(5)}
\otimes \sigma_1^{(9)} \nfr{chargeconj}
changes sign on all charges, as well as the ``family label'', and
commutes with the GSO projection operators.

Similarly, given any incoming ``electron'' or ``positron'' state with
certain $SO(14)$ and $SO(4)$ charges $\{ \bar{q}_l \}$, ``family label''
$f$, four-momentum $p$ and spin $s$, we define the outgoing state with
the same quantum numbers by
$$ {\Bbb V}^{\pm}_{\rm out} (-k,s;\{ \bar{q}_l \},f) = {\Bbb T} \left(
{\Bbb V}^{\pm}_{\rm in} (k,s; \{ \bar{q}_l \} ; f ) \right)^{*} \ , \efr
where
$$ {\Bbb T} \equiv - e^{-i \pi \varphi_{\rm c}} \ \Gamma_{SO(14)} \,
\Gamma_{SO(4)} \, \sigma_3^{(\overline{17})} \, {\Bbb C} \,
\Gamma^0 \nfr{inout}
also commutes with all GSO projection operators. The definitions
\chargeconj\ and \inout\ are chosen so as to reproduce standard field
theory results for the pair annihilation and brehmsstrahlung
tree-level amplitudes, see below.

We may now compute the tree-level amplitudes of various processes, using
the formula (in the notation of section \S 2):
$$\eqalignno{ & T^{\rm tree}(\lambda_1, \dots, \lambda_{N_{\rm out}} \vert
\lambda_{N_{\rm out} + 1 } , \dots, \lambda_{N_{\rm out} +
N_{\rm in}} ) \ = \ - {4 \pi^3 \over \alpha' \kappa^2} \int
\prod_{i=1}^{N_{\rm tot}-3} {\rm d}^2 z_i
& \numali \cr
&  \wew{  \left| \prod_{i=1}^{N_{\rm tot}-3}
(\eta_{z_i} \vert b) \prod_{i=1}^{N_{\rm tot}} c(z_i) \right|^2
\prod_{A=1}^{N_B + N_{FP} -2} \Pi(w_A) \ {\cal V}_{\langle \lambda_1
\vert } (z_1,\bar{z}_1) \dots
{\cal V}_{\vert \lambda_{N_{\rm tot}} \rangle } (z_{N_{\rm
tot}},\bar{z}_{N_{\rm tot}} )} \ . \cr} $$
For the elastic scattering of ``electrons/positrons'' with photons we
find in the limit $s \rightarrow \infty$, $t \rightarrow 0$ (where $s$
and $t$ are the usual Mandelstam variables) the correct result
[\Ref{Veneziano}]
$$ \langle \gamma \ e^{\pm} \vert T \vert \gamma \ e^{\pm} \rangle_{\rm
tree} = - \kappa^2 {s^2 \over t} \efr
if we make the identification
$$ (N_f)^2 = - {\kappa^2 \sqrt{\alpha'} \over \pi^2 } e^{-i \pi \varphi_{\rm
c}} \ . \efr
For the pair-annihilation and brehmsstrahlung processes we recover the
standard field theory results
$$ T^{\rm tree} (\gamma \vert  e^{\pm} \ e^{\mp} )
= \pm e u^T (p_1,s_1) C \gamma^{\mu} \epsilon_{\mu} u(p_2,s_2) \efr
$$ T^{\rm tree} (e^{\pm} \ \gamma \vert  e^{\pm} ) =
\pm e u^{\dagger} (-p_1,s_1) \gamma^0 \gamma^{\mu} \epsilon_{\mu}
u(p_2,s_2) \ , \efr
where $e = -|e|$ is the $U(1)$ charge of the ``electron'', if we
identify
$$ |e| = {\kappa \over \sqrt{2 \alpha'}}\ .\efr

\appendix{Proof of the equality of the two expressions for the
$\II\alpha\beta$ function}
\setchap{AppIIfunct}
In this appendix we present a proof of the simplest of the identities
in theta functions that we encountered in subsection \S \sectccorr ,
arising from the requirement of Lorentz covariance
of the genus one correlators appearing in the one-loop three-point
amplitude. The other identities needed in that
computation can be proven in a similar way.

The identity we want to prove can be stated as the fact that on the
torus
$$ L\left[{}^\alpha_\beta\right] (z,z_1,z_2\vert\tau)\ =\
 R\left[{}^\alpha_\beta\right] (z,z_1,z_2\vert\tau) \ ,
\efr
where
$$\eqalignno{& L\left[{}^\alpha_\beta\right] (z,z_1,z_2\vert\tau) =
\del_z\log{E(z,z_1)\over E(z,z_2)} +2 {\omega(z)\over 2\pi i}
\del_\nu \log \Teta\alpha\beta (\nu\vert\tau)\vert_{\nu=\frac12\nu_{12}}
&\nameali{lrdef} \cr
&R\left[{}^\alpha_\beta\right] (z,z_1,z_2\vert\tau) =
{E(z_1,z_2)\over E(z,z_1) E(z,z_2) }\left( {\Teta\alpha\beta(\nu_z
-\frac12\nu_{z_1} -\frac12\nu_{z_2} \vert\tau )\over
\Teta\alpha\beta (\frac12\nu_{12}\vert\tau) } \right)^2 \cr
& \nu_z = \int^z {\omega\over 2\pi i} \cr
& \nu_{12} = \int^{z_1}_{z_2} {\omega\over 2\pi i}\ . \cr}
$$
The proof consists in showing that $L$ and $R$,
considered as meromorphic one-forms in $z$ (for every fixed value of $z_1$,
$z_2$ and $\tau$), have the same periodicity properties,
zeros, poles and residues
at the poles. Then
$$
{ L\left[{}^\alpha_\beta\right] (z,z_1,z_2\vert\tau) \over
 R\left[{}^\alpha_\beta\right] (z,z_1,z_2\vert\tau) }
\nfr{Eoneidentity}
is a single-valued,
globally holomorphic function of $z$, that is a constant. Since $L$
and $R$ have the same residues, the constant is $1$, thus proving the
identity.
So, we need to study the periodicity properties,
zeros and poles of $R$ and $L$ as
one-forms in $z$.

First of all, using the formul\ae\ given in Appendix \Appconv, we can
rewrite $L$ and $R$ as follows
$$\eqalignno{& L\left[{}^\alpha_\beta\right] (z,z_1,z_2\vert\tau) =
{\omega(z)\over 2\pi i} \bigl( \del_\nu \log \Theta_1 (\nu\vert\tau)
\vert_{\nu=\nu_z-\nu_{z_1}} \ -&\numali\cr
&\qquad\qquad\qquad\qquad\ \left.\del_\nu\log \Theta_1 (\nu\vert\tau)
\vert_{\nu_z - \nu_{z_2}} +2 \del_\nu \log \Teta\alpha\beta
(\nu\vert\tau)\vert_{\nu=\frac12\nu_{12}} \right)\cr
&R\left[{}^\alpha_\beta\right] (z,z_1,z_2\vert\tau) =
{\omega(z)\over 2\pi i} \, {\Theta_1(\nu_{z_1} -\nu_{z_2} \vert\tau )
\Theta^\prime_1(0\vert\tau) \over \Theta_1(\nu_z-\nu_{z_1} \vert\tau)
\Theta_1(\nu_z-\nu_{z_2} \vert\tau) }\ \times\cr
&\qquad\qquad\qquad\qquad\ \left( {\Teta\alpha\beta(\nu_z
-\frac12\nu_{z_1} -\frac12\nu_{z_2} \vert\tau )\over
\Teta\alpha\beta (\frac12\nu_{12}\vert\tau) } \right)^2\ .\cr}
$$

Now, using formula \thetashift\ one can show that $L$ and $R$ are both
single-valued on the torus, that is, under the shift $\nu_z \rightarrow
\nu_z +m\tau +n$ ($m,n$ integers) they are invariant.

As a function of $z$, it is obvious from eq.~\lrdef\ that both $L$ and $R$
have poles at $z=z_1$ and $z=z_2$, with residues $+1$ and
$-1$ respectively.

Thus, what is left to be proven is that $L$ and $R$ have the same
zeros. For $R$, a zero in $z$ can come only from the factor
$$
\left( \Teta\alpha\beta(\nu_z
-\frac12\nu_{z_1} -\frac12\nu_{z_2} \vert\tau )\right)^2\ .
\efr
Using formula \thetazero\ and the fact that $\Theta_1(m\tau+n\vert\tau)=0$,
we get that $R$ has a double zero when
$$
\nu_z -\frac12\nu_{z_1} -\frac12 \nu_{z_2} -\alpha\tau +\beta +
m\tau + n\ =\ 0
\nfr{Ezerocond}  
or
$$
\nu_z\ =\ \frac12\nu_{z_1} +\frac12 \nu_{z_2} +\alpha\tau -\beta -
m\tau - n\ \equiv\ \nu_0 -m\tau -n\ .
\efr

\sjump
We now look for the zeros of $L$. First it is convenient
to write
$$
\del_\nu \log \Teta\alpha\beta (\nu\vert\tau) \ =\ \del_\nu \log\Theta_1
(\nu-\alpha\tau+\beta\vert\tau) - 2\pi i \alpha
\efr
and to introduce
$$
x\ =\ \nu_z-\nu_0\ .
\efr
Thus
$$\eqalignno{
&L\left[{}^\alpha_\beta\right] (z,z_1,z_2\vert\tau) =
{\omega(z)\over 2\pi i} \left( \del_\nu \log \Theta_1 (\nu\vert\tau)
\vert_{\nu=x-\frac12\nu_{12} +\alpha\tau-\beta}\ - \right.&\numali\cr
&\qquad\qquad \left. \del_\nu\log \Theta_1 (\nu\vert\tau)
\vert_{\nu=x+\frac12\nu_{12} + \alpha\tau-\beta} +2 \del_\nu \log
\Theta_1(\nu\vert\tau)\vert_{\nu=\frac12\nu_{12}-\alpha\tau+\beta}
-4\pi i \alpha\right)\ .\cr}
$$
Now, as long as $2\alpha$ and $2\beta$ are integers, it follows from
\thetashift\ that
$$\eqalignno{
& \del_\nu \log\Theta_1(\nu\vert\tau) \vert_{\nu=x+\frac12\nu_{12} +
\alpha\tau -\beta} = \del_\nu\log\Theta_1 (\nu\vert\tau) \vert_{\nu=x
+\frac12 \nu_{12} -\alpha\tau+\beta} - 4\pi i \alpha \cr
& \del_\nu \log\Theta_1(\nu\vert\tau) \vert_{\nu=x-\frac12\nu_{12} +
\alpha\tau -\beta} = - \del_\nu\log\Theta_1 (\nu\vert\tau) \vert_{\nu=-x
+\frac12 \nu_{12} -\alpha\tau+\beta} & \numali  \cr } $$
from which we get
$$\eqalignno{
L\left[{}^\alpha_\beta\right] (z,z_1,z_2\vert\tau)& =
{\omega(z)\over 2\pi i} \left( -\del_\nu \log \Theta_1 (\nu\vert\tau)
\vert_{\nu=-x+\frac12\nu_{12} -\alpha\tau+\beta}\ - \right.
&\nameali{Ezeroleft}\cr
&\left. \del_\nu\log \Theta_1 (\nu\vert\tau)\vert_{\nu=x+\frac12\nu_{12}
- \alpha\tau+\beta} +2 \del_\nu \log \Theta_1(\nu\vert\tau)
\vert_{\nu=\frac12\nu_{12}-\alpha\tau+\beta} \right)\ .\cr}
$$
Thus $L$ vanishes when $x=0$, and then, by periodicity on the torus,
whenever equation \Ezerocond\ holds.
Since $L$ is a meromorphic one-form on the torus, the number of zeros
(counted with multiplicities) must be equal to the
number of poles, that is 2.
Now, the quantity in the bracket in eq.~\Ezeroleft\ is an even function
of $x$, so the zero $x=0$ must be at least of second order.
This shows that the zero is precisely second
order and also, that there are no other zeros.

Thus, we have proven that the one-forms $L(z)$ and $R(z)$
have the same zeros, poles and residues at the poles. This concludes
our proof of the identity \Eoneidentity.

\references
\beginref
\Rref{BK}{Z.~Bern and D.~Kosower, Nucl.Phys. {\bf B379} (1992) 451.}
\Rref{FP}{S.~Ferrara and M.~Porrati, Phys.Lett. {\bf B288} (1992) 85.}
\Rref{KLT}{H.~Kawai, D.C.~Lewellen and S.-H.H.~Tye, Nucl.Phys.
{\bf B288} (1987) 1.}
\Rref{Koste}{V.A.~Kostelecky, O.~Lechtenfeld, W.~Lerche, S.~Samuel and
S.~Watamura, Nucl.Phys. {\bf B288} (1987) 173.}
\Rref{Atick}{J.J.~Atick, L.J.~Dixon and A.~Sen, Nucl.Phys. {\bf B292}
(1987) 109;\newline J.J~Atick and A.~Sen, Phys.Lett. {\bf 186B} (1987) 339,
Nucl.Phys. {\bf B286} (1987) 189, Nucl.Phys. {\bf B293} (1987) 317.}
\Rref{Verlinde}{E.~Verlinde and H.~Verlinde, Phys.Lett. {\bf B192}
(1987) 95.}
\Rref{PDV1}{P.~Di Vecchia, M.L.~Frau, K.~Hornfeck, A.~Lerda, F.~Pezzella
and S.~Sciuto, Nucl.Phys. {\bf B322} (1989) 317.}
\Rref{PDV2}{P.~Di Vecchia, invited talk at the Workshop on
String quantum Gravity and Physics at the Planck scale, Erice,
June 1992, Int.Journ.Mod.Phys. {\bf A}.}
\Rref{Verlinde2}{E.~Verlinde and H.~Verlinde, Nucl.Phys. {\bf B288}
(1987) 357.}
\Rref{Anto}{I.~Antoniadis, C.~Bachas, C.~Kounnas and P.~Windey,
Phys.Lett. {\bf 171B} (1986) 51; \newline
I.~Antoniadis, C.~Bachas and C.~Kounnas, Nucl.Phys. {\bf B289} 87;
\newline I.~Antoniadis and C.~Bachas, Nucl.Phys. {\bf B298} (1988) 586.}
\Rref{Kaplu}{V.S.~Kaplunovsky, Nucl.Phys. {\bf B307} (1988) 145,
\hbox{hep-th/9205070}.}
\Rref{FMS}{D.~Friedan, E.~Martinec and S.~Shenker, Nucl.Phys. {\bf B271}
(1986) 93.}
\Rref{Pott}{V.A.~Kostelecky and R.~Potting, Nucl.Phys. {\bf B359} (1991)
545, preprint \hbox{hep-ph/9211116}.}
\Rref{Sonoda}{H.~Sonoda, Nucl.Phys. {\bf B326} (1989) 135.}
\Rref{Berera}{A.~Berera, Nucl.Phys. {\bf B411} (1994) 157.}
\Rref{Weinberg}{S.~Weinberg, ``Radiative Corrections in String Theory",
talk given at the APS meeting 1985, in DPF Conf. 1985, 850.}
\Rref{Hoker}{K.~Aoki, E.~D'Hoker and D.H.~Phong, Nucl.Phys. {\bf B342}
(1990) 149;\newline
E.~D'Hoker and D.H.~Phong, Phys.Rev.Lett. {\bf 70} (1993) 3692, preprint
\hbox{hep-th/9404128}, preprint \hbox{hep-th/9410152}.}
\Rref{Weisberger}{J.L.~Montag and W.I.~Weisberger, Nucl.Phys. {\bf B363}
(1991) 527.}
\Rref{GSW}{M.B.~Green, J.H.~Schwarz and E.~Witten, {\sl Superstring Theory},
Cambridge University Press, 1987.}
\Rref{Phong}{For a review, see
E.~D'Hoker and D.H.~Phong, Rev.Mod.Phys. {\bf 60} (1988) 917.}
\Rref{Kaj}{G.~Cristofano, R.~Marotta and K.~Roland, Nucl.Phys. {\bf
B392} (1993) 345.}
\Rref{Us}{A.~Pasquinucci and K.~Roland, in preparation.}
\Rref{Bluhm}{R.~Bluhm, L.~Dolan and P.~Goddard, Nucl.Phys. {\bf B309}
(1988) 330.}
\Rref{PDV3}{P.~di~Vecchia, in ``{\it Physics and Mathematics of Strings}'',
V.~Knizhnik Memorial Volume, World Scientific, Singapore, 1990.}
\Rref{Fischler}{W.~Fischler and L.~Susskind, Phys.Lett. {\bf B171}
(1986) 383, Phys.Lett. {\bf B173} (1986) 262.}
\Rref{Tseytlin}{A.A.~Tseytlin, in ``{\sl Trieste Superstrings 1989\/}'',
487-549, and references cited therein.}
\Rref{Kaj2}{K.~Roland, Phys.Lett. {\bf B312} (1993) 441.}
\Rref{Martinec}{E.~Martinec, Nucl.Phys. {\bf B281} (1987) 157.}
\Rref{Veneziano}{D.~Amati, M.~Ciafaloni and G.~Veneziano, Phys.Lett.
{\bf B197} (1987) 81.}
\Rref{IZ}{C.~Itzykson and J.-B.~Zuber, ``{\sl Quantum Field Theory}",
McGraw-Hill, New York, 1980.}
\endref
\ciao
